\documentclass[letterpaper,11pt]{article}
\usepackage[utf8]{inputenc}
\usepackage{amsmath}
\usepackage{amsthm}
\usepackage{mathrsfs}
\usepackage{fullpage}
\usepackage{color}
\usepackage{bbold}
\newtheorem{theorem}{Theorem}[section]
\newtheorem{corollary}{Corollary}[section]
\newtheorem{lemma}[theorem]{Lemma}

\newtheorem{claim}[theorem]{Claim}
\theoremstyle{definition}

\newtheorem{definition}{Definition}
\usepackage{algorithm}
\usepackage{graphicx}
\usepackage{url}
\usepackage{subcaption}
\definecolor{Blue}{rgb}{0.0,0.2,0.6}
\usepackage{hyperref}
\hypersetup{
    linktocpage=true,
    colorlinks=true,				
    linkcolor=Blue,				
    citecolor=Blue,				
    urlcolor=Blue,			
}

\newcommand{\dist}{\mathop{\mathrm{dist}}}
\newcommand{\LC}{\mathop{\mathrm{L}}}
\newcommand{\dH}{\mathop{\mathrm{dist}_{\text{H}}}}
\newcommand{\dV}{\mathop{\mathrm{dist}_{\text{V}}}}
\newcommand{\gE}{\mathop{g_{\text{E}}}}
\newcommand{\gL}{\mathop{g_{\text{L}}}}

\title{Smooth Sensitivity for Geo-Privacy}

\author{
Yuting Liang\thanks{Computer Science and Engineering, Hong Kong University of Science and Technology. \href{mailto:yliangbs@cse.ust.hk}{yliangbs@cse.ust.hk}}
\and
Ke Yi\thanks{Computer Science and Engineering, Hong Kong University of Science and Technology. \href{mailto:yike@cse.ust.hk}{yike@cse.ust.hk}}
}
\date{}
\begin{document}
\maketitle
\begin{abstract}
Suppose each user $i$ holds a private value $x_i$ in some metric space $(U, \dist)$, and an untrusted data analyst wishes to compute $\sum_i f(x_i)$ for some function $f : U \rightarrow \mathbb{R}$ by asking each user to send in a privatized $f(x_i)$.  This is a fundamental problem in privacy-preserving population analytics, and the local model of differential privacy (LDP) is the predominant model under which the problem has been studied.  However, LDP requires any two different $x_i, x'_i$ to be $\varepsilon$-distinguishable, which can be overly strong for geometric/numerical data.  On the other hand, Geo-Privacy (GP) stipulates that the level of distinguishability be proportional to $\dist(x_i, x_i')$, providing an attractive alternative notion of personal data privacy in a metric space. However, existing GP mechanisms for this problem, which add a uniform noise to either $x_i$ or $f(x_i)$, are not satisfactory.  In this paper, we generalize the smooth sensitivity framework from Differential Privacy to Geo-Privacy, which allows us to add noise tailored to the hardness of the given instance.  We provide definitions, mechanisms, and a generic procedure for computing the smooth sensitivity under GP equipped with a general metric. Then we present three applications: one-way and two-way threshold functions, and Gaussian kernel density estimation, to demonstrate the applicability and utility of our smooth sensitivity framework.
\end{abstract}

\section{Introduction}
Differential Privacy (DP) is one of the most widely adopted models for personal privacy protection, where it has been deployed by large companies and government agencies \cite{appledp2017,ding2017collecting,erlingsson2014rappor,abowd2018us}. It has received extensive attention from the privacy research community, and many useful tools have been developed. 
However, a major issue with the DP definition is that it can be overly strong for geometric/numerical data, especially in the local model.  Consider a local-DP setting where each user $i$ has a private value $x_i \in \mathbb{R}$ (e.g., annual income), and an untrusted data curator wishes to collect them.  This problem is unsolvable under local-DP, which requires the message from a user to be $\varepsilon$-distinguishable between \textit{any} two different values $x_i$ may take.  Since the domain $\mathbb{R}$ is unbounded, the user must add an infinite amount of noise to $x_i$.

\begin{table*}
    \caption{A comparison of different approaches for computing ${1\over n} \sum_{i=1}^n f(x_i)$.}
    \label{tab:comparison}
\centering
\resizebox{\textwidth}{!}{
    \begin{tabular}{c|cccc}
    \hline
         & \vtop{\hbox{\strut LDP: }\hbox{\strut $f(x_i) + \mathrm{Lap}\left({\Delta_f\over \varepsilon}\right)$}}  & \vtop{\hbox{\strut GP:}\hbox{\strut $f\left(x_i+\mathrm{Lap}({1\over \varepsilon})\right)$}}  & \vtop{\hbox{\strut GP:}\hbox{\strut $f(x_i)+\mathrm{Lap}\left({K_f\over \varepsilon}\right)$}}  & \vtop{\hbox{\strut GP:}\hbox{\strut $f(x_i) + \mathrm{Noise}\left({B_f^*(x_i) \over \varepsilon}\right)$}}  \\
         \hline
         \hline
    Distinguishability & $\varepsilon$ & $\varepsilon\cdot \dist(x_i,x_i')$ & $\varepsilon\cdot \dist(x_i,x_i')$ & $\varepsilon\cdot \dist(x_i,x_i')$\\
    \hline
    Noise & a posterior & a priori & a posterior &a posterior  \\
    \hline
    Bias & $0$ & $>0$ for nonlinear $f$ & 0 & 0\\
    \hline
    Variance & ${1\over \varepsilon^2n} \Delta_f^2$ & ${1\over n^2} \sum_i (E[f(x_i+\mathrm{Lap}({1\over \varepsilon}))]-f(x_i))^2$ & ${1\over \varepsilon^2 n} K_f^2$ & ${1\over (\varepsilon n)^2} \sum_i B_f^*(x_i)^2$  \\ 
    \hline
    \end{tabular}}
\end{table*}

\paragraph{Geo-Privacy}
An elegant solution to addressing the issue above is Geo-Privacy (GP) \cite{chatzikokolakis2013broadening,andres2013geo,liang2023concentrated}.  Its key argument is that the level of distinguishability should not be the same for all pairs of inputs; instead, it shall be proportional to their distance.  More generally, it employs a metric space $(U,\dist)$, and requires the messages be $(\varepsilon\cdot \dist(x_i, x_i'))$-distinguishable on any pair of inputs $x_i,x_i'\in U$ (formal definition is given in Section \ref{sec:prelim}).  Note that local-DP is actually a special case of GP by taking $\dist(\cdot,\cdot)$ to be the discrete metric $\dist(x_i,x_i')=\mathbb{1}\{x_i\ne x'_i\}$, but for geometric/numerical data, the Euclidean distance is usually a more appropriate choice.  Under the Euclidean metric, it has been shown that simply adding $\mathrm{Lap}(1/\varepsilon)$, i.e., a Laplace noise of scale $1/\varepsilon$, to $x_i$ satisfies $\varepsilon$-GP \cite{andres2013geo}.  

While most of the work under GP studies how to collect the raw data $\{x_i\}_i$, in this paper we are interested in computing $\sum_i f(x_i)$ or ${1 \over n} \sum_i f(x_i)$ for some function $f:U \rightarrow \mathbb{R}$.  In fact, the latter is a more common problem in privacy-preserving population analytics, and most DP work falls under this topic.  As an example, suppose $x_i$ represents the annual income of user $i$ and an data analyst is interested in estimating the proportion of users with income higher than some threshold $T$. Then $f$ is the threshold function $f(x):=\mathbb{1}\{x > T\}$.  Another example is kernel density estimation (KDE).  Assuming the Gaussian kernel, the estimate at point $t$ is ${1\over n} \sum_i f_t(x_i)$ where $f_t(x) := e^{-{\|x-t\|^2} \over 2h^2}$.  Under local-DP, both problems can be solved by asking each user to send in $f(x_i) + \mathrm{Lap}(\Delta_f/\varepsilon)$.  Here, $\Delta_f := \sup_{x,x'}|f(x)-f(x')|$ is known as the global sensitivity of $f$, which is equal to $1$ for both problems.  This leads to an unbiased estimator of ${1 \over n} \sum_i f(x_i)$ with variance $1/(\varepsilon^2 n)$.  However, under GP equipped with the Euclidean metric, there is potential to do much better: Suppose all income numbers are much higher than $T$, then the users may all report $1$ (with high probability) without much impact on their privacy.  Existing GP mechanisms fall under the following two categories.

\paragraph{Noise a priori}
The first solution is that each user submits a noisy $\tilde{x}_i = x_i + \mathrm{Lap}(1/\varepsilon)$, which satisfies GP, and the data analyst computes ${1 \over n} \sum_{i} f(\tilde{x}_i)$. 
However, there are two issues with this \textit{noise-a-priori} solution.  First, although releasing $\tilde{x}_i$ satisfies GP, this  reveals more information than necessary for the purpose of estimating ${1 \over n} \sum_i f(x_i)$.  In the high-income example above, a person with income much higher than $T$ would be much more comfortable with a ``yes'' (w.h.p.) answer than revealing their noisy income.  The second issue is that, although $E[\tilde{x_i}] = x_i$, we no longer have 
$E[\mathbb{1}\{\tilde{x}_i>T\}] = \mathbb{1}\{x_i>T\}$.  There is an inherent bias depending on $x_i - T$ (more details given in Section \ref{sec:threshold}). Note that since the bias depends on the private $x_i$, it cannot be removed.  More generally, this is an issue for every nonlinear function $f$ such as the KDE function.  The consequence is that the data analyst would get a biased estimator for ${1\over n}\sum_i f(x_i)$.  A bias in the estimator would always remain regardless of $n$, the number of users that participate.  In contrast, in most statistical analyses, we would like to have a consistent estimator that approaches the true answer as $n\rightarrow \infty$.

\paragraph{Noise a posterior}
Thus, in this paper we are more interested in \textit{noise-a-posterior} approaches that add noise to $f(x_i)$ instead of $x_i$.  The GP counterpart  \cite{chatzikokolakis2013broadening} of the local-DP solution mentioned above is to use $K_f$ in place of $\Delta_f$, where $K_f$ is the (global) Lipschitz constant of $f$, defined as the infimum of $K$ such that $|f(x)-f(x')|\le K\cdot \dist(x,x')$ for all $x,x'\in U$.  This resolves both issues above.  A technical issue, however, is that the threshold function is not Lipschitz due to the sudden jump at $x=T$.  A common technique to get around this technicality to smooth out the discontinuities of $f$.  For the threshold function, we turn the hard threshold into a soft threshold, by turning the sudden jump into a line of slope $1/\tau$ (see Figure \ref{fig:1way_thres_cons} for an illustration).  This means that values within a region of width $\tau$ centered around $T$ are counted partially. 
This results in $K_f=1/\tau$, so a noise of scale $1/(\varepsilon \tau)$ suffices. 

However, this simple approach suffers from the same issue as the local-DP solution in that it adds the same amount of noise to $f(x_i)$ for every $x_i$.  It does not take advantage of the relaxed privacy notion of GP (under Euclidean metric), which allows faraway points to be distinguished more easily.  Thus, we would like to have a mechanism that adds noise depending on $f'(x_i)$, which measures how sensitive $f(\cdot)$ is at $x_i$.  However, directly using $f'(x_i)$ as the noise scale violates GP: Observe that for the soft threshold function we have $f'(x) = 0$ for $x\not\in [T-\tau/2, T+\tau/2]$.  Clearly, returning a deterministic answer does not satisfy DP or GP.

\subsection{Smooth Sensitivity for Geo-Privacy}
The global Lipschitz constant $K_f$ in GP plays a similar role as the global sensitivity $\Delta_f$ in local-DP.  Both take a one-size-fits-all approach to adding noise to $f(x_i)$.  On the other hand, in the central-DP model, we note plenty of techniques that add instance-specific noise that can be much smaller than the worst case on typical instances.  In particular, the smooth sensitivity framework \cite{nissim2007smooth} has enjoyed much success, with applications in mean estimation, PCA analysis, and machine learning \cite{bun2019average, gonem2018smooth, zafarani2020differentially, sun2020differentially}.  However, this framework critically depends on the notion of ``instances differing by $k$ records'', so it is not applicable to the local model, where each user just holds a single record.

In this paper, we generalize the smooth sensitivity framework from DP to GP, by providing definitions, mechanisms, and a generic procedure for computing the smooth sensitivity of Lipschitz functions.  Our key observation is that ``instances differing by $k$ records'' are essentially what the Hamming metric embodies, so we ``just'' need to generalize the original smooth sensitivity framework from the Hamming metric to an arbitrary metric.  However, in the process we have to overcome a number of technical challenges:

\begin{enumerate}
    \item The output domain of the Hamming metric is discrete while that of a general metric is continuous.  This requires generalizing the growth function and showing that the generalized smooth sensitivity maintains the desirable properties. 
    \item Since the privacy guarantee of GP is different and more general than that of the DP, we must re-analyze the privacy guarantees of various noise distributions under GP.
    \item While computing the smooth sensitivity over the Hamming metric takes exponential time for many problems \cite{nissim2007smooth}, we show that it can often be computed analytically for a closed-form $f$ over $\mathbb{R}^d$. 
\end{enumerate}

Table \ref{tab:comparison} gives a general comparison of the different approaches for computing ${1\over n} \sum_i f(x_i)$.  In terms of privacy, local-DP provides a uniform distinguishability guarantee while GP provides a distance-dependent one, which are not comparable.  The noise-a-posterior approaches are preferred, which not only yield unbiased estimators, but also offer additional privacy, especially for functions with a small output domain like the threshold function.  For the variance, we will see that $B^*_f(x) \le K_f$ for all $f$ and $x$, where $B^*_f(x)$ is the smooth sensitivity of $f$ at $x$, so our smooth sensitivity-based approach is always better than the standard GP mechanism based on $K_f$.  However, how it compares with local-DP and noise-a-priori GP will depend on $f$ and the actual instance $\{x_i\}_i$.

To demonstrate the applicability and compare the utility of our smooth sensitivity framework with the other methods, we present three applications: one-way and two-way threshold functions, and Gaussian kernel density estimation.   
We show how the smooth sensitivity can be calculated in each application, and demonstrate the advantage of our method via experimental evaluation on some some real data sets.

\subsection{Related Work}
In addition to smooth sensitivity, central-DP has a powerful toolbox of instance-specific noise-addition techniques, including the Exponential Mechanism \cite{mcsherry2007mechanism},  Propose-Test-Release \cite{dwork2009differential}, the inverse sensitivity \cite{asi2020instance,fang2022shifted}, and instance-optimal truncation \cite{huang2021instance,dong2022r2t}.  
However, there are very few techniques for GP besides the basic Laplace \cite{andres2013geo} and Gaussian mechanisms \cite{liang2023concentrated}.  Since many of the aforementioned DP tools also rely on the notion of ``instances differing by $k$ records'', our ideas could be potentially be useful for migrating them to GP as well.

For the specific applications we consider, there are some related prior works under DP.  The threshold query is a case of range counting; the latter refers to the problem where the goal is to count the number of points inside some range $\mathcal{R}\subset\mathbb{R}^d$. In the central-DP model, answering one range counting query is trivial, so the majority of the works study the problem of how to answer a class of range counting queries \cite{cormode2019answering,huang2021approximate,muthukrishnan2012optimal,wang2019answering}.
In the local-DP model, since $\Delta_f=1$, one can either add a Laplace noise of scale $1/\varepsilon$ as mentioned above, or use randomized response \cite{warner1965randomized, duchi2018minimax, wang2019collecting}.  Both achieve the same variance $\Theta({1\over \varepsilon^2 n})$ up to a constant-factor difference.  On the other hand, our smooth sensitivity method under GP adds an instance-specific noise.  In particular, when most $x_i$'s are far away from $T$, its variance can be much lower than $O({1 \over \varepsilon^2 n})$.

Gaussian kernel density estimation has been studied in the central-DP model \cite{hall2013differential, wagner2023fast}. These works are mainly concerned with the private release of the KDE as a functional, in the sense that infinitely many queries can be handled by the private mechanism without breaking the privacy guarantee. In particular, \cite{wagner2023fast} also considered and improved the efficiency of computing such queries by using locality sensitive quantization. Here, in our work we are mainly concerned with providing good utility in the query output for the GP model and leave the design of private functionals under GP as future work.

\section{Preliminaries}
\label{sec:prelim}
\subsection{Location-scale families of distributions}
For a real-valued function $f:U\rightarrow \mathbb{R}$, we are interested in private mechanisms of the form $M(x):=f(x)+\sigma(x)\cdot Z$, where $Z$ is a random variable from a suitable distribution and $\sigma(x)$ is a scaling factor which might depend on $x$. Since $M(x)$ is a linear transformation of $Z$, it is convenient to work with the location-scale families of distributions, which are closed under linear transformations. Let $h(z)$ denote the pdf of $Z$ from a location-scale distribution, then $a+bZ$ has pdf $\frac{1}{b}h(\frac{z-a}{b})$ for $b>0$. 

\paragraph{Laplace Distribution.}
The Laplace distribution with location $a$ and scale $b$ has pdf
\[
h(y;a,b) = \frac{1}{2b}\cdot e^{-\frac{|y-a|}{b}}, \; \forall y\in \mathbb{R}.
\]
We denote this family as $\mathrm{Lap}(a,b)$. A random variable $Z\sim \mathrm{Lap}(a,b)$ has mean $a$ and variance $2b^2$.

\paragraph{Generalized Cauchy Distribution.}
For $b, p> 0$ and $p\theta>1$, the generalized Cauchy distribution \cite{Rider1957GeneralizedCD} has pdf
\[
    h(y;a,b,p,\theta) = \frac{1}{b}\cdot\frac{c_{\beta,\theta}}{(1+|\frac{y-a}{b}|^{p})^{\theta}}, \; \forall y\in \mathbb{R}.
\]
where $c_{p,\theta}=\frac{p \Gamma(\theta)}{2\Gamma(1/p)\Gamma(\theta-1/p)}$. 
We denote this distribution family as $\mathrm{GenCauchy}(a,b,p,\theta)$. The distribution reduces to a standard Cauchy distribution for $p=2$ and $\theta=1$, which has undefined mean and variance. For $Z\sim \mathrm{GenCauchy}(a,b,p>3,\theta=1)$, we have $\mathbb{E}[Z]=a$ and the $t$-th central moment of $|Z-a|$ exists for $p>t+1$ \cite{bronstein2005quasi}. In particular, for $p=4$ and $\theta=1$, $Z$ has variance $\mathrm{Var}(Z)=1$. (We cannot find a reference on how to draw from this distribution, so we provide one in Appendix~\ref{appendix:gen_cauchy4}.)

\paragraph{Student's $t$-Distribution.}
The location-scale student's $t$-distribution with $\nu >0$ degrees of freedom has pdf
\[
    h(y;a,\tau,\nu)=\frac{1}{\sqrt{\nu}\tau}\cdot\frac{c_{\nu}}{\left(1+\frac{1}{\nu}\left(\frac{y-a}{\tau}\right)^2\right)^{(\nu+1)/2}}, \; \forall y\in \mathbb{R}.
\]
where $c_{\nu}=\frac{\Gamma(\frac{\nu+1}{2})}{\sqrt{\pi}\cdot\Gamma(\frac{\nu}{2})}$. We denote this family as $\mathcal{T}_{\nu}(a,\tau)$. For $\nu>2$, a random variable $Z\sim\mathcal{T}_{\nu}(a,\tau)$ has expectation $\mathbb{E}[Z]=a$ and variance $\mathrm{Var}(Z)=\tau^2\frac{\nu}{\nu-2}$. Note that the generalized Cauchy distribution coincides with the student's $t$ distribution for the special case where $p=2$, $b=\sqrt{\nu}\tau$ and $\theta=(\nu+1)/2$. 
\subsection{Differential Privacy}
\subsubsection{Definition and mechanism}
Differential privacy is defined using the notion of \textit{neighboring} inputs, where $x, x'\in U$ are referred to as neighbors, denoted $x\sim x'$, if they differ by one record.  Let $\dH(x,x')$ be the Hamming distance between $x$ and $x'$.  Then $x\sim x'$ iff $\dH(x,x')=1$.
\begin{definition} [Differential Privacy \cite{dwork2006calibrating}]
    Fix $\varepsilon, \delta \ge 0$. A randomized mechanism $M:U\rightarrow V$ is $(\varepsilon,\delta)$-differentially private, if for all measurable subsets $S\subseteq V$ and all $x\sim x'$,
    \[
        \Pr[M(x)\in S] \le e^{\varepsilon}\Pr[M(x')\in S] + \delta
    .\]
\end{definition}
    When $U=\chi^n$ for $n>1$, the input $x$ is an $n$-tuple containing $n$ records from $\chi$. We refer to this as the \textit{central} model with abbreviation $(\varepsilon,\delta)$-DP. If the input is just one record, we refer to it as the \textit{local} model \cite{erlingsson2014rappor}, with abbreviation $(\varepsilon,\delta)$-LDP. When $\delta=0$, the mechanism is said to satisfy pure DP, and usually the parameter $\delta$ will be omitted in this case.  The  $\delta>0$ case is often referred to as approximate DP.

There is a simple mechanism for releasing the function value $f(x)$ under pure DP, which involves drawing from a Laplace distribution centered at $f(x)$.
\begin{lemma} [Laplace Mechanism \cite{dwork2006calibrating}]
    \label{lm:lap_mech}
    For the real-valued function $f:U\rightarrow V\subseteq \mathbb{R}$, the mechanism which on input $x$ releases $M(x):=f(x)+\frac{\Delta_f}{\varepsilon} \cdot Z$, where $\Delta_f:=\sup_{z\sim z'}|f(z)-f(z')|$ and $Z\sim \mathrm{Lap}(0,1)$, satisfies $\varepsilon$-DP.
\end{lemma}

\subsubsection{The Smooth Sensitivity Framework for Differential Privacy}
The quantity $\Delta_f$ in Lemma~\ref{lm:lap_mech} is usually referred to as the (global) sensitivity of the function $f$. It is defined over all pairs of neighboring inputs $z\sim z'$ and can be very large. The goal of the smooth sensitivity framework \cite{nissim2007smooth} is to allow adding noise with a magnitude smaller than $\Delta_f$. In particular, the noise magnitude should adapt to hardness of the given input.

\begin{definition} [Pointwise sensitivity\footnote{The original definition was termed ``local sensitivity'', as to contrast with ``global sensitivity''. Here, we refer to it as pointwise sensitivity to avoid confusion with the local model, and also the customary usage of local to describe behavior of a function in a constrained neighborhood.} \cite{nissim2007smooth}]
    Fix $x\in U$. The pointwise sensitivity of a function $f:U\rightarrow \mathbb{R}^d$ at $x$ is defined as
    \[
        \Delta_f(x):=\sup_{x':\dH(x,x')=1}\|f(x)-f(x')\|_1.
    \]
\end{definition}
We abuse the notation and write the global sensitivity as $\Delta_f = \sup_{x\in U} \Delta_f(x)$.  While adding noise of scale $\Delta_f(x)$ to $f(x)$ does not satisfy DP, it turns out that using a smooth upper bound of $\Delta_f(x)$ does.

\begin{definition} [Smooth upper bound on pointwise sensitivity \cite{nissim2007smooth}]
    Fix $\gamma > 0$. A function $B:U\rightarrow\mathbb{R}_{>0}$ is a $\gamma$-smooth upper bound on the pointwise sensitivity of $f$ if:
    \begin{enumerate}
        \item $\forall x\in U: B(x)\ge \Delta_f(x)$
        \item $\forall x, x'\in U,  \dH(x,x')=1: B(x)\le e^{\gamma}\cdot B(x')$.
    \end{enumerate} 
\end{definition}
 The smooth sensitivity, defined below, was shown by \cite{nissim2007smooth} to be the smallest function which satisfies the smooth upper bound definition above. 
\begin{definition} [Smooth sensitivity \cite{nissim2007smooth}]
    Fix $\gamma > 0$. The $\gamma$-smooth sensitivity of the function $f$ is defined as
    \[
        B_{f,\gamma}^*(x) := \max_{z\in U} \left(\Delta_f(z)\cdot e^{-\gamma\cdot\dH(x,z)}\right).
    \]
\end{definition}
The (generalized) Cauchy and Laplace distributions were shown to be suitable distributions for the smooth sensitivity mechanism.
\begin{lemma} [Cauchy and Laplace Mechanisms using Smooth Sensitivity \cite{nissim2007smooth}]
    Fix $\gamma > 0$. Let $f:U\rightarrow \mathbb{R}$, and let $B:U\rightarrow \mathbb{R}_{>0}$ be a $\gamma$-smooth upper bound on the pointwise sensitivity of $f$. Let $M$ be the mechanism which on input $x$ releases $M(x)=f(x)+\frac{B(x)}{\eta}\cdot Z$. Then
    \begin{enumerate}
        \item $M$ is $\varepsilon$-DP, for $\gamma\le \frac{\varepsilon}{2(p+1)}$, $\eta = \frac{\varepsilon}{2(p+1)}$, and $Z$ drawn from a distribution with pdf $h(z)\propto \frac{1}{1+|z|^p}$ where $p>1$;
        \item $M$ is $(\varepsilon,\delta)$-DP, for $\gamma\le \frac{\varepsilon}{2\ln(2/\delta)}$ with $\delta\in (0,1)$, $\eta = \frac{\varepsilon}{2}$, and $Z\sim \mathrm{Lap}(0,1)$.
    \end{enumerate}
\end{lemma}
\subsection{Geo-Privacy}
Geo-Privacy was first introduced in \cite{andres2013geo,chatzikokolakis2013broadening}.  We adopt the more general version defined in \cite{liang2023concentrated}.
Let $(U,\dist)$ be a metric space. We write $x\sim_{\Upsilon} x'$ or $x'\in \mathcal{B}(x,\Upsilon;U)$ if $\dist(x,x') \le \Upsilon$, where $\mathcal{B}(x_0,r;U):=\{z\in U: \dist(z,x_0)\le r\}$ denotes the ball of radius $r$ centred at $x_0$. 

\begin{definition} [$(\varepsilon,\delta,\Upsilon)$-GP \cite{liang2023concentrated}]
\label{def:eps_delta_Delta_GP}
        Fix $\varepsilon, \delta \ge 0$ and $\Upsilon\in \mathbb{R}_{>0}\cup\{\infty\}$. A randomized mechanism $M:U\rightarrow V$ is $(\varepsilon,\delta,\Upsilon)$-GP, if for all measurable subsets $S\subseteq V$ and all $x \sim_{\Upsilon} x'$,
    \[
        \Pr[M(x)\in S] \le e^{\varepsilon\cdot \dist(x,x') }\Pr[M(x')\in S] + \delta.
    \]
\end{definition}

This general definition incorporates many DP models as special cases \cite{liang2023concentrated}.  Taking $\dist(\cdot,\cdot)$ to be the discrete metric, $(\varepsilon,\delta, \infty)$-GP coincides with $(\varepsilon,\delta)$-LDP.  Taking $\dist(\cdot,\cdot)$ to be the Hamming metric, $(\varepsilon,0,\infty)$-GP coincides with $(\varepsilon,0)$-DP, while $(\varepsilon,\delta,O(\log(1/\delta)/\varepsilon))$-GP is equivalent to $(\varepsilon,\delta)$-DP.  While our definitions and general mechanisms will apply to an arbitrary metric, we describe how to compute the smooth sensitivity for GP assuming $U\subseteq \mathbb{R}^d$ and taking $\dist(\cdot,\cdot)$ as the Euclidean distance, which is the most interesting setting of GP.  This corresponds to the local model where each user holds a point $x_i$ in Euclidean space and wishes to privatize $f(x_i)$ for some function $f:U\rightarrow V$. Note that unlike DP, the $\varepsilon$ in GP is no longer a unit-less constant.  Instead, it measures the privacy loss per unit distance.  For example, if we want to make two income numbers that are $\$10,000$ apart $1$-distinguishable, then we should set $\varepsilon = 1/\$10,000$.

If $V$ is also a metric space equipped with distance function $\dV(\cdot,\cdot)$, we have the following GP mechanism if $f$ is $K$-Lipschitz, i.e., $\dV(f(x),f(x'))\le K\cdot \dist(x,x')$ for all $x, x'\in U$.

\begin{lemma} [Laplace Mechanism for GP \cite{chatzikokolakis2013broadening}]
\label{lm:lapmech_GP}
Let $f:U\rightarrow V$ be $K$-Lipschitz for $K>0$. Then the mechanism which on input $x$, draws a $y\in V$ with pdf $\propto e^{-\frac{\varepsilon}{K}\dV(f(x),y)}$, is $(\varepsilon,0,\infty)$-GP.
\end{lemma}

\section{Smooth Sensitivity for Geo-Privacy}

\subsection{Defining Smooth Sensitivity for GP}

\subsubsection{Local Lipschitzness}
\begin{definition}[$\Lambda$-local Lipschitzness\footnote{Also called \textit{Lipschitz in the small} \cite{luukkainen1979rings, beer2015locally, aggarwal2023some}, see these for some variants of local Lipschitzness. \cite{beer2015locally} also showed that every uniformly continuous real-valued function can be approximated to arbitrary precision by these functions.} \cite{luukkainen1979rings}]
Let $(U,\dist), (V,\dV)$ be metric spaces. Fix $K\ge 0$ and $\Lambda\in \mathbb{R}_{>0}\cup\{\infty\}$. A function $f:U\rightarrow V$ is $\Lambda$-locally $K$-Lipschitz if for all $x\sim_{\Lambda} x'$, 
\[\dV(f(x),f(x'))\le K\cdot \dist(x,x').\]
\end{definition}

When $\Lambda=\infty$, we refer to $f$ as (globally) $K$-Lipschitz.
From the definition above and Lemma~\ref{lm:lapmech_GP}, we immediately get:
\begin{corollary}
\label{cor:lapmech_GP_approx}
    Let $f:U\rightarrow V$ be $\Lambda$-locally $K$-Lipschitz. A mechanism which on input $x$ draws a $y\in V$ with pdf $\propto e^{-\frac{K}{\varepsilon}\dV(f(x),y)}$, is $(\varepsilon,0,\Lambda)$-GP.
\end{corollary}

The following can be considered the GP counterpart of pointwise sensitivity.
\begin{definition} [Pointwise Lipschitzness]
    Let $(U,\dist), (V,\dV)$ be metric spaces. Fix $\Lambda\in \mathbb{R}_{>0}\cup\{\infty\}$. A function $f:U\rightarrow V$ is $\Lambda$-locally Lipschitz at $x$ with constant $K$ if for all $x'\in \mathcal{B}(x,\Lambda;U)$,
    \[\dV(f(x),f(x'))\le K\cdot \dist(x,x').\]
    We denote by $\LC_{f,\Lambda}(x)$ the infimum of such $K$.
\end{definition}
Thus, $f$ is $\Lambda$-locally Lipschitz with constant $\sup_{x\in U} \LC_{f,\Lambda}(x)$, and $f$ is (globally) Lipschitz with constant $\sup_{x\in U} \LC_{f,\infty}(x)$.

\subsubsection{Smooth Upper Bounds}

 Let $g:U\times U\rightarrow \mathbb{R}_{> 0}$ be a function defined by $g(x,x'):=g_0(\dist(x,x'))$ for some monotonically increasing function $g_0:\mathbb{R}_{\ge 0}\rightarrow \mathbb{R}_{\ge 1}$ which satisfies $g_0(0)=1$ and $g_0(r_1+r_2)\le g_0(r_1)\cdot g_0(r_2)$ for $r_1,r_2\ge 0$. We refer to such functions $g$ as \textit{smooth growth functions}. Some examples are $\gE:(x,x')\mapsto e^{\gamma\dist(x,x')}$ and $\gL:(x,x')\mapsto 1+\gamma\dist(x,x')$, for any $\gamma > 0$. 
 
\begin{definition} [Smooth upper bound on pointwise Lipschitz constant]
\label{def:smoothupper_plc}
    Let $\Lambda\in \mathbb{R}_{>0}\cup\{\infty\}$ and let $g$ be a smooth growth function. A function $B:U\rightarrow \mathbb{R}_{\ge 0}$ is a $g$-smooth upper bound on the pointwise Lipschitz constant of a $\Lambda$-locally Lipschitz function $f:U\rightarrow \mathbb{R}$ if it satisfies:
    \begin{enumerate}
        \item $\forall x\in U: B(x)\ge \LC_{f,\Lambda}(x)$; and
        \item $\forall x, x'\in U: B(x)\le g(x,x')\cdot B(x')$.
    \end{enumerate}
\end{definition}

Condition (2) can also be stated with the looser requirement that it holds for $x\sim_{\lambda} x'$ for any $\lambda \ge \Lambda$. A discussion is included in Appendix~\ref{appendix:smooth_upper_finite}.

\begin{definition} [Smooth sensitivity for GP]
For $\Lambda$-locally Lipschitz function $f:U\rightarrow \mathbb{R}$ and smooth growth function $g$, the \textit{$g$-smooth sensitivity} of $f$ is a function $B^*:U \rightarrow \mathbb{R}_{\ge0}$ defined by
\begin{equation}
\label{eqn:smooth_sens_GP}
B^*(x; f, g, U, \Lambda):=\sup_{z\in U} \frac{\LC_{f,\Lambda}(z)}{g(x,z)}.
\end{equation}
\end{definition}
We show that the smooth sensitivity function given above indeed satisfies the smooth upper bound conditions of Definition~\ref{def:smoothupper_plc}.
\begin{lemma} 
Let $\Lambda\in \mathbb{R}_{>0}\cup\{\infty\}$. For $\Lambda$-locally Lipschitz function $f:U\rightarrow \mathbb{R}$, the function $B^*$ given in equation~\eqref{eqn:smooth_sens_GP}
is a $g$-smooth upper bound for $\LC_{f,\Lambda}$.
\end{lemma}
\begin{proof}
    First, $B^*(x) \ge \frac{\LC_{f,\Lambda}(x)}{g(x, x)} = \LC_{f,\Lambda}(x)$. Next, we show $B^*(x)\le g(x,x')\cdot B^*(x')$ for all $x, x'$. Fix $x, x' \in U$ and any $z\in U$. 
    Then 
    \begin{align*}
    g(x',z)&=g_0(\dist(x',z))\le g_0\left(\dist(x,x')+\dist(x,z)\right)\\ 
    &\le g(x,x')\cdot g(x,z) \implies g(x',z)/g(x,z)\le g(x,x').
    \end{align*}
    Thus,
    \begin{align*}
        \frac{\LC_{f,\Lambda}(z)}{g(x, z)} &= \frac{\LC_{f,\Lambda}(z)}{g(x', z)}\cdot \frac{g(x', z)}{g(x, z)} \le \frac{\LC_{f,\Lambda}(z)}{g(x', z)}\cdot g(x,x') \\
        &\le \sup_{z\in U} \left(\frac{\LC_{f,\Lambda}(z)}{g(x', z)}\cdot g(x,x')\right) = B^*(x')\cdot g(x,x').
    \end{align*}
   Taking $\sup$ over $z\in U$ on the left hand side gives the desired inequality.

\end{proof}

Next, we show that $B^*(x; f,g,U,\Lambda)$ is also the smallest $g$-smooth upper bound for $\LC_{f,\Lambda}$, for any smooth growth function $g$.
\begin{lemma}
\label{lm:smoothsens_smallest_bound}
For $\Lambda$-locally $K$-Lipschitz $f:U\rightarrow \mathbb{R}$ and any smooth growth function $g$, let $B$ be any $g$-smooth upper bound for $\LC_{f,\Lambda}$ and let $B^*$ be the $g$-smooth sensitivity function given in equation~\eqref{eqn:smooth_sens_GP} for $\LC_{f,\Lambda}$. Then $\forall x\in U: B(x)\ge B^*(x)$.
\end{lemma}
\begin{proof}
Fix any $x\in U$. From the conditions for smooth upper bound, for any $z\in U$
    \[
    B(x) \ge \frac{B(z)}{g(z,x)} \ge \frac{\LC_{f,\Lambda}(z)}{g(z,x)}.
    \]
    Taking $\sup$ over $z\in U$ on the right hand side gives $B(x)\ge B^*(x)$.
\end{proof}

\subsection{GP Mechanisms with Smooth Upper Bounds}
In this subsection, we give smooth sensitivity-based mechanisms for GP.

\subsubsection{Pure GP}

\begin{lemma}[Generalized Cauchy mechanism for GP]
\label{lm:gp_1d}
    Fix $\varepsilon, \gamma > 0$, let $\Lambda\in \mathbb{R}_{>0} \cup \{\infty\}$. Suppose $B(\cdot)$ is a $\gE(\cdot;\gamma)$-smooth upper bound on $\LC_{f,\Lambda}$ of a function $f:U\rightarrow \mathbb{R}$. Then the mechanism $M$ which on input $x$ releases $M(x):=f(x)+\frac{B(x)}{\eta}\cdot Z$, where $Z\sim \mathrm{GenCauchy}(0,1,p>1,\theta\ge 1)$, is $(\varepsilon,0,\Lambda)$-GP, where $\varepsilon=\max(\gamma,(p\theta-1)\gamma)+(p-1)^{\frac{p-1}{p}}\theta\eta$.
\end{lemma}
\begin{proof}
    Fix $x\sim_{\Lambda} x'\in U$. Write $a:=f(x), a':=f(x'), b:=\frac{B(x)}{\eta}$ and $b':=\frac{B(x')}{\eta}$. Then $m(x)(y) = \frac{1}{b} h\left(\frac{y-a}{b}\right)$ and $m(x')(y) = \frac{1}{b'} h\left(\frac{y-a'}{b'}\right)$. For $S\subseteq \mathbb{R}$, 
    \begin{align*}
    &{}\Pr[M(x)\in S] = \int_S m(x)(y) dy = \int_S \frac{1}{b} h\left(\frac{y-a}{b}\right) dy \\
    &= \int_S \frac{\frac{1}{b}h\left(\frac{y-a}{b}\right)}{\frac{1}{b'}h\left(\frac{y-a'}{b'}\right)}\cdot \frac{1}{b'}h\left(\frac{y-a'}{b'}\right) dy = \int_S \underbrace{\frac{\frac{1}{b}h\left(\frac{y-a}{b}\right)}{\frac{1}{b'}h\left(\frac{y-a'}{b'}\right)}}_{l(y;x,x')}\cdot m(x')(y) dy.
    \end{align*}
Thus, it suffices to show that $l(y;x,x'):=\frac{\frac{1}{b}h\left(\frac{y-a}{b}\right)}{\frac{1}{b'}h\left(\frac{y-a'}{b'}\right)}=\frac{\frac{1}{b}h\left(\frac{y-a}{b}\right)}{\frac{1}{b'}h\left(\frac{y-a}{b'}\right)}\cdot \frac{\frac{1}{b'}h\left(\frac{y-a}{b'}\right)}{\frac{1}{b'}h\left(\frac{y-a'}{b'}\right)} =: l_1(y;b,b')\cdot l_2(y;a,a',b')
$ is bounded pointwise by $e^{\varepsilon \cdot \dist(x, x')}$.
Below we separately bound the two terms.
\begin{align*}
    l_1(y;b,b'):=\frac{\frac{1}{b}h\left(\frac{y-a}{b}\right)}{\frac{1}{b'}h\left(\frac{y-a}{b'}\right)} = \frac{\frac{1}{b}\frac{1}{\left(1+|\frac{y-a}{b}|^p\right)^{\theta}}}{\frac{1}{b'}\frac{1}{\left(1+|\frac{y-a}{b'}|^p\right)^{\theta}}} = \frac{b'}{b}\frac{\left(1+|\frac{y-a}{b'}|^p\right)^{\theta}}{\left(1+|\frac{y-a}{b}|^p\right)^{\theta}}.
\end{align*}
If $b'\ge b$ or $y=a$, then $l_1(y;b,b')\le b'/b$. Otherwise, 
\[l_1(y;b,b')\le \frac{b'}{b}\cdot \frac{|\frac{y-a}{b'}|^{p\theta}}{|\frac{y-a}{b}|^{p\theta}} = \frac{b'}{b}\cdot \left(\frac{b}{b'}\right)^{p\theta} = \left(\frac{b}{b'}\right)^{{p\theta}-1}.\]
To bound $l_2(y;a,a',b')$ we will make use of the continuous function $\phi_c:\mathbb{R}_{\ge 0}\rightarrow \mathbb{R}_{\ge 0}$ defined by $\phi_c(z):=\ln(1+z^c)$, where $\forall z: |\phi_c'(z)| \le c $ for $c >1$ was shown in \cite{nissim2007smooth}. Here, we use a slightly tighter version.
\begin{claim}\label{clm:phi_gamma_bound}
For $c > 1$, the function $\phi_c:\mathbb{R}_{\ge 0}\rightarrow \mathbb{R}_{\ge 0}$ defined by $\phi_c(z):=\ln(1+z^c)$ satisfies $\forall z: |\phi_c'(z)|\le (c-1)^{\frac{c-1}{c}}$.
\end{claim}
\begin{proof} We have $\phi_c'(z)=\frac{cz^{c-1}}{1+z^c}$, and 
\begin{align*}
&\phi_c''(z) = \frac{d}{dz} \left( \frac{cz^{c-1}}{1+z^c}\right) = \frac{-c^2 z^{2(c-1)}}{(1+z^c)^2}+\frac{(c-1)c z^{c-2}}{1+z^c}\\
&= \frac{(1+z^c)c(c-1)z^{c-2}-c^2z^{2c-2}}{(1+z^c)^2}=\frac{z^{c-2}(c(c-1)-cz^c)}{(1+z^c)^2}.
\end{align*}
Since $\phi_c'$ is continuous on $\mathbb{R}_{\ge 0}$, it suffices to check $|\phi_c'|$ on the critical points, which are $\{0,(c-1)^{1/c}\}$ where the latter is found by setting $\phi_c''(z)$ to zero. Then we have $|\phi_c'(0)|=0$ and $|\phi_c'((c-1)^{1/c})|=(c-1)^{\frac{c-1}{c}}> 0$. I.e., $|\phi_c'(z)|\le (c-1)^{\frac{c-1}{c}}$.
\end{proof}
From above we get $\sup_{z\ge 0}|\phi_{p}'(z)|\le (p-1)^{\frac{p-1}{p}}$. Then, as a consequence of the mean value theorem $|\phi_{p}(z_1)-\phi_{p}(z_2)|\le (p-1)^{\frac{p-1}{p}} |z_1-z_2|$ for all $z_1,z_2 \in \mathbb{R}_{\ge 0}$. Let $z_1:= \frac{y-a}{b'}$, $z_2:= \frac{y-a'}{b'}$. Then
\begin{align*}
    l_2(y;a,a',b') &:= \frac{\frac{1}{b'}h\left(\frac{y-a}{b'}\right)}{\frac{1}{b'}h\left(\frac{y-a'}{b'}\right)}=\frac{\left(1+|\frac{y-a'}{b'}|^p\right)^{\theta}}{\left(1+|\frac{y-a}{b'}|^p\right)^{\theta}}\\
    \left|\ln(l_2(y;a,a',b'))\right| &= \theta\left|\phi_p(z_2)-\phi_p(z_1)\right| \le \theta(p-1)^{\frac{p-1}{p}} |z_2-z_1|\\ 
    &= \theta(p-1)^{\frac{p-1}{p}} \frac{|a-a'|}{b'}.
\end{align*}
Thus,
\begin{align*}
    &{}|\ln(l(y;x,x'))| \le |\ln(|l_1(y;b,b')|)|+\left|\ln(|l_2(y;a,a',b')|)\right|\\
    &\le \max(\ln(b'/b), (p\theta-1) \ln(b/b'))+(p-1)^{\frac{p-1}{p}}\theta \frac{|a-a'|}{b'}\\
    &\le \max(1,p\theta-1) \cdot\ln(e^{\gamma\dist(x,x')}) + (p-1)^{\frac{p-1}{p}}\theta \frac{|f(x)-f(x')|}{B(x')/\eta}\\
    &\le \max(1,p\theta-1) \cdot \ln(e^{\gamma\dist(x,x')}) + (p-1)^{\frac{p-1}{p}}\theta \frac{\eta B(x')\dist(x, x')}{B(x')}\\
    &\le \max(1,p\theta-1) \cdot \gamma\dist(x,x') + (p-1)^{\frac{p-1}{p}}\theta\eta\cdot \dist(x,x')\\
    &= (\max(\gamma,(p\theta-1)\gamma)+(p-1)^{\frac{p-1}{p}}\theta\eta)\cdot\dist(x, x').
\end{align*}
\end{proof}

For functions with outputs in $V\subseteq \mathbb{R}^m$, we can apply the lemma above to each coordinate. This is the same as setting $\dV = \ell_1$.
\begin{corollary}\label{cor:l1_cauchy}
    Fix $\varepsilon, \gamma > 0$, let $\Lambda\in \mathbb{R}_{>0} \cup \{\infty\}$. Let $f:U\rightarrow \mathbb{R}^m$ equipped with the $\ell_1$ metric $\|\cdot\|_1$. Suppose $B(\cdot)$ is a $\gE(\cdot;\gamma)$-smooth upper bound on $\LC_{f,\Lambda}$. Then the mechanism $M$ which on input $x$ releases $M(x):=f(x)+\frac{B(x)}{\eta}\cdot Z$, where $Z=[Z_1,\dotsb,Z_m]^T$ and each $Z_j\sim_{iid} \mathrm{GenCauchy}(0,1,p>1,\theta\ge 1)$, is $(\varepsilon,0,\Lambda)$-GP, where $\varepsilon=\max(m\gamma,m(p\theta-1)\gamma)+(p-1)^{\frac{p-1}{p}}\theta\eta$.
\end{corollary}
The proof is similar to that in Lemma~\ref{lm:gp_1d}, where we bound the terms $l_1$ and $l_2$, which are now products of $m$ terms. The interested reader can find the details in Appendix~\ref{appendix:cor_l1_cauchy}.

\begin{lemma} [Student's $t$ mechanism for GP]
\label{lm:gp_1d_t}
    Fix $\varepsilon, \gamma > 0$, let $\Lambda\in \mathbb{R}_{>0} \cup \{\infty\}$. Suppose $B(\cdot)$ is a $\gE(\cdot;\gamma)$-smooth upper bound on $\LC_{f,\Lambda}$ of a function $f:U\rightarrow \mathbb{R}$. Then the mechanism $M$ which on input $x$ releases $M(x):=f(x)+\frac{B(x)}{\eta}\cdot Z$, where $Z\sim \mathcal{T}_\nu(0,1)$ for $\nu > 1$, is $(\varepsilon,0,\Lambda)$-GP, where $\varepsilon=\nu\gamma+\frac{\nu+1}{2\sqrt{\nu}}\eta$.
\end{lemma}
The proof follows similar arguments as those in the proof of Lemma~\ref{lm:gp_1d}, and can be found in Appendix~\ref{appendix:proofs_ssmech}.

\paragraph{Remark.} By using $\dist=\dH$ in Lemmas \ref{lm:gp_1d} and \ref{lm:gp_1d_t}, one obtains standard $\varepsilon$-DP mechanisms. 
Here we have slightly tighter constants for the Cauchy and student's $t$ mechanisms than previously obtained in \cite{nissim2007smooth, bun2019average}. In particular, here we have with the $\mathrm{GenCauchy}(0,1,p>1,1)$ distribution, $\varepsilon=(p-1)\gamma+(p-1)^{(p-1)/p}\eta$ vs. $\varepsilon=(p+1)\gamma+(p+1)\eta$ in \cite{nissim2007smooth}, and $\varepsilon=\nu\gamma+\frac{\nu+1}{2\sqrt{\nu}}\eta$ with the $\mathcal{T}_{\nu}(0,1)$ distribution vs. $\varepsilon=(\nu+1)\gamma+\frac{\nu+1}{2\sqrt{\nu}}\eta$ in \cite{bun2019average}. As an example, plugging in $\gamma=0.1$, $\eta=0.5$ and $p=4=\nu$, Lemma \ref{lm:gp_1d} gives $\varepsilon \approx 1.44$ (vs. $\varepsilon=3$ by \cite{nissim2007smooth}), and Lemma~\ref{lm:gp_1d_t} gives $\varepsilon = 1.025$ (vs. $\varepsilon=1.125$ by \cite{bun2019average}).

\subsubsection{Approximate GP}
\begin{lemma} [Laplace mechanism for approximate GP]
\label{lm:laplace}
Let $\varepsilon, \delta, \gamma > 0$, $\Lambda\in\mathbb{R}_{>0}\cup\{\infty\}$. Let $g_L$ be the function defined by $\gL(x,x')=1+\gamma\cdot \dist(x,x')$. Let $B(\cdot)$ be a $\gL$-smooth upper bound on $\LC_{f,\Lambda}$ of a function $f:U\rightarrow \mathbb{R}$. Then the mechanism $M$ which on input $x$ releases $M(x):=f(x)+\frac{B(x)}{\eta}\cdot Z$, where $Z\sim \mathrm{Lap}(0,1)$, is $(\varepsilon=\eta+\gamma\ln(1/\delta),\delta, \Lambda)$-GP.
\end{lemma}
\begin{proof}
    Fix $x\sim_{\Lambda} x'\in U$, let $Y\sim \mathcal{M}(x)$ be a random variable, where $\mathcal{M}(x)$ is the distribution of $M(x)$. Write $a:=f(x), a':=f(x'), b:=\frac{B(x)}{\eta}$ and $b':=\frac{B(x')}{\eta}$. Let $l:\mathbb{R}\rightarrow\mathbb{R}_{\ge 0}$ be defined by $l(y)=\frac{m(x)(y)}{m(x')(y)}$. It suffices to show that $\ln(l(Y)) \le \varepsilon \dist(x,x')$ except on a set of measure at most $\delta$.
    \begin{align*}
        \ln(l(y)) &= \ln\left(\frac{m(x)(y)}{m(x')(y)}\right) = \ln\left(\frac{\frac{1}{2b}e^{-\frac{|y-a|}{b}}}{\frac{1}{2b'}e^{-\frac{|y-a'|}{b'}}}\right) \\
        &= \ln(b'/b) - {\frac{1}{b}|y-a|+\frac{1}{b'}|y-a'|}
        = \ln(b'/b) + {\frac{1}{b}\left(\frac{b}{b'}|y-a'|-|y-a|\right)}.
    \end{align*}
    If $b\le b'$, then 
    \begin{align*}
    \ln(l(y))&\le \ln(b'/b)+\left(|y-a'|-|y-a|\right)/b \\
    &\le \ln\left(1+\gamma\cdot\dist(x,x')\right) + |a-a'|/b\\
    &\le \gamma\dist(x,x')+\eta|f(x')-f(x)|/B(x') \le (\gamma+\eta)\dist(x,x').
    \end{align*}
    For $b > b'$,
    \begin{align*}
        \ln(l(y)) &\le \ln(b'/b) + {\frac{1}{b}\left(\frac{b}{b'}|y-a-(a'-a)|-|y-a|\right)} \\
        &\le {\frac{1}{b}\left((\frac{b}{b'}-1)|y-a|+\frac{b}{b'}|a'-a|\right)}\\
        &= {\frac{1}{b}(\frac{b}{b'}-1)|y-a|+\frac{\eta}{B(x')}|f(x')-f(x)|}\\
        &\le {\frac{1}{b}(\frac{b}{b'}-1)|y-a|+\frac{\eta}{B(x')}B(x')\cdot\dist(x, x')} \\
        &\le \frac{\eta}{B(x)}(\gamma\dist(x, x')|y-f(x)) + \eta\cdot\dist(x,x').\\
        \ln(l(Y)) &\le {\frac{\eta}{B(x)}\left(\gamma\dist(x,x')\left|Y-f(x)\right|\right)+\eta\cdot\dist(x,x')}\\
        &= {\frac{\eta}{B(x)}\left(\gamma\dist(x,x')\left|\frac{B(x)}{\eta}\cdot Z\right|\right)+\eta\cdot\dist(x,x')}\\
        &= {\gamma\dist(x,x')|Z|}+{\eta\cdot\dist(x,x')}.
    \end{align*}
Now since $|Z|\le \ln(1/\delta)$ with probability at least $1-\delta$, the desired inequality follows.
\end{proof}

\paragraph{Remark.} The original smooth sensitivity framework for DP \cite{nissim2007smooth} only considered using $\gE$.  However, there are some technicalities in their proof when using $\gE$ with the Laplace mechanism, as observed in \cite{bun2019average}.  This is why we generalized the framework to all smooth growth functions $g$.  In particular, we found that $\gL$ works better with the Laplace mechanism yielding a clean result for approximate GP.  Since GP under the Hamming metric degenerates to DP, Lemma \ref{lm:laplace} also gives a clean Laplace-based smooth sensitivity mechanism for approximate DP, which could be of independent interest.

\subsection{Computing Smooth Sensitivity}
\label{sec:smoothsens_C1}
In this section, we provide a generic procedure for computing smooth sensitivity for real-valued functions.  For now on we assume $U\subseteq \mathbb{R}^d$ and equipped with the Euclidean metric.  First, we show that if $U \subset \mathbb{R}^d$, we can treat it as if it were equal to $\mathbb{R}^d$.

\begin{lemma} \label{lm:approx_smooth}
    For $\Lambda$-locally Lipschitz $f:U\rightarrow V$, let $\{G_r\}$ be a family of functions indexed by $r\in \mathbb{R}_{\ge 0}$, where for all $x\in U$, $G_r:U\rightarrow \mathbb{R}_{\ge 0}$ satisfies $\LC_{f,\Lambda}(x)\le G_0(x)$ and for any $x'$ with $\dist(x,x')=\Delta\le \Lambda: G_r(x)\le G_{r+\Delta}(x')$. Then the function $B_G:x\mapsto \sup_{r\ge 0} \frac{G_r(x)}{g_0(r)}$ is a $g$-smooth upper bound for $\LC_{f,\Lambda}$.
\end{lemma}

\begin{proof}
    Fix $x\in U$. Then $B_G(x)\ge G_0(x)\ge \LC_{f,\Lambda}(x)$ by assumption. For any $x'\sim_{\Lambda}x$, let $\Delta:=\dist(x,x')$.
    \begin{align*}
        \frac{G_r(x)}{g_0(r)} &\le \frac{G_{r+\Delta}(x')}{g_0(r)} = 
        \frac{G_{r+\Delta}(x')}{g_0(r+\Delta)}
        \cdot \frac{g_0(r+\Delta)}{g_0(r)}
        \le \frac{G_{r+\Delta}(x')}{g_0(r+\Delta)}
        \cdot g_0(\Delta) \\
        &\le \sup_{r\ge\Delta} \frac{G_{r}(x')}{g_0(r)} g_0(\Delta) \le \sup_{r\ge 0} \frac{G_{r}(x')}{g_0(r)} g_0(\Delta) = B_G(x')\cdot g_0(\Delta).
    \end{align*}
    Taking $\sup$ over $r\ge 0$ on the left hand side, we get $B_G(x) = \sup_{r\ge 0} \frac{G_r(x)}{g_0(r)} \le B_G(x')\cdot g(x,x')$.
\end{proof}

For $r\in \mathbb{R}_{\ge 0}$, let ${\mathcal{B}}(x_0,r):=\{z\in \mathbb{R}^d: \|z-x_0\|\le r\}$, 
let $G_r:x\mapsto \sup_{z\in {\mathcal{B}}(x,r)}{\LC_{f,\Lambda}(z)}$. Then for $x'\in {\mathcal{B}}(x,\Lambda)$ with $\dist(x,x')=\|x-x'\|=\Delta \le \Lambda$, $G_r(x)=\sup_{z\in {\mathcal{B}}(x,r)} {\LC_{f,\Lambda}(z)}\le \sup_{z\in {\mathcal{B}}(x',r+\Delta)} {\LC_{f,\Lambda}(z)}$ since ${\mathcal{B}}(x,r)\subseteq {{\mathcal{B}}(x',r+\Delta)}$. I.e., $G_r$ satisfies the conditions of the lemma above.
Now, for $x\in U$:
\begin{align*}
B_G(x)&:=\sup_{r\ge 0} \frac{G_r(x)}{g_0(r)} = \sup_{r\ge 0} \frac{\sup_{z\in {\mathcal{B}}(x,r)}{\LC_{f,\Lambda}(z)}}{g_0(r)} \\
&= \sup_{z\in\mathbb{R}^d} \frac{\LC_{f,\Lambda}(z)}{g_0(\dist(z,x))}=:B^*(x;f,g,\mathbb{R}^d,\Lambda).
\end{align*}

We've ``lifted'' the calculation of smooth sensitivity from being dependent on $U$ (taking $\sup$ over $z\in U$) to be (nearly) unconstrained. Note that $U$ can be any subset of $\mathbb{R}^d$ and in particular, need not be concave (e.g. $U=[u]\times[l,r]\times\mathbb{R}\subset \mathbb{R}^3$). When we work in the local model, we might have limited knowledge about $U$, except only that data are represented as tuples in $\mathbb{R}^d$. 
From now on we need not worry about the actual shape of $U$ and will just compute $B^*$ on $\mathbb{R}^d$.

Let $f:U\subseteq \mathbb{R}^d\rightarrow \mathbb{R}$ be a differentiable function, where $U$ is equipped with the $\ell_2$ metric. Fix $x\in\mathbb{R}^d$. For $d=1$, the function $\psi_x :\mathbb{R}\rightarrow \mathbb{R}_{\ge 0}$ defined by $\psi_x(w):=\frac{f(w)-f(x)}{w-x}$ is continuous and differentiable except at $w=x$, where it's understood that $\lim_{w\rightarrow x}\left|\frac{f(w)-f(x)}{w-z}\right|=|f'(x)|$. For $d \ge 2$, we can similarly define $\psi_x: w\mapsto \frac{f(w)-f(x)}{\|w-x\|}$, then $\lim_{w\rightarrow x}|\psi_x| = \|\nabla f(x)\|$ where $\nabla f(x)$ denotes the gradient of $f$ at $x$.

Thus, $|\psi_x|$ achieves max at one of the following: a stationary point in $\mathcal{B}(x,\Lambda)\setminus \{x\}$, a boundary point, or at $x$. I.e., we can compute
\[
    {\LC_{f,\Lambda}}(x) = \max_{w\in \mathcal{B}(x,\Lambda)}\frac{|f(w)-f(x)|}{\|w-x\|}.
\]
If $f$ has undefined derivative at a finite set of points, we just need to separately check the objective values of $|\psi(\cdot)|$ at these points. Then, we can write
\[
    B^*(x)=\max_{w\neq z:\|w-z\|\le \Lambda}\left(\frac{|f(w)-f(z)|}{\|w-z\|}\cdot \frac{1}{g(x,z)}\right).
\]
Let $\Psi: (z,w) \mapsto \psi_z(w)\cdot \frac{1}{g(x,z)}$. 
We have the follow general procedure for computing $B^*$, assuming $g_0$ is differentiable on $\mathbb{R}_{\ge 0}$, which involves identifying the critical and boundary points of $f$. Below, $\frac{d}{dw}$ computes the vector of derivatives w.r.t. to each component of $w$. Similarly, $\frac{\partial}{\partial z}$ computes the vector of partial derivatives w.r.t. to each component of $z$. Let $S_0:=\{w:f'(w) \text{\; undefined}\}$,
\begin{align*}
S_1 :=& \left\{(w,z): \frac{d}{dw}\psi_z(w)=0, \frac{\partial}{\partial z} \Psi(z,w)=0 \right\} \cup \left\{(w\in S_0, z): \frac{\partial}{\partial z} \Psi(z,w)=0\right\}, \\
S_2:=& \left\{w:\frac{d}{dw}\psi_x(w)=0\right\}, \\
S_3:=& \left\{z:\frac{d}{dz}\left(\frac{\|\nabla f(z)\|}{g(x,z)}\right)=0\right\},\\ 
S_4:=& \left\{(w,z): \|w-z\|=\Lambda, \frac{\partial}{\partial z} \Psi(z, w)=0 \right\}\;\text{($S_4:=\emptyset$ if $\Lambda = \infty$)}.
\end{align*}
Then $B^*(x)$ can be computed as the max of the following candidate values.
\begin{enumerate}
    \item $w\neq z, z\neq x$: $\max_{(w,z)\in S_1\cup S_4} |\Psi(w,z)|$.
    \item $w\neq z, z = x$: $\max_{w\in S_0\cup S_2} |\psi_x(w)|$.
    \item $w = z, z\neq x$: $\max_{z\in S_3} \frac{\|\nabla f(z)\|}{g(x,z)}$.
    \item $w = z = x$: $\|\nabla f(x)\|$.
\end{enumerate}

\section{Applications}
\label{sec:app}
In section, we discuss three applications in the smooth sensitivity framework for GP: one-way and two-way threshold functions, and Gaussian kernel density estimation. The model for GP we adopt in these applications is the local model, where each user privatizes their own data before sending it to an aggregator. 

\subsection{Threshold Functions}
\label{sec:threshold}
Let $f:U\subseteq \mathbb{R} \rightarrow \{0,1\}\subset \mathbb{R}$ be the function defined by $x\mapsto \mathbb{1}{\{x>T\}}$, i.e., $f(x)=1$ if $x>T$, else $f(x)=0$. We assume $T>0$ without loss of generality. Let $A=A(x_1,\dotsb,x_n):=\frac{1}{n}\sum_{i\in[n]}f(x_i)$ denote the aggregation of the function values from  the $n$ users.

\subsubsection{Noise a priori}
In the noise-a-priori solution, each user $i$ sends $M_0(x_i):=x_i+Z_i$ to the aggregator, where $Z_i\sim \mathrm{Lap}(\frac{1}{\varepsilon})$ for $i\in [n]$, and the aggregator computes $\tilde{A}:=\frac{1}{n}\sum_{i\in [n]} f(M_0(x_i)) = \frac{1}{n}\sum_{i\in [n]} \mathbb{1}{\{M_0(x_i)>T\}}$. 

Below we analyze the mean squared error (MSE) of this solution. Write $y_i=f(x_i)$ and $\tilde{y}_i=f(M_0(x_i))$, then
\[
    \tilde{y}_i-y_i = \begin{cases}
        1, & x_i+Z_i > T \text{\;and\;} x_i \le T \\
        -1,   & x_i+Z_i \le T \text{\;and\;} x_i > T\\
        0, & \text{otherwise}.
    \end{cases}
\]
Then,
\begin{align*}
    \Pr\left[\tilde{y_i}-y_i=1|x_i\le T\right] 
    &=\Pr[Z_i>T-x_i|T-x_i\ge 0] 
    =1-\Pr[Z_i\le T-x_i]
    =\frac{1}{2}e^{-\varepsilon(|x_i-T|)},\\
    \Pr\left[\tilde{y_i}-y_i=-1|x_i > T\right] &= \Pr[Z_i\le T-x_i|T-x_i<0] = \frac{1}{2} e^{-\varepsilon|x_i-T|}.
\end{align*}
Thus, 
\[\mathbb{E}\left[|\tilde{y_i}-y_i|^2\right] = 1^2\cdot \frac{1}{2} e^{-\varepsilon|x_i-T|} = \frac{1}{2} e^{-\varepsilon|x_i-T|}.\]
Let $I_0:=\{i: x_i\le T\}, I_1:=\{i: x_i>T\}$, $I_{m,k}=\{(i,j):i\in I_m, j\in I_k\}$ for $m, k\in \{0, 1\}$.
The MSE is
\begin{align*}
    &{}\mathbb{E}\left[(\tilde{A}-A)^2\right] = \mathbb{E}\left[\left(\frac{1}{n}\sum_i \left(\tilde{y}_i-y_i\right)\right)^2\right]
    = \frac{1}{n^2}\mathbb{E}\left[\sum_i (\tilde{y}_i-y_i)^2+\sum_{i\neq j}(\tilde{y}_i-y_i)(\tilde{y}_j-y_j)\right] \\
    &= \frac{1}{n^2}\sum_i \mathbb{E}\left[(\tilde{y}_i-y_i)^2\right] + \frac{1}{n^2}\sum_{i\neq j}\mathbb{E}(\tilde{y}_i-y_i)\cdot \mathbb{E}(\tilde{y}_j-y_j)\\
    &=\frac{1}{2n^2} \sum_i e^{-\varepsilon|x_i-T|} + \frac{1}{n^2}\sum_{(i,j)\in I_{0,0}\cup I_{1,1}:i\neq j} \frac{1}{4}e^{-\varepsilon(|x_i-T|+|x_j-T|)} 
    - \frac{1}{n^2}\sum_{(i,j)\in I_{0,1}\cup I_{1,0}:i\neq j} \frac{1}{4}e^{-\varepsilon(|x_i-T|+|x_j-T|)}.
\end{align*}
We see that the error of this noise-a-priori solution depends on $\{x_i-T\}_i$. 
Let $n_0:=\left|I_0\right|$, $n_1:=\left|I_1\right|$. Then $\left|(i,j)\in I_{0,0}\cup I_{1,1}:i\neq j\right|=\frac{n_0(n_0-1)}{2}+\frac{n_1(n_1-1)}{2}=\frac{n(n-1)}{2}-n_0\cdot n_1$, while $\left|(i,j)\in I_{0,1}\cup I_{1,0}:i\neq j\right|$ $= n_0\cdot n_1$. 
Note that the second sum dominates the error, unless the $x_i$'s are approximately symmetrically distributed around $T$. In the scenario where $O(n)$ of the $x_i$'s are within $O(1/\varepsilon)$ distance of the threshold, the error is $\Omega(1)$, i.e., it does not reduce to $0$ as $n\rightarrow \infty$.

\subsubsection{Noise a posterior via global Lipschitz constant}
\begin{figure}[htbp]
     \centering
      \begin{subfigure}[t]{0.3\linewidth}
            \centering
            \includegraphics[width=\textwidth]{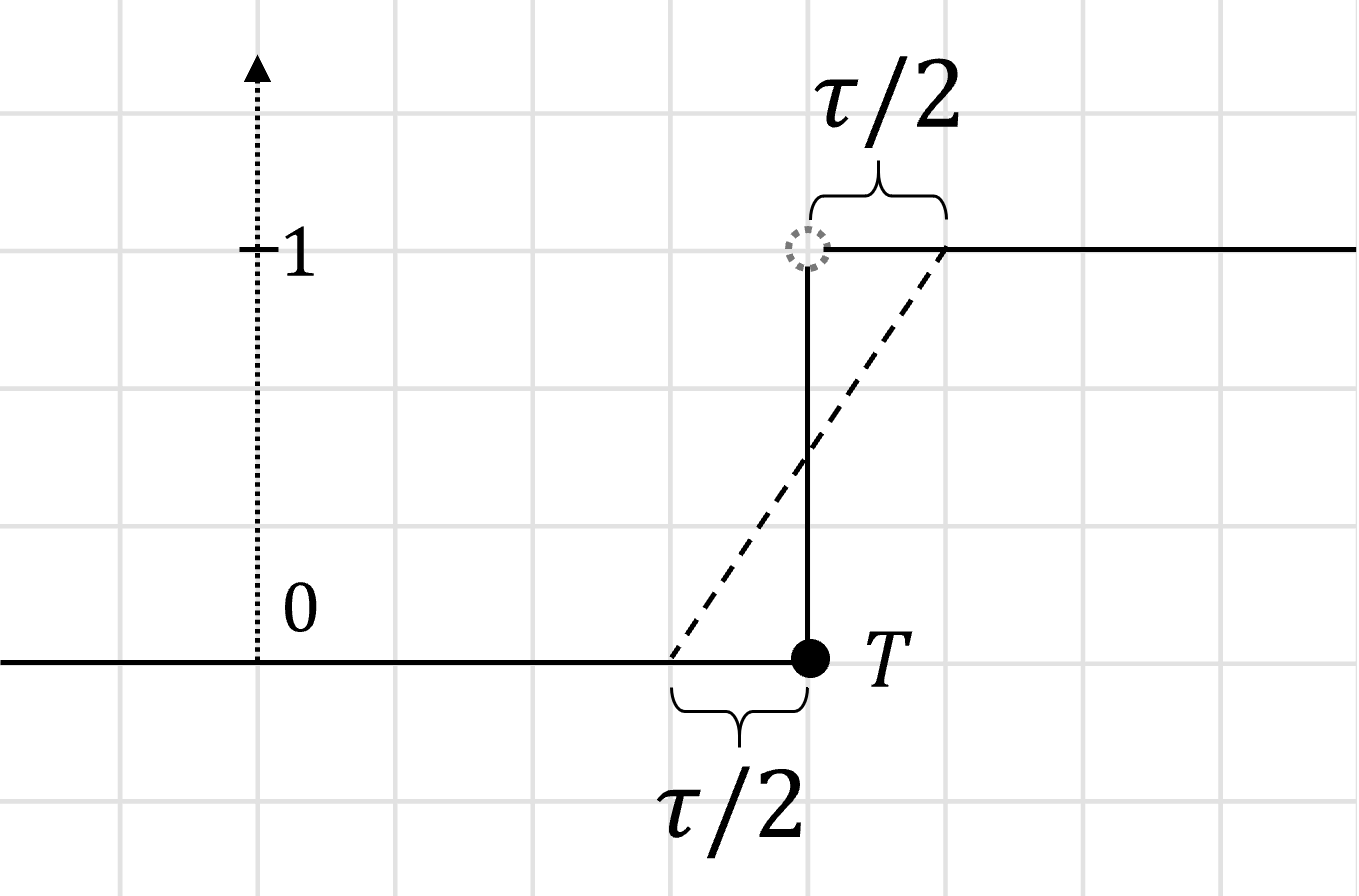}
            \subcaption{Construction of $f_{\tau}$.}
            \label{fig:1way_thres_cons}
         \end{subfigure}
         \;
               \begin{subfigure}[t]{0.3\linewidth}
            \centering
            \includegraphics[width=\textwidth]{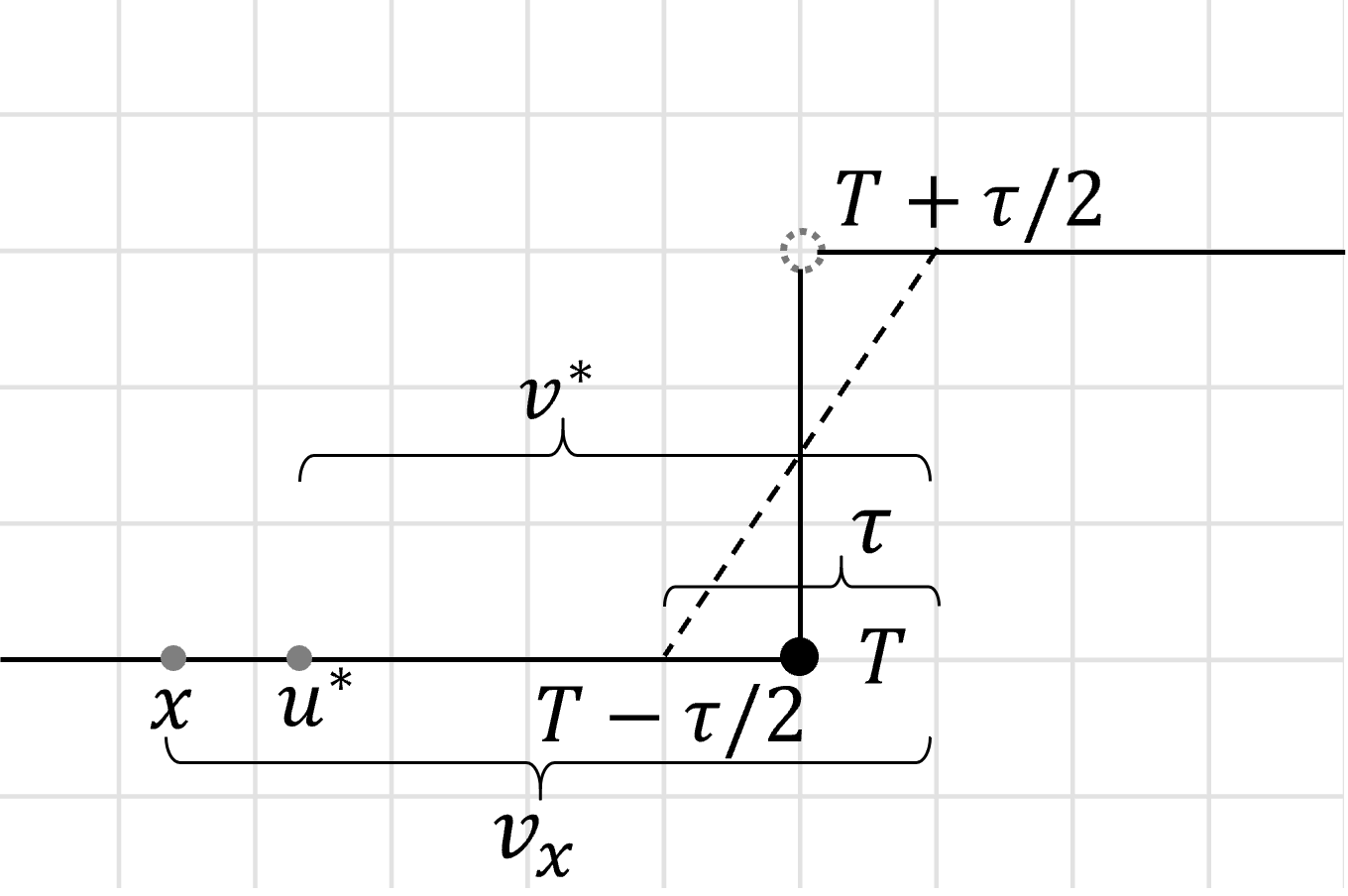}
            \subcaption{Illustration for Lemma~\ref{lm:1way_thres_sens}.}
            \label{fig:1way_thres_vx}
         \end{subfigure}
         \caption{$f_{\tau}$ for one-way threshold, where $T$ is the threshold.}
    \vskip -.01in
\end{figure}

The (hard) threshold function is not Lipschitz at points near $T$.  Thus we use a soft threshold function given by (see Fig.~\ref{fig:1way_thres_cons} for an illustration)
\[
    f_{\tau}(x) = \begin{cases}
        0, & x-T < -\frac{\tau}{2}\\
        \frac{1}{\tau}(x-T)+\frac{1}{2}, & -\frac{\tau}{2}\le x-T \le \frac{\tau}{2}\\
        1, & x-T > \frac{\tau}{2}
    \end{cases}
\]
where $\tau$ is the width of the region around $T$ in which points are partially counted.  This soft threshold function may actually be preferred in practice, as it is more robust to small variations in the data.

It's easy to see that $f_{\tau}$ is (globally) $\frac{1}{\tau}$-Lipschitz w.r.t. to $|\cdot|$, since the slope between any two points $f(x)$ and $f(x')$ is at most $\frac{1}{\tau}$.  Then a baseline noise-a-posterior solution is to release $f(x_i) + \mathrm{Lap}(0,1/(\varepsilon\tau))$ by Lemma \ref{lm:lapmech_GP}.

\subsubsection{Noise a posterior via smooth sensitivity}

\begin{lemma}
\label{lm:1way_thres_sens}
For $\gE:(x,x')\mapsto e^{\gamma\dist(x,x')}$, the $\gE$-smooth sensitivity for $f_{\tau}$ is
\[
    B^*(x) = \begin{cases}
        \max\left(\frac{1}{|x-T|+\tau/2},\frac{1}{\tau}e^{-\gamma(|x-T|-\tau/2)}
        \right), & |x-T| > \frac{\tau}{2}\\
        \frac{1}{\tau}, & |x-T| \le \frac{\tau}{2}
    \end{cases}.
\]
\end{lemma}
\begin{proof}
    Fix $x\in U$. The case where $|x-T|\le \frac{\tau}{2}$ is trivial so assume $|x-T|\ > \frac{\tau}{2}$; specifically, assume $x < T-\frac{\tau}{2}$ (the case where $x > T+\frac{\tau}{2}$ can be shown using symmetric arguments). For $z \notin [T-\frac{\tau}{2},T+\frac{\tau}{2}]$, its pointwise sensitivity is $\frac{1}{|z-T|+\tau/2}$. We only need to consider $z\in [x,T-\frac{\tau}{2}]$; the points to the left of $x$ has lower pointwise sensitivity than that of $x$, and those to the right of $T-\frac{\tau}{2}$ is farther away from $x$ and has pointwise sensitivity at most that of $T-\frac{\tau}{2}$.

    Let $v_x:=|x-T|+\tau/2$ be the distance of $x$ to the point $T+\tau/2$. Consider the function $\varphi: v\in [\tau,v_x] \mapsto \frac{1}{v} e^{-\gamma(v_x-v)}$. Write $v^*:=|u^*-T|+\tau/2$. Then $|u^*-x|=|x-T|+\tau/2-(|u^*-T|+\tau/2)=v_x-v^*$ (see Fig.~\ref{fig:1way_thres_vx} for an illustration). Thus, it suffices to find the value $v\in [\tau,v_x]$ which maximizes $\frac{1}{v} e^{-\gamma(v_x-v)}$. Observe $\varphi$ is continuous on $ [\tau,v_x]$ and has stationary point $v_0$ given by
    \begin{align*}
        0=\frac{d}{dv} \varphi(v) &= -\frac{1}{v^2}e^{-\gamma(v_x-v)} + \frac{\gamma e^{-\gamma(v_x-v)}}{v} 
        = \frac{e^{-\gamma(v_x-v)}}{v}\left(\gamma-\frac{1}{v}\right) 
        \implies v_0=\frac{1}{\gamma}.
    \end{align*}
    Moreover, observe that $\frac{d}{dv} \varphi(v) \le 0$ for $v\le \frac{1}{\gamma}$ and $\frac{d}{dv} \varphi(v) \ge 0$ for $v\ge \frac{1}{\gamma}$. I.e., $v_0=\frac{1}{\gamma}$ is a local minimum. Thus, $\varphi$ is maximized at either end point, with function value $\varphi(v_x)=\frac{1}{v}=\frac{1}{|x-T|+\tau/2}$, or $\varphi(\tau)=\frac{1}{\tau}e^{v_x-\tau}=\frac{1}{\tau}e^{|x-T|+\tau/2-\tau}=\frac{1}{\tau}e^{|x-T|-\tau/2}$.
\end{proof}
Note: In the proof above, for $x<T-\tau/2$, we have the sets $S_0=\{T-\tau/2,T+\tau/2\}$, $S_1=\{(w,z):f(w)=f(z)\}\cup\{(T+\tau/2,T+\tau/2-1/\gamma)\}$, $S_2=\{w: w<T-\tau/2\}$ and $S_3=\{z: |z-T| \ge \tau/2\}$ corresponding to the procedure described in Section~\ref{sec:smoothsens_C1}. In particular, $S_0$ gives candidate value $\frac{1}{|T-x|+\tau/2}$ (from the point $T+\tau/2$), and $S_3$ gives candidate value $\frac{1}{\tau}e^{-|x-T|+\tau/2}$ (from the point $T-\tau/2$).

By Lemma~\ref{lm:gp_1d}, $\varepsilon$-GP is achieved if each user $i$ sends $M_{\nu}(x_i):=f_{\tau}(x_i)+ Z_i$, 
where $Z_i\sim \mathcal{T}_{\nu}(0, \frac{B^*(x_i)}{\eta})$, for appropriately chosen $\gamma>0$, $\nu>2$ and $\eta \le \frac{2\sqrt{\nu}}{\nu+1}\varepsilon$. It is clear that  $M_{\nu}(x_i)$ is an unbiased estimator of $f_{\tau}(x_i)$.  Thus $\frac{1}{n}\sum_i M_{\nu}(x_i)$ is an unbiased estimator of $\frac{1}{n}\sum_i f_{\tau}(x_i)$ with variance $O(\sum_i B^*(x_i)^2 /(\varepsilon n)^2)$. 

\paragraph{Remark.} 
The method in this subsection easily extends to bounded range functions of the form $x\mapsto \mathbb{1}\{l\le x \le r\}$ for $l<r\in \mathbb{R}$, where points in the region $[l-\tau/2,r+\tau/2]$ get assigned an interpolated value in $[0,1]$. The interested reader can refer to Appendix~\ref{appendix:smoothsens_1Drange} for the details.

\subsection{Two-Way Threshold Functions}
For $T_1, T_2 > 0$, let $f:U\subseteq \mathbb{R}^2 \rightarrow \{0, 1\}$ be the function defined by $x\mapsto \mathbb{1}{\{x>(T_1,T_2)\}}:=\mathbb{1}{\{x_1>T_1\}}\cdot \mathbb{1}{\{x_2>T_2\}}$. Note that a two-way threshold function cannot be solved by simply solving two one-way threshold functions since there could be correlation in the underlying $2$-dimensional distribution. We are again interested in the aggregate value $A=A(x_1,\dotsb,x_n):=\frac{1}{n}\sum_{i\in[n]}f(x_i)$ from the $n$ users.

\begin{figure}[htbp]
     \centering
      \begin{subfigure}[t]{0.3\linewidth}
            \includegraphics[width=\textwidth]{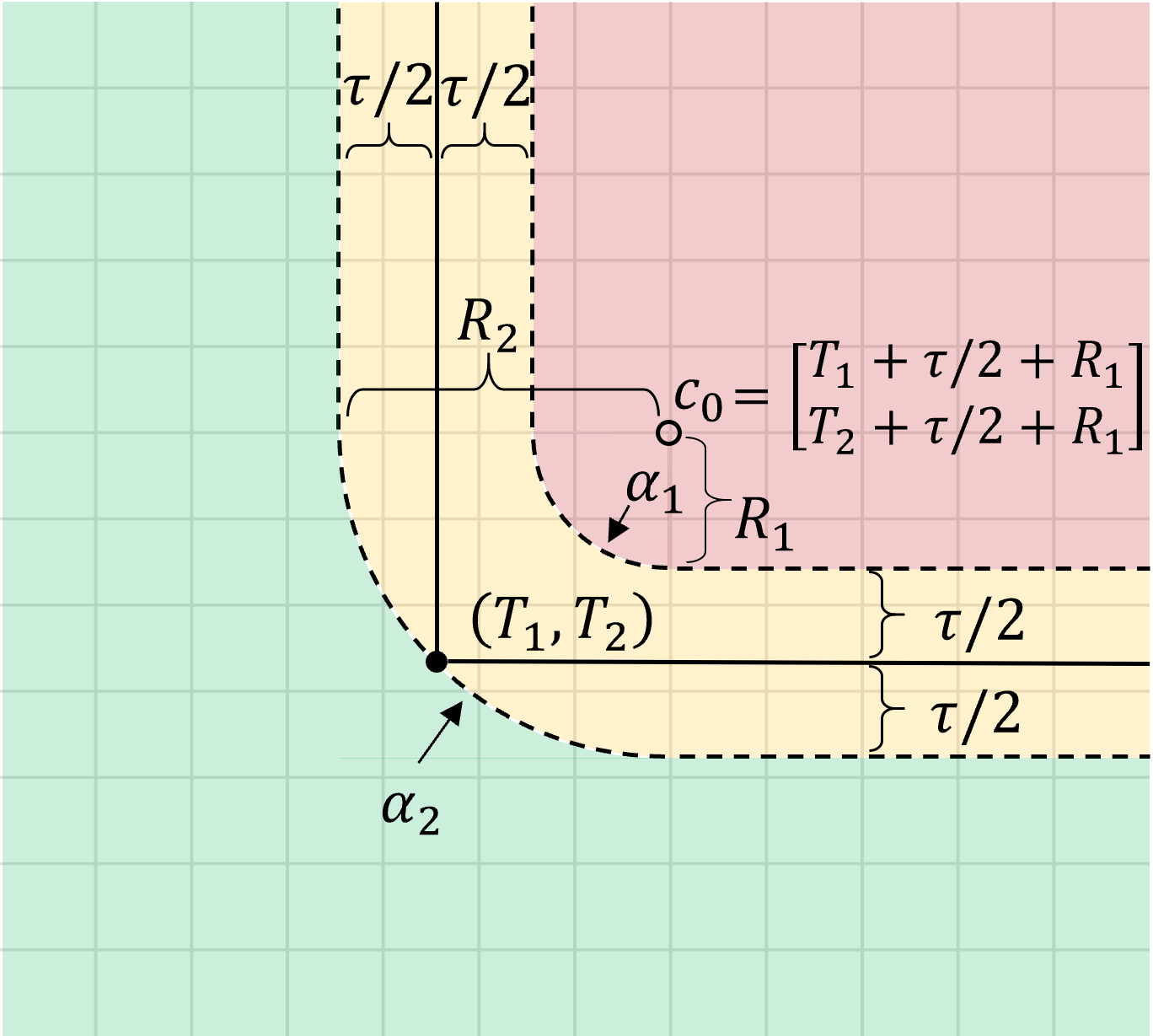}
          \subcaption{Construction of the transition band.}
          \label{fig:2way_thres_cons}
         \end{subfigure}
         \;
         \begin{subfigure}[t]{0.3\linewidth}
            \includegraphics[width=\textwidth]{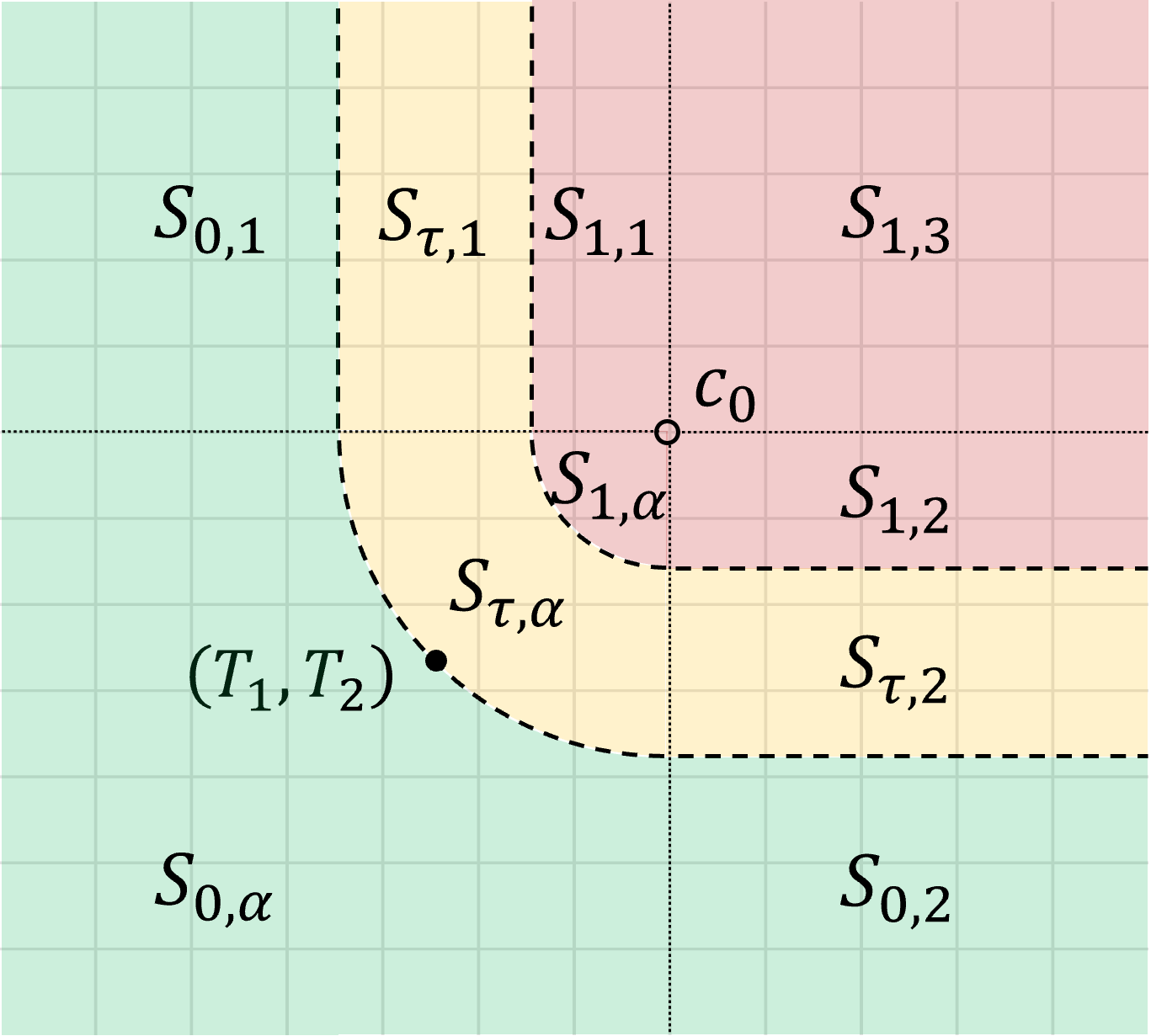}
         \subcaption{Sub-regions for computing smooth sensitivity.} 
         \label{fig:2way_thres_subregions}
         \end{subfigure}
         \vskip -.05in
         \caption{$f_{\tau}$ for two-way threshold: red region corresponds to function value $1$; green region corresponds to function value $0$; yellow region corresponds to transition band.}
    \vskip -.05in
\end{figure}

As in the case of the (one-way) threshold function, the two-way threshold function is also not Lipschitz at its boundary. We approximate $f$ with a Lipschitz function which involves a band of transition and points falling into this band gets assigned an interpolated value in $[0,1]$. We divide $\mathbb{R}^2$ into different regions using two circles. Let $0<\tau<2\min(T_1,T_2)$ be the width of the transition band. Let $R_1:=\frac{\tau}{\sqrt{2}}$, $R_2:=R_1+\tau=\left(\frac{1}{\sqrt{2}}+1\right)\tau$. Let $c_0:=(T_1+\frac{\tau}{2}+R_1,T_2+\frac{\tau}{2}+R_1)$.  For the circle centered at $c_0$ of radius $R_1$, let $\alpha_1$ denote the arc which has end points at $(T_1+\frac{\tau}{2},T_2+\frac{\tau}{2}+R_1)$ and $(T_1+\frac{\tau}{2}+R_1,T_2+\frac{\tau}{2})$. For the circle centered at $c_0$ of radius $R_2$, let $\alpha_2$ denote the arc which has end points at $(T_1-\frac{\tau}{2},T_2+\frac{\tau}{2}+R_1)$ and $(T_1+\frac{\tau}{2}+R_1,T_2-\frac{\tau}{2})$. The the transition band is composed of the area between the arcs $\alpha_1$ and $\alpha_2$, and the points (see Fig.~\ref{fig:2way_thres_cons})
\begin{align*}
&{}\left\{(p_1,p_2):p_1 \ge T_1+\frac{\tau}{2}+R_1, T_2-\frac{\tau}{2} \le p_2 \le T_2+\frac{\tau}{2}\right\}\\
\cup &{}\left\{(p_1,p_2): T_1-\frac{\tau}{2}\le p_1 \le T_1+\frac{\tau}{2}, p_2 \ge T_2+\frac{\tau}{2}+R_1\right\}.
\end{align*}
Let
\begin{align*}
    S_{1,\alpha} := &\left\{(p_1,p_2): T_1+\frac{\tau}{2}+R_1 > p_1 > T_1+\frac{\tau}{2}, \; T_2+\frac{\tau}{2}+R_1 > p_2 > T_2+\frac{\tau}{2}\text{\; and\; } \|(p_1,p_2)-c_0\| < R_1 \right\}\\
    S_{1, 1} := &\left\{(p_1,p_2): T_1 + \frac{\tau}{2} + R_1 > p_1 > T_1 + \frac{\tau}{2} \text{\; and\; }p_2 > T_2 + \frac{\tau}{2} + R_1 \right\} \\
    S_{1, 2} :=&\left\{(p_1,p_2): T_2 + \frac{\tau}{2} + R_1 > p_2 > T_2 + \frac{\tau}{2}\text{\; and\; } p_1 > T_1 + \frac{\tau}{2} + R_1 \right\} \\
    S_{1, 3} :=&\left\{(p_1,p_2): (p_1, p_2) > \left(T_1 + \frac{\tau}{2} + R_1, T_2 + \frac{\tau}{2} + R_1\right)\right\} \\
    S_{0,\alpha} := &\left\{(p_1,p_2): (p_1,p_2) \le \left(T_1+\frac{\tau}{2}+R_1,T_2+\frac{\tau}{2}+R_1\right) \text{\; and\; } \|(p_1,p_2)-c_0\| > R_2\right\}\\
    S_{0, 1} :=&\left\{(p_1,p_2): p_1 \le T_1 - \frac{\tau}{2} \text{\; and\; } p_2 > T_2 + \frac{\tau}{2} + R_1\right\} \\
    S_{0, 2} :=&\left\{(p_1,p_2): p_2 \le T_2 - \frac{\tau}{2} \text{\; and\; } p_1 > T_1 + \frac{\tau}{2} + R_1\right\} \\
    S_{\tau, \alpha} := &\left\{(p_1,p_2): (p_1,p_2) \le \left(T_1+\frac{\tau}{2}+R_1,T_2+\frac{\tau}{2}+R_1\right) \text{\; and\; } R_1 \le \|(p_1,p_2)-c_0\| \le R_2\right\}\\
    S_{\tau, 1} := &\left\{(p_1,p_2): T_1 - \frac{\tau}{2} \le p_1 \le T_1 + \frac{\tau}{2}\text{\; and\; } p_2 > T_2 + \frac{\tau}{2} + R_1\right\}\\
    S_{\tau, 2} := &\left\{(p_1,p_2): T_2 - \frac{\tau}{2} \le p_2 \le T_2 + \frac{\tau}{2}
    \text{\; and\; } p_1 > T_1 + \frac{\tau}{2} + R_1\right\}.
\end{align*}
Then our approximation function $f_\tau$ is given by:
\begin{align*}
    f_\tau(x_i) = \begin{cases}
        1, &x_i\in S_{1,\alpha}\cup S_{1,1} \cup S_{1,2} \cup S_{1,3} \\
        0, &x_i\in S_{0,\alpha}\cup S_{0,1} \cup S_{0,2}\\
        \frac{R_2-\|x_i-c_0\|}{\tau}, &x_i\in S_{\tau,\alpha}\\
        \frac{x_{i,1}-(T_1-\tau/2)}{\tau}, &x_i\in S_{\tau,1}\\
        \frac{x_{i,2}-(T_2-\tau/2)}{\tau}, &x_i\in S_{\tau,2}.
    \end{cases}
\end{align*}

\begin{claim}
\label{clm:2way_thres_lipschitz}
The function $f_{\tau}$ defined above is $\frac{1}{\tau}$-Lipschitz on $\mathbb{R}^2$ (w.r.t. $\|\cdot\|$).
\end{claim}
The proof is deferred to Appendix~\ref{appendix:2way_thres_lipschitz}.

\begin{lemma}
For $\gE(x,x')\mapsto e^{\gamma\dist(x,x')}$, the $\gE$-smooth sensitivity of $f_{\tau}$ at $z=(z_1,z_2)$ can be computed as
\begin{flalign*}
B^*(z) = 
\begin{cases}
\max\left(\frac{1}{T_1+\tau/2-z_1},\frac{1}{\tau}e^{-\gamma(T_1-\tau/2-z_1)}\right), \;\;\;\;\;  \hfill z\in S_{0,1} \\
\max\left(\frac{1}{T_2+\tau/2-z_2},\frac{1}{\tau}e^{-\gamma(T_2-\tau/2-z_2)}\right),\;\;\;\;\;  \hfill z\in S_{0,2}  \\
\max\left(\frac{1}{\|z-c_0\|-R_1},\frac{1}{\tau}e^{-\gamma(\|z-c_0\|-R_2)}\right), \;\;\;\;\; \hfill z\in S_{0,\alpha}  \\
\max\left(\frac{1}{z_1-(T_1-\tau/2)},\frac{1}{\tau}e^{-\gamma(z_1-(T_1+\tau/2))}\right), \;\;\;\;\;  \hfill  z\in S_{1,1}  \\
\max\left(\frac{1}{z_2-(T_2-\tau/2)},\frac{1}{\tau}e^{-\gamma(z_2-(T_2+\tau/2))}\right), \;\;\;\;\; \hfill  z\in S_{1,2}  \\
\max\left(\frac{1}{z_1-(T_1-\tau/2)}, \frac{1}{\tau}e^{-\gamma(z_1-(T_1+\tau/2))}, \frac{1}{z_2-(T_2-\tau/2)},
\frac{1}{\tau}e^{-\gamma(z_2-(T_2+\tau/2))}\right), \;\;\;\;\; \hfill z\in S_{1,3}  \\
\max\left(\frac{1}{R_2-\|z-c_0\|},\frac{1}{\tau}e^{-\gamma(R_1-\|z-c_0\|)}\right), \;\;\;\;\; \hfill z\in S_{1,\alpha}  \\
\;\;\frac{1}{\tau}, \;\;\;\;\; \hfill \;  z\in S_{\tau,1}\cup S_{\tau,2}\cup S_{\tau,\alpha}
\end{cases}
\end{flalign*}
\end{lemma}
\begin{proof}
The case where $z$ falls inside the transition band is clear. Assume $z$ falls outside the transition band, and let $z_t$ be the nearest point on the transition band to $z$. The line joining $z$ and $z_t$ gives a vertical cross-section which corresponds to the one-way threshold function. Then, the smooth sensitivity can be computed as in Lemma~\ref{lm:1way_thres_sens}. It can be verified that each case of $B^*$ above computes a one-way threshold smooth sensitivity, where depending on the region that $z$ is in, the nearest point $z_t$ is computed differently (see Fig.~\ref{fig:2way_thres_subregions}). E.g., if $z\in S_{0,1}$, then $z_t=(T_1-\tau/2, z_2)$. The rest can be verified in the same manner.
\end{proof}

\subsection{Gaussian Kernel Density Estimation}
In this subsection, we are interested in computing the kernel density estimate at some given point $t\in \mathbb{R}^d$. Specifically, the kernel of interest is the Gaussian kernel $\kappa_t:\mathbb{R}^d \rightarrow \mathbb{R}_{\ge 0}$ defined by $x\mapsto e^{-\frac{\|x-t\|^2}{2h^2}}$, where $h>0$ is a given bandwidth. We would like to compute $A=A(x_1,\dotsb,x_n):=\frac{1}{n}\sum_{i\in[n]}\kappa_t(x_i)$.

We will reduce the smooth sensitivity calculation from an optimization problem with variables in $\mathbb{R}^d$ to one with variables in $\mathbb{R}$. We first present some technical results.

\begin{claim}
\label{clm:gauss_1d_lipschitzconstant}
    For $t\in \mathbb{R}$, $h>0$, let $\kappa_t:\mathbb{R}\rightarrow \mathbb{R}_{\ge 0}$ be the function $x\mapsto e^{-\frac{(x-t)^2}{2h^2}}$.  Then $|\kappa_t(x) - \kappa_{t}(x')| \le \frac{e^{-1/2}}{h}|x-x'|$ for all $x, x'\in \mathbb{R}$.
\end{claim}
\begin{proof}
    It suffices to show $\sup_{x\in \mathbb{R}}\left|\frac{d}{dx} \kappa_t(x)\right| \le \frac{e^{-1/2}}{h}$, where the inequality then follows from the mean value theorem. The derivation can be found in Appendix~\ref{appendix:lipschitz_claim_1dgauss}.
\end{proof}

For $t\in \mathbb{R}^d$, $\kappa_t:\mathbb{R}^d\rightarrow \mathbb{R}_{\ge 0}$, we can again re-parameterize the function as $c\mapsto \frac{1}{h}e^{-c^2/2}$. In particular, for $x, x'\in \mathbb{R}^d$, let $c=\|x-t\|/h$, $c'=\|x'-t\|/h$. Then if $\left|\frac{1}{h}e^{-c^2/2}-\frac{1}{h}e^{-c'^2/2}\right|\le L|c-c'|$,
\begin{align*}
    \left|e^{-\frac{\|x-t\|}{2h^2}|}-e^{-\frac{\|x'-t\|}{2h^2}}\right|&=|e^{-c^2/2}-e^{-c'^2/2}|\le hL|c-c'|\\
    &= L\left|\|x-t\|-\|x'-t\|\right|\le L \|x-x'\|.
\end{align*}
Thus, we also have $|\kappa_t(x)-\kappa_t(x')|\le \frac{e^{-1/2}}{h}\|x-x'\|$ for $x, x'\in \mathbb{R}^d$, i.e., $\kappa_t$ is $\frac{e^{-1/2}}{h}$-Lipschitz.
Next, we show that to calculate the smooth sensitivity $B^*(x)=\max_{w\neq z} \left(\frac{|\kappa_t(w)-\kappa_t(z)|}{\|w-z\|}\cdot \frac{1}{g(x,z)}\right)$, we need only consider pairs $(w, z)$ which are on the line connecting $x$ and $t$ (for $x \neq t$).

\begin{lemma} \label{lm:gaussker_md_to_1d} Let $U\subseteq \mathbb{R}^d$ be equipped with the $\ell_2$ metric. Fix $t\in \mathbb{R}^d$. For $f:U \rightarrow V \subseteq \mathbb{R}$, suppose there is $f_0: \mathbb{R}_{\ge 0} \rightarrow \mathbb{R}$ such that $f(x) = f_0(\|x-t\|)$ for all $x\in U$. Suppose for $x \neq t$, we have $(w^*,z^*)$ such that $B^*(x) = \frac{|f(w^*)-f(z^*)|}{\|w^*-z^*\|g(x,z^*)}$. Then there is a pair $(w_L,z_L)$ on the line connecting $x$ and $t$ such that $\frac{|f(w_L)-f(z_L)|}{\|w_L-z_L\|g(x,z_L)}\ge B^*(x)$.
\end{lemma}
\begin{proof} Let $z_L:= c_z x + (1-c_z)t$, $w_L:= c_w x + (1-c_w)t$ where $c_z := \frac{\|z^*-t\|}{\|x-t\|}$ and $c_w := \frac{\|w^*-t\|}{\|x-t\|}$. We will show $\|z_L-t\|=\|z^*-t\|$, $\|w_L-t\|=\|w^*-t\|$, $\|w_L-z_L\|\le \|w^*-z^*\|$ and $\|z_L-x\|\le \|z^*-x\|$, whence the stated inequality follows. The full derivations can be found in Appendix~\ref{appendix:smoothsens_KDE}.
\end{proof}

Now we can calculate $B^*(x)$ for $x\neq t$ using the parameterization in the lemma above.

\begin{lemma} 
\label{lm:gauss_smoothsens_xneqt}
Fix $t \in \mathbb{R}^d$, $h > 0$ and $\gamma > 0$. For the function $\kappa_t: x\mapsto e^{-\frac{\|x-t\|^2}{2h^2}}$ and smooth growth function $\gE: (x,x')\mapsto e^{\gamma\|x-x'\|}$, we have for $x\neq t$, the $\gE$-smooth sensitivity
    \begin{align*}
        B^*(x) =  \max &\left(\max_{(c_w,c_z)\in S_1} \frac{c}{h}\cdot\frac{c_w e^{-c_w^2c^2/2}}{e^{\gamma|c_z-1|ch}}, \max_{c_w\in S_2}\frac{|e^{-c_w^2c^2/2}-e^{-c^2}|}{|c_w-1|ch}, \max_{c_z\in S_3}\frac{c}{h}\cdot \frac{c_z e^{-c_z^2c^2/2}}{e^{\gamma|c_z-1|ch}}, \frac{c}{h}e^{-c^2/2}\right), 
    \end{align*}
where
\begin{align*}
    S_1 = &\left\{(c_w,c_z>1)\in \mathbb{R}^2_{\ge 0}: c_w = \phi_1(-\gamma), \phi_0(c_w,c_z) = 0\right\} \\
     &\cup \left\{(c_w,c_z<1)\in \mathbb{R}^2_{\ge 0}: c_w = \phi_1(\gamma), \phi_0(c_w,c_z) = 0\right\}\\
    S_2 = &\left\{c_w\in \mathbb{R}_{\ge 0}: \phi_0(c_w,1)=0 \right\} \cup \left\{c_w=0\right\}\\
    S_3 = &\left\{c_z\in \mathbb{R}_{\ge 0}: c_z = \phi_1(-\gamma), c_z > 1 \right\} 
    \cup \left\{c_z\in \mathbb{R}_{\ge 0}: c_z = \phi_1(\gamma), c_z < 1 \right\},
\end{align*}
and
\begin{align*}   
&{}\phi_0: (c_w, c_z) \mapsto e^{-c_w^2c^2/2}-(1+c^2(c_w-c_z)c_w)e^{-c_w^2c^2/2},\\
&{}\phi_1: \gamma \mapsto \frac{h}{2c}\left(\gamma + \sqrt{\gamma^2+4/h^2}\right).
\end{align*}
\end{lemma}

\begin{proof}
Let $c:=\|x-t\|/h>0$. Write $z_L=c_zx+(1-c_z)t$, $w_L=c_wx+(1-c_w)t$, for $c_z, c_w\in \mathbb{R}$. Then $\|z_L-t\|=\|c_z(x-t)\|=|c_z|\cdot\|x-t\|=|c_z|ch$, $\|w_L-t\|=|c_w|ch$, $\|w_L-z_L\|=|c_w-c_z|ch$ and $\|x-z_L\|=|c_z-1|ch$. Thus, instead of working with $w, z\in \mathbb{R}^d$, we can work with $c_w, c_z\in \mathbb{R}$ by Lemma~\ref{lm:gaussker_md_to_1d}, where
\[
    \frac{|\kappa_t(w_L)-\kappa_t(z_L)|}{\|w_L-z_L\| g(x,z_L)} = \frac{|e^{-c_w^2c^2/2}-e^{-c_z^2c^2/2}|}{(|c_w-c_z|ch)e^{\gamma|c_z-1|ch}}.
\]
The rest of the (rather lengthy) calculations follow the generic procedure outlined in Section~\ref{sec:smoothsens_C1}, which we defer to Appendix~\ref{appendix:lm_gauss_smoothsens_xneqt}.
\end{proof}
For the case where $x=t$, we have:
\begin{lemma} 
\label{lm:gauss_smoothsens_xeqt}
Fix $t \in \mathbb{R}^d$, $\gamma > 0$ and $h > 0$. For the function $\kappa_t: x\mapsto e^{-\frac{\|x-t\|^2}{2h^2}}$ and smooth growth function $\gE: (x,x')\mapsto e^{\gamma\|x-x'\|}$, we have for $x = t$, the $\gE$-smooth sensitivity
    \begin{align*}
        B^*(t) =  \max &{}\left(\max_{(c_w,l_z)\in S_1} \frac{l_z}{h}\cdot\frac{c_w l_z e^{-c_w^2l_z^2/2}}{e^{\gamma l_z h}}, \max_{l_w\in S_2}\frac{|e^{-l_w^2/2}-1|}{l_w h}, \max_{l_z\in S_3}\frac{l_z}{h}\cdot \frac{e^{-l_z^2/2}}{e^{\gamma l_z h}}\right),
        \end{align*}
where
\begin{align*}
    S_1 = &\left\{(c_w,l_z>0)\in \mathbb{R}^2: \phi_0(c_w,\phi_1(c_w)) = 0, l_z=\phi_1(c_w) \right\} \\
    S_2 = &\left\{l_w\in \mathbb{R}_{\ge 0}: l_w = \sqrt{-2 W^{-1}(-e^{-1/2}/2)-1}\right\}\\
    S_3 = &\left\{l_z\in \mathbb{R}_{\ge 0}: l_z = \phi_1(1) \right\},
\end{align*}
$W^{-1}(\cdot)$ is the Lambert $W$ function and
\begin{align*}    
&{}\phi_0: (c_w, l_z) \mapsto e^{-l_z^2/2}-(1+l_z^2(c_w-1)c_w)e^{-c_w^2l_z^2/2}, \\
&{}\phi_1: c_w \mapsto \frac{h}{2c_w^2}\left(-\gamma + \sqrt{\gamma^2+4c_w^2/h^2}\right).
\end{align*}
\end{lemma}
The derivations for this lemma can be found in Appendix~\ref{appendix:smoothsens_KDE}.

\subsection{Smooth Sensitivity for Other Queries}
When $U\subseteq \mathbb{R}^d$, in order to efficiently compute the smooth sensitivity, we've shown in the previous subsection, that we can reduce the dimensions of the variables involved. In particular, 
we showed that this is possible for functions which depend on the coordinates only through their Euclidean distance to some reference point $t\in \mathbb{R}^d$. In fact, this is also possible for functions of the form $f(x)=f_0(\varphi(x))$, where $\varphi(x):=\beta_0 + \langle \beta, x\rangle$ for some $\beta_0\in \mathbb{R}$ and $\beta\in \mathbb{R}^d$.
\begin{lemma} 
\label{lm:compute_smooth_others}
Fix $\beta_0\in \mathbb{R}$, $\beta = [\beta_1,\beta_2,\dotsb,\beta_d]\in \mathbb{R}^d$ such that $\|\beta\|\neq 0$. 
Let $f:\mathbb{R}^d \rightarrow \mathbb{R}$ be defined by $f(x)=f_0(\varphi(x))$, where $f_0: \mathbb{R} \rightarrow \mathbb{R}$ and $\varphi(x):=\beta_0 + \langle \beta, x\rangle$.
    Let $(w^*,z^*)$ be such that $B^*(x)=\frac{|f(w^*)-f(z^*)|}{\|w^*-z^*\|g(x,z^*)}$. 
    Then, there is $(w_a, z_c)$ where $z_c:=x+\frac{c}{\|\beta\|}\beta$, $w_a:=z_c+\frac{a}{\|\beta\|}\beta$ for some $a, c\in \mathbb{R}$ such that $\frac{|f(w_a)-f(z_c)|}{\|w_a-z_c\|g(x,z_c)}\ge B^*(x)$.
\end{lemma}

Thus, we can work with the variables $w_a,z_z\in\mathbb{R}$ instead of $w,z\in \mathbb{R}^d$. The proof is deferred to Appendix~\ref{appendix:lm_compute_smooth_others}.

\section{Experiments}

In this section, we evaluate the performance of our smooth sensitivity-based mechanisms for the three applications we described in Section~\ref{sec:app}. The evaluation is performed on two real-world datasets: the California household income dataset extracted from the 2022 CENSUS \cite{census2022cali}, and the New York motor vehicle collisions dataset \cite{nyc2024mvc}. We adopt the Euclidean metric\footnote{If one desires to measure privacy using the ratios of inputs, e.g., income numbers that are 2x apart shall be $1$-distinguishable, then we can first take $\log_2$ of the $x_i$'s and still use Euclidean metric with $\varepsilon =1$. I.e. $\dist(x_i,x_i')=|\log_2(x_i)-\log_2(x_i')|.$} for GP.  

We compare our mechanism against the following baseline GP mechanisms: 
\begin{description}
    \item[$M_0$:] $M_0(x_i) = f(x_i+Z_i),\;\; Z_i$ is drawn from $d$-dimensional Planar Laplace distribution with scale ${1}/{\varepsilon}$ for $x_i\in \mathbb{R}^d$.
 \item[$M_1$:] $ M_1(x_i) =  \exp(-\frac{(\|x_i-t\|+Z_i)^2}{2h^2}), \;\;Z_i\sim \mathrm{Lap}\left(0, {1}/{\varepsilon}\right).$
\item[$M_2$:] $ M_2(x_i) = f(x_i)+Z_i, \;\;Z_i\sim \mathrm{Lap}\left(0, {K_f}/{\varepsilon}\right).$
\end{description}
1) $M_0$ is the noise-a-priori method that privatizes each $x_i$ and the data analyst then applies $f$; 2) $M_1$ is another version of the noise-a-priori method that is only applicable to Gaussian KDE, which privatizes the distance of each $x_i$ to the reference point $t$; 3) $M_2$ is the baseline noise-a-posterior method that privatizes $f(x_i)$ by adding noise proportional to the global Lipschitz constant $K_f$.
Our mechanism $M_3$ is given by
\begin{description}
\item[$M_3$:]  $M_3(x_i)= f(x_i)+Z_i,\;\;  Z_i\sim \mathcal{T}_{\nu}(0,B^*(x_i)/\eta)$,
\end{description}
where $\varepsilon = \nu \gamma + \frac{\nu+1}{2\sqrt{\nu}}\eta$. We set $\nu=3, \nu\gamma=\frac{1}{3}\varepsilon$ in our experiments. 
For reference, we also provide experimental results under the LDP model with parameter $\epsilon'$, i.e., adding $1/\epsilon'$ Laplace noise to $f(x_i)$.
\begin{description}
\item[$M_{\mathrm{LDP}}$:] $M_{\mathrm{LDP}}(x_i)= f(x_i)+Z_i,\;\; Z_i\sim \mathrm{Lap}\left(0, {1}/{\epsilon'}\right)$.
\end{description}
We set $\epsilon'=\varepsilon\cdot C$.  Note that LDP provides a uniform distinguishability of $\epsilon'$ while GP's distinguishability is $\varepsilon \cdot \dist(x_i, x_i')$.  So for $\dist(x_i, x_i') < C$, GP has stronger privacy; whereas for $\dist(x_i, x_i') > C$, LDP is stronger. The code for all experiments can be found at: \url{https://github.com/PublicRepo2024/SmoothGP}. 
 
We measure the error using the mean squared error (MSE) and the aggregated squared error (ASE). The MSE is the mean of squared error over $m$ repetitions of the experiment. The ASE is computed as the mean of the total squared error over a set of queries. Specifically, for queries parameterized by a value in $\mathbb{R}^2$ (i.e., the threshold $T=(T_1,T_2)$ in two-way thresholds, and the query point $t$ in Gaussian KDE), we evaluate the mechanisms over an evenly spaced $k\times k$ grid $G:=\{v_{j,l}: 0\le j,l\le k\}$. The ASE across the grid is
\[
\mathrm{ASE}(M)=\frac{1}{(k+1)\times (k+1)} \sum_{j=0}^{k} \sum_{l=0}^{k} (\tilde{A}_{j,l}(M)-A_{j,l})^2
\]
where $A_{j,l}:=\frac{1}{n}\sum_{i\in[n]}f(x_i;v_{j,l})$, and $\tilde{A}_{j,l}(M)$ is the estimate produced by mechanism $M$ for the query corresponding to $v_{j,l}$.

\begin{figure*}[htbp]
     \centering
         \begin{subfigure}[t]{0.323\linewidth}%
            \centering
            \includegraphics[width=\textwidth]{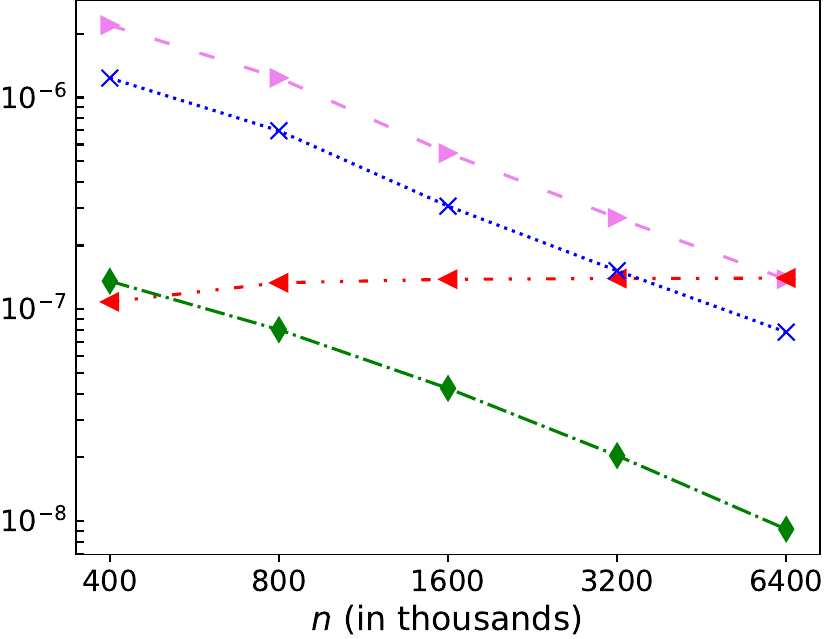}
             \vskip -.08in
             \subcaption{\scriptsize{$T=\$10000, \varepsilon ={1}/{\$800}, \\
             {\;\;\;\;\;\;\;} C=\$1200$.}}
            \;
            \label{fig:1way_thres_expT10klp}
         \end{subfigure}
        \hfill
         \begin{subfigure}[t]{0.323\linewidth}%
            \centering
            \includegraphics[width=\textwidth]{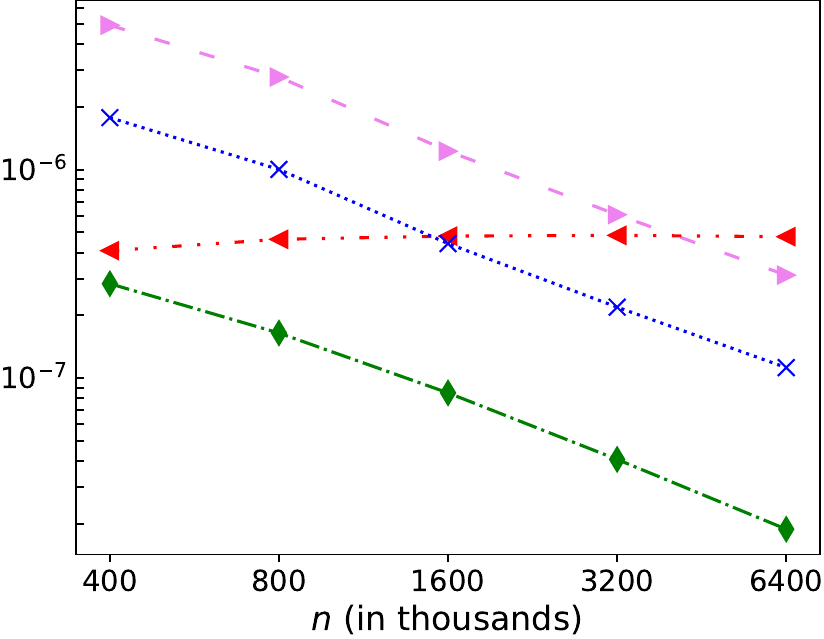}
            \vskip -.08in
            \subcaption{\scriptsize{$T=\$10000, \varepsilon ={1}/{\$1200}, \\
            {\;\;\;\;\;\;\;} C=\$1200$.}}
            \;
            \label{fig:1way_thres_expT10kmp}
         \end{subfigure}
         \hfill
         \begin{subfigure}[t]{0.323\linewidth}%
            \centering
            \includegraphics[width=\textwidth]{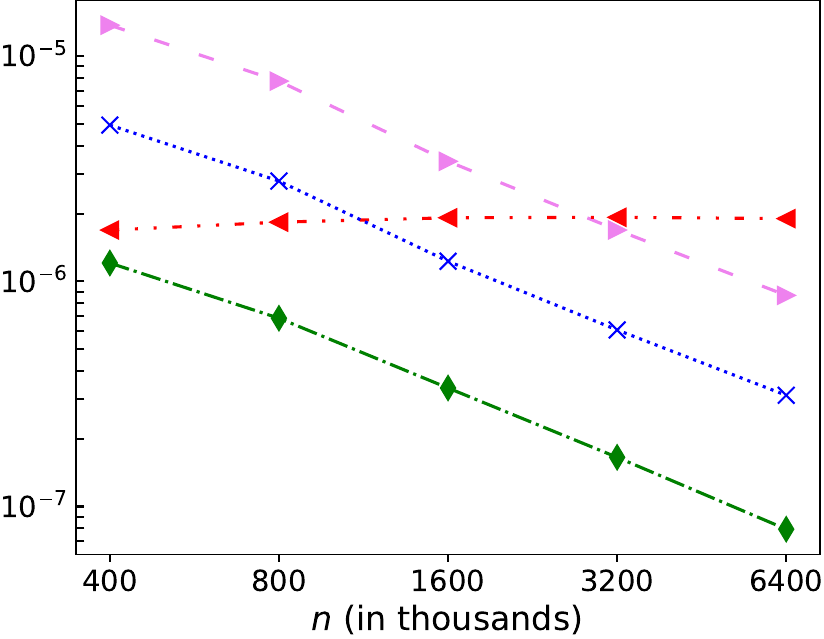}
            \vskip -.08in
            \subcaption{\scriptsize{$T=\$10000, \varepsilon ={1}/{\$2000}, \\
            {\;\;\;\;\;\;\;} C=\$1200$.}}
            \label{fig:1way_thres_expT10khp}
         \end{subfigure}

            \begin{subfigure}[t]{0.323\linewidth}%
            \centering
            \includegraphics[width=\textwidth]{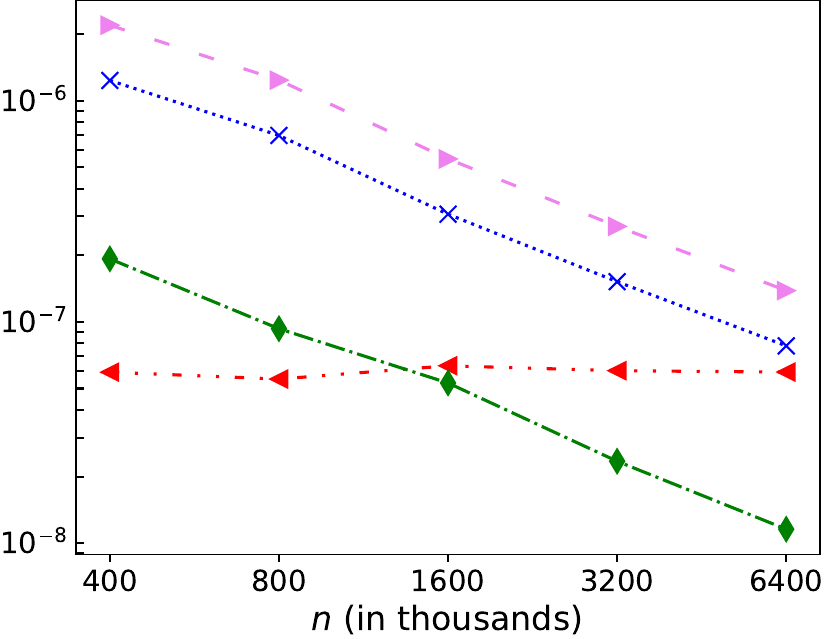}
             \vskip -.08in
            \subcaption{\scriptsize{$T=\$500000, \varepsilon ={1}/{\$16000},\\
            {\;\;\;\;\;\;\;}C=\$24000$.}}
            \label{fig:1way_thres_expT500klp}
         \end{subfigure}
        \hfill
         \begin{subfigure}[t]{0.323\linewidth}%
            \centering
            \includegraphics[width=\textwidth]{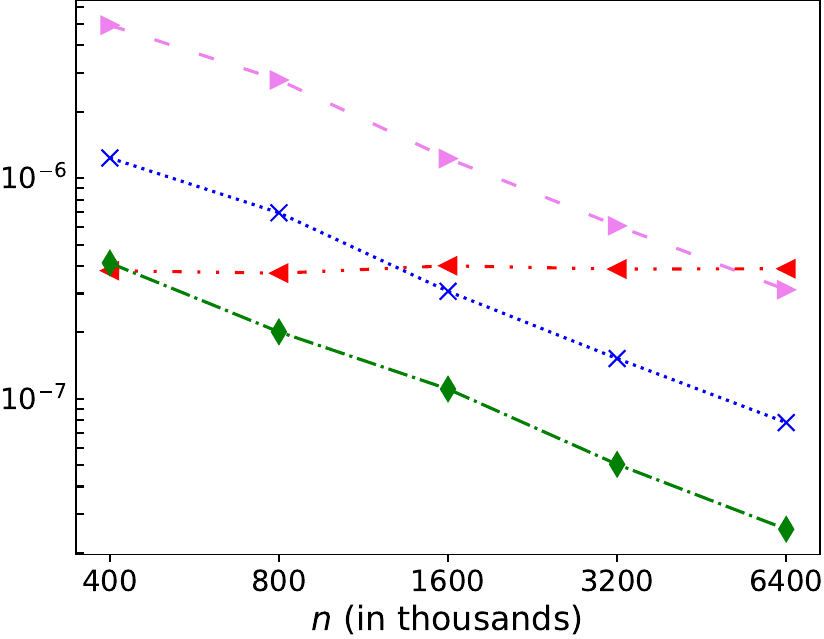}
            \vskip -.08in
            \subcaption{\scriptsize{$T=\$500000, \varepsilon =1/\$24000,\\
            {\;\;\;\;\;\;\;}C=\$24000$.}}
            \label{fig:1way_thres_expT500kmp}
         \end{subfigure}
         \hfill
         \begin{subfigure}[t]{0.323\linewidth}%
            \centering
            \includegraphics[width=\textwidth]{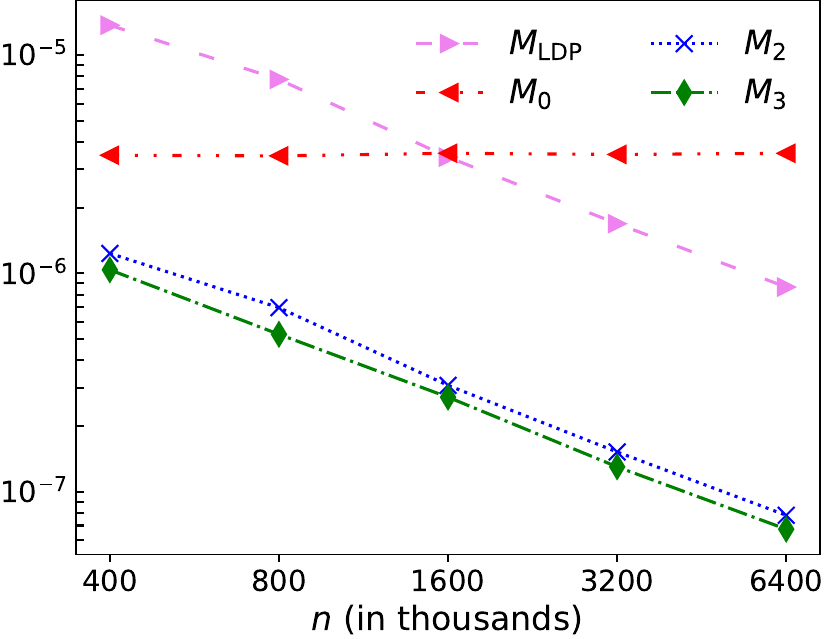}
            \vskip -.08in
            \subcaption{\scriptsize{$T=\$500000, \varepsilon =1/\$40000, \\
            {\;\;\;\;\;\;\;}C=\$24000$.}}
            \label{fig:1way_thres_expT500khp}
         \end{subfigure}%
        \vskip -.08in
         \caption{MSE for one-way threshold query. Threshold amounts in USD.}
         \label{fig:1way_thres_hhincome}
    \vskip -.08in
\end{figure*}
\paragraph{Threshold Query.}
In this experiment, the query is $A=\frac{1}{n}\sum_i f_{\tau}(x_i)$ where $f_{\tau}$ is the soft threshold function as defined in Figure \ref{fig:1way_thres_cons}.  We set $\tau=\min(0.2T,2/\varepsilon)$, i.e., values that fall within a distance of (at most) $10\%$ of $T$ to $T$ are assigned an interpolated value in $[0,1]$.

We draw samples of household incomes from the California household income dataset according to the household weights provided in the dataset. 
We examine the error w.r.t. to changes in $n$, $\varepsilon$ and threshold levels in Fig.~\ref{fig:1way_thres_hhincome}. We measure the MSE (computed with $f_{\tau}$) over $m=500$ repetitions. In each subplot, the MSE w.r.t. various sample sizes ($n$) are reported. Across the columns, privacy strength is varied from low to high (values of $\varepsilon$ ordered from large to small, specified per unit $\$1$USD), while each row corresponds to a different threshold level ($T$). 

For $T=\$10,000$ in Fig.~\ref{fig:1way_thres_expT10klp}-\ref{fig:1way_thres_expT10khp} and $T=\$500,000$ in Fig.~\ref{fig:1way_thres_expT500klp}-\ref{fig:1way_thres_expT500khp}, most of $x_i$'s are far from the ``area of action'' --- the area around the threshold. For plots corresponding to these threshold levels, we see a clear advantage of $M_3$ over $M_2$. $M_0$ also performs well in these scenarios for large values of $\varepsilon$, since there are very few points near the threshold and adding a relatively small noise to $x_i$ has nearly no effect on changing $f(x_i)$ for most $x_i$'s; however, the error of $M_0$ does not benefit from larger sample sizes due to its inherent bias.
Overall, we see that all mechanisms have reduced error as $\varepsilon$ increases; the reduction in $M_2$ is not obvious because $\varepsilon$ only increases by a factor of $2.5$ from the right column to the left. On the other hand, this shows that with only a modest increase in $\varepsilon$, $M_3$ is able to gain an order of magnitude improvement in performance.
We also provide experimental results on $T=\$100,000$ and $T=\$1,000,000$ in Fig.~\ref{fig:1way_thres_hhincome2} of Appendix~\ref{appendix:add_1way_exp}.

\begin{figure}[h]
     \centering
         \begin{subfigure}[t]{0.23\linewidth}%
            \centering
            \includegraphics[width=\textwidth]{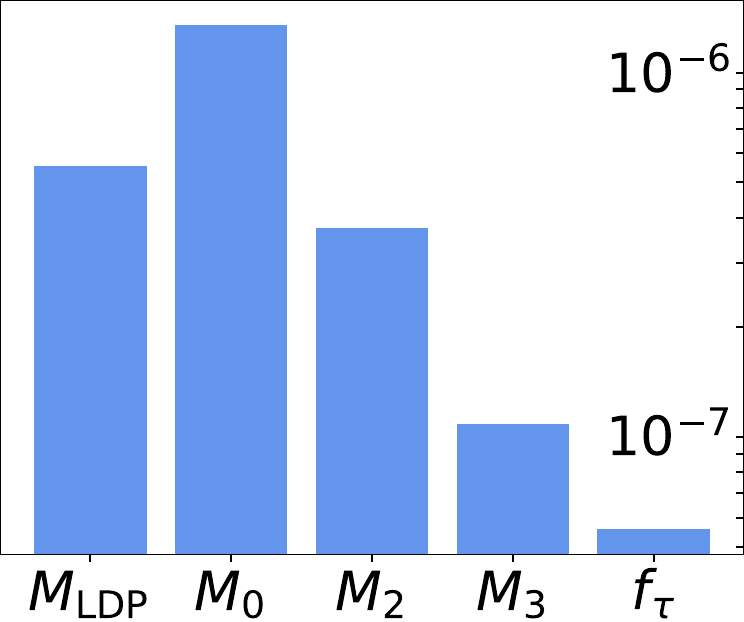}%
             \vskip -.08in            
             \subcaption{$\varepsilon={1}/{\$8000}$.}
            \label{fig:2way_thres_lp}%
         \end{subfigure}%
        \;\;
         \begin{subfigure}[t]{0.23\linewidth}%
            \centering
            \includegraphics[width=\textwidth]{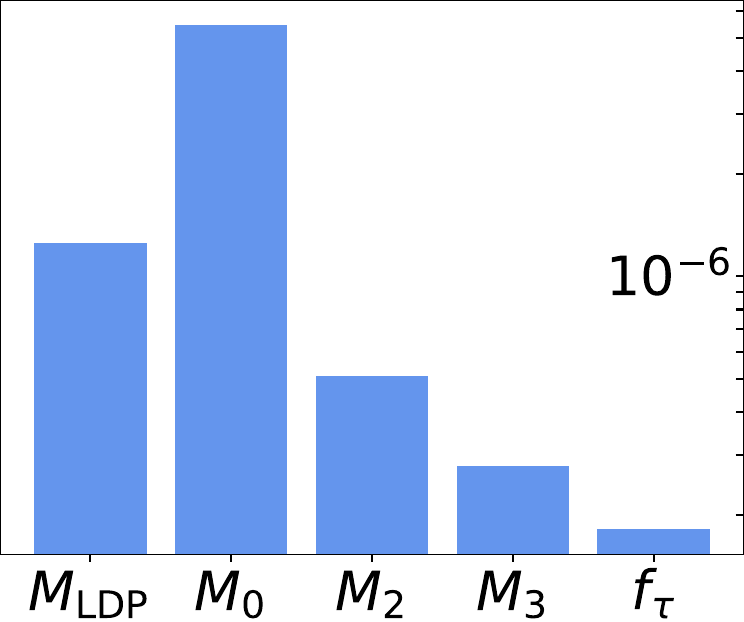}%
            \vskip -.08in
            \subcaption{$\varepsilon=1/\$12000$.}
            \label{fig:2way_thres_mp}%
         \end{subfigure}%
         \;\;
            \begin{subfigure}[t]{0.23\linewidth}%
            \centering
            \includegraphics[width=\textwidth]{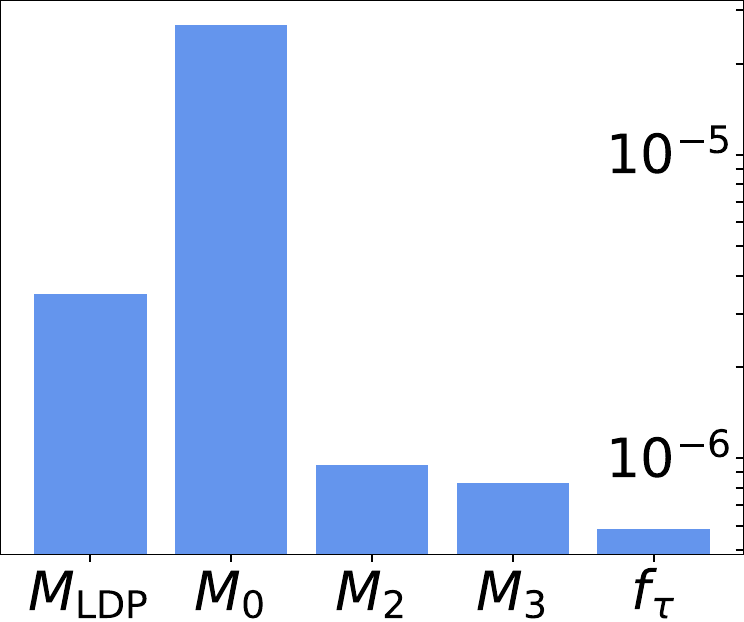}%
            \vskip -.08in
    \subcaption{$\varepsilon=1/\$20000$.}
            \label{fig:2way_thres_hp}%
         \end{subfigure}%
         
         \medskip
         
            \begin{subfigure}[t]{0.23\linewidth}%
            \centering
            \includegraphics[width=\textwidth]{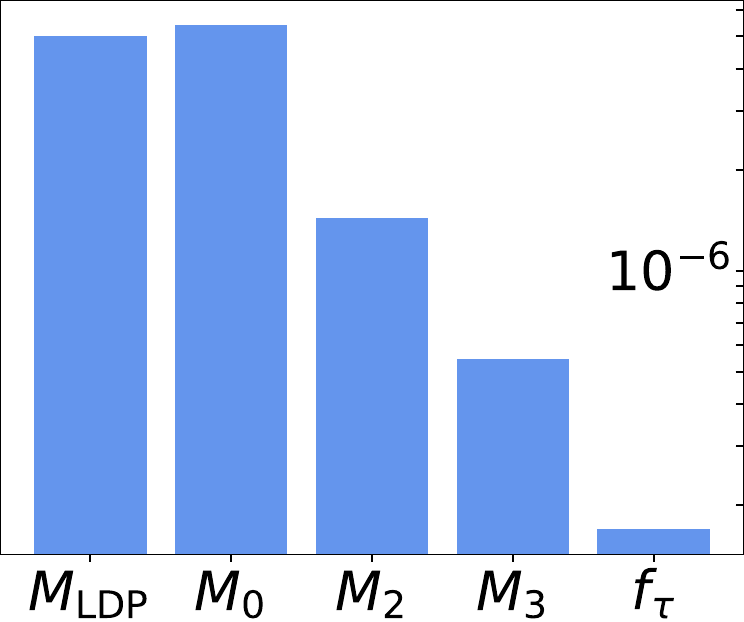}%
            \vskip -.08in
            \subcaption{$n=400000$.}%
            \label{fig:2way_thres_smalln}%
         \end{subfigure}
         \;
            \begin{subfigure}[t]{0.23\linewidth}%
            \centering
            \includegraphics[width=\textwidth]{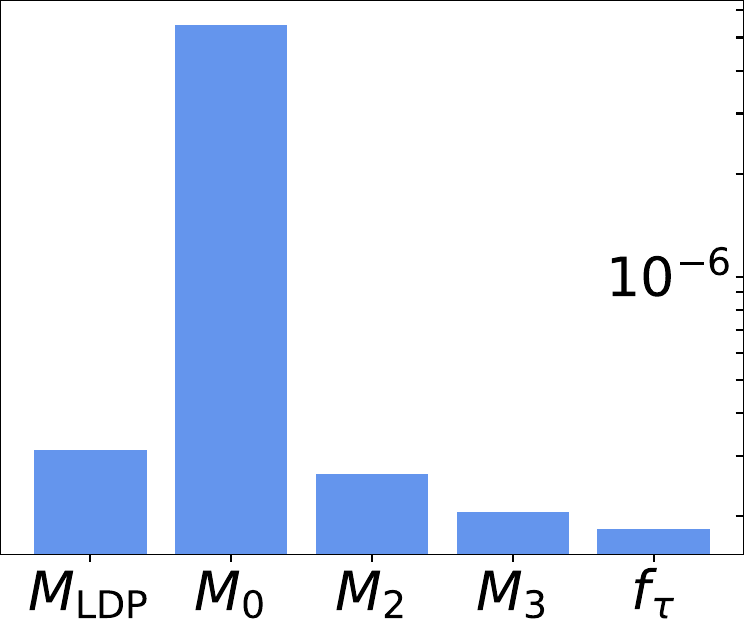}%
            \vskip -.08in
            \subcaption{$n=6400000$.}%
            \label{fig:2way_thres_largen}%
         \end{subfigure}
         \caption{Two-way threshold query. ASE computed on $33\times 33$ grid. Default parameters: $\varepsilon=1/\$12000$ and  $n=1600000$, corresponding to  (\subref{fig:2way_thres_mp}). $C=\$12000$.}
         \label{fig:2way_thres_hhincomedebt}
    \vskip -.08in
\end{figure}

\paragraph{Two-way Threshold Query.}
In this experiment, the query is $A=\frac{1}{n}\sum_{i\in[n]} \mathbb{1}\{x_{i,1}>T_1\}\mathbb{1}\{x_{i,2}>T_2\}$ for given thresholds $(T_1,T_2)$. We set $\tau=\min(0.2\|(T_1,T_2)\|,2/\varepsilon)$ for $f_{\tau}$. 
We generate household income and debt data using a correlated log-normal model with correlation ratio $\rho = 0.4$ (a correlation ratio between assets/income and debt of $>0.4$ has been observed in recent years \cite{mason2018income,kuhn2017great}), where the parameters for income ($\mu_1$, $\sigma_1$ for mean and standard deviation of the log values) are fitted from the California income dataset at various percentiles; the parameters used for debt are $\mu_2=\ln(1.791)+\mu_1, \sigma_2=\sigma_1$ where $1.791$ is the household debt-to-income ratio for California in 2022 \cite{frs2024efa}. We evaluate the mechanisms across an evenly spaced grid 
where $\$10,000\le T_1,T_2\le \$1,000,000$ and report the mean of the ASE over 50 repetitions in Fig.~\ref{fig:2way_thres_hhincomedebt}.

In the first row, we examine the error w.r.t. to the privacy level, where we fix $n=1600000$. We see that as the value of $\varepsilon$ increases from right to left, the error is reduced for all the mechanisms. The reduction is more prominent in $M_0$ and $M_3$. In the second row, we examine the error w.r.t to sample size $n$, where we fix $\varepsilon=1/\$12000$. We see a reduction in the error in $M_2$ and $M_3$ as the sample size $n$ increases (nearly converging to that for the non-private $f_{\tau}$ for the larger $n$), while $M_0$ does not benefit from it; in fact, there seems to be an increase in error for $M_0$ (due to increase in the bias) for the larger value of $n$.
Overall, we see that $M_3$ performs the best over the entire set of queries. The analysis is similar to that for the one-way threshold experiments: if many points are far away from the threshold $(T_1,T_2)$, then $M_0$ and $M_3$ would perform well. Otherwise, $M_2$ would perform the best, but $M_3$ would have similar performance as well. Thus, it would be preferable to use $M_3$ for this query, when little is known about the distribution of distances of the $x_i$'s relative to the threshold.

\begin{figure*}[htbp]
        \centering
            \includegraphics[width=0.9\textwidth]{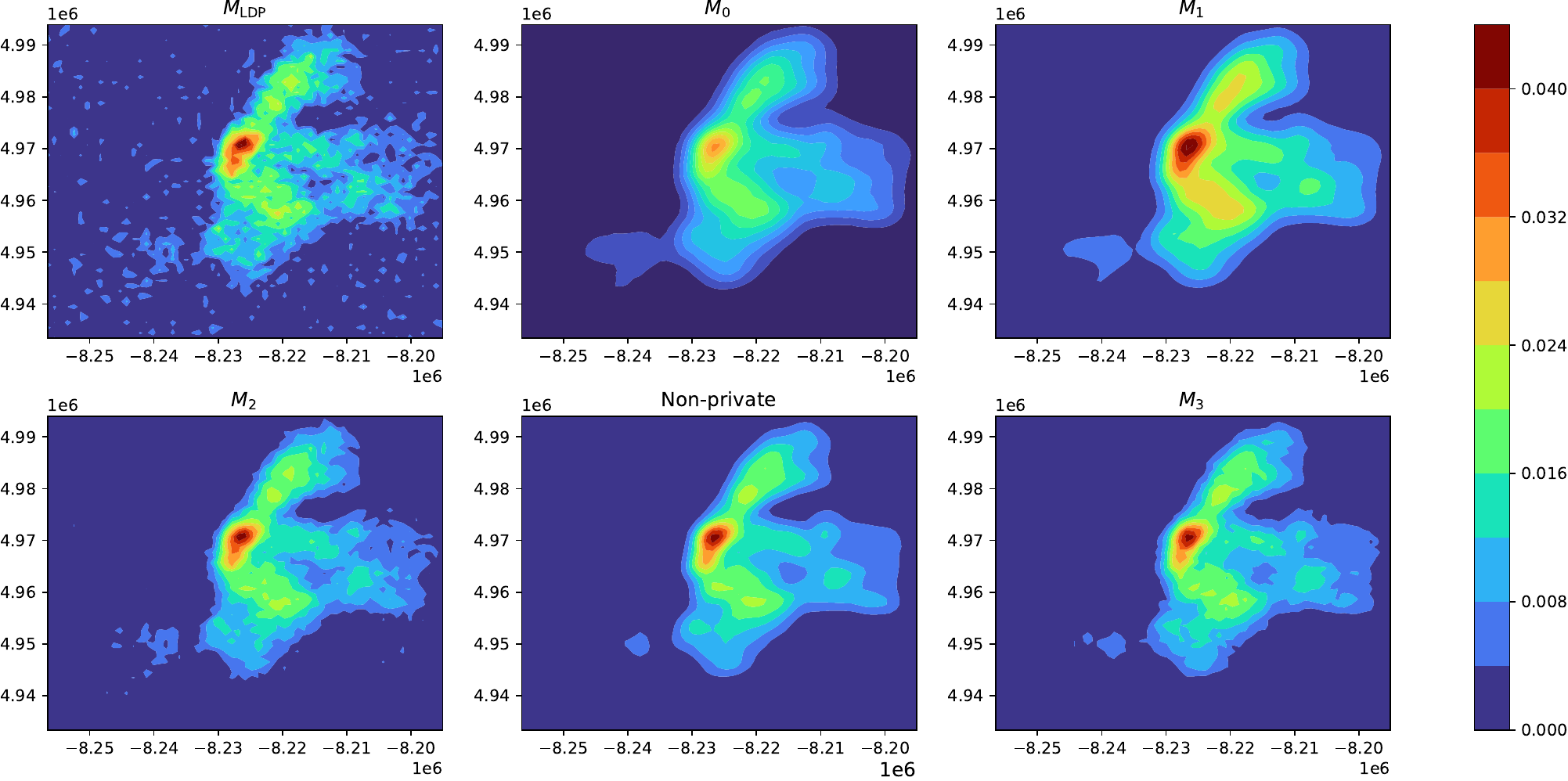}%
         \vskip -.08in
         \caption{KDE on New York motor vehicle collision dataset, computed on $60\times 60$ grid.  $\varepsilon=1/1000\mathrm{m}, h=w, n=200000$; corresponding error plot given in Fig.~\ref{fig:kde_bar_ref}.}
         \label{fig:kde_nymvc_ref}
    \vskip -.01in
\end{figure*}
\paragraph{Gaussian KDE}
\begin{figure}[htbp]
     \centering
           \begin{subfigure}[t]{0.23\linewidth}%
            \centering
            \includegraphics[width=\textwidth]{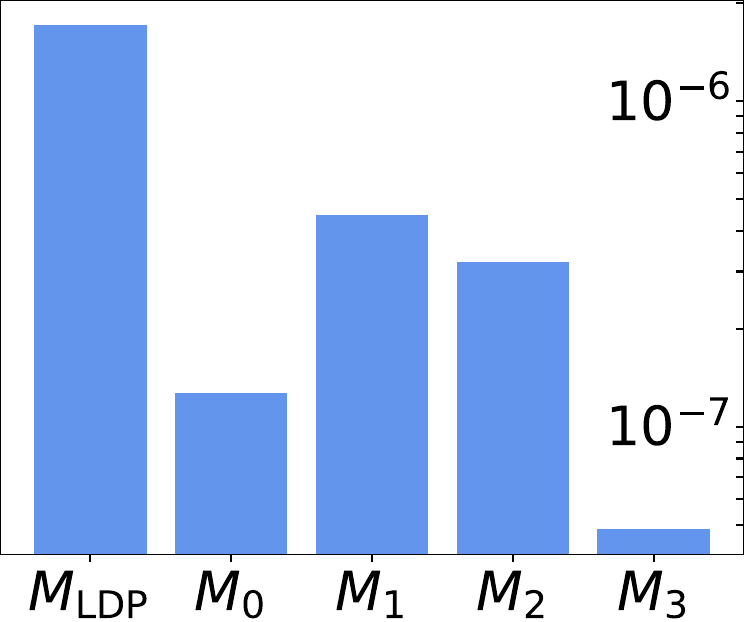}%
            \vskip -.08in
            \subcaption{$\varepsilon=1/500\mathrm{m}$.}%
            \label{fig:kde_bar_lp}
         \end{subfigure}
         \;
            \begin{subfigure}[t]{0.23\linewidth}%
            \centering
            \includegraphics[width=\textwidth]{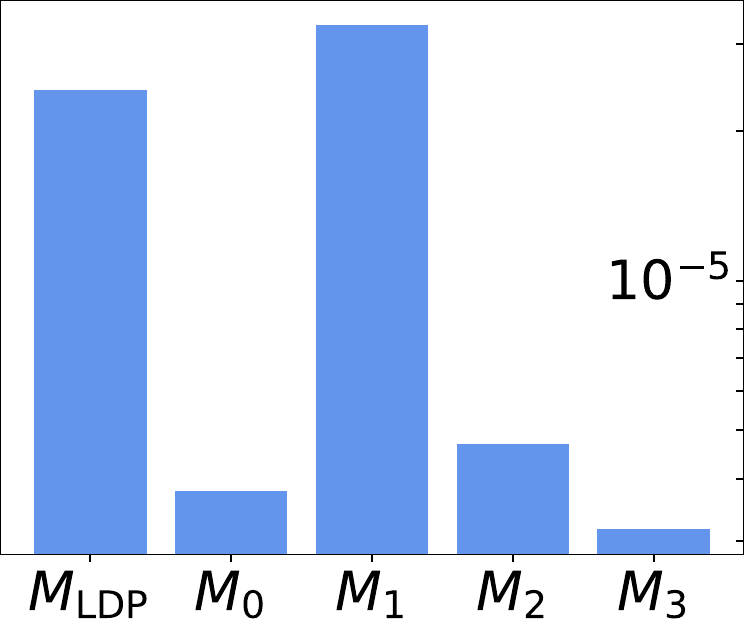}%
            \vskip -.08in
            \subcaption{$\varepsilon=1/2000\mathrm{m}$.}%
            \label{fig:kde_bar_hp}
         \end{subfigure}
         
         \medskip
         
         \begin{subfigure}[t]{0.23\linewidth}%
            \centering
            \includegraphics[width=\textwidth]{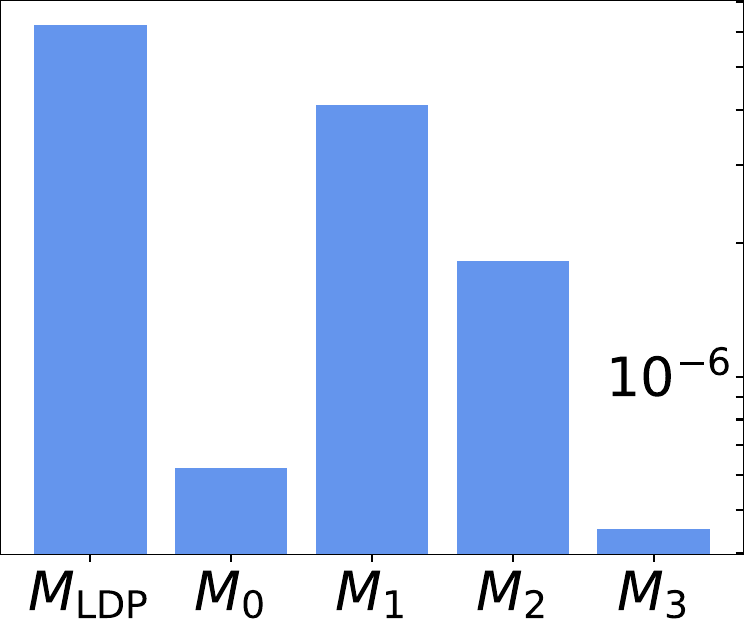}%
             \vskip -.08in            
             \subcaption{$h=0.8w$}
            \label{fig:kde_bar_smallh}
         \end{subfigure}%
\;\;
         \begin{subfigure}[t]{0.23\linewidth}%
            \centering
            \includegraphics[width=\textwidth]{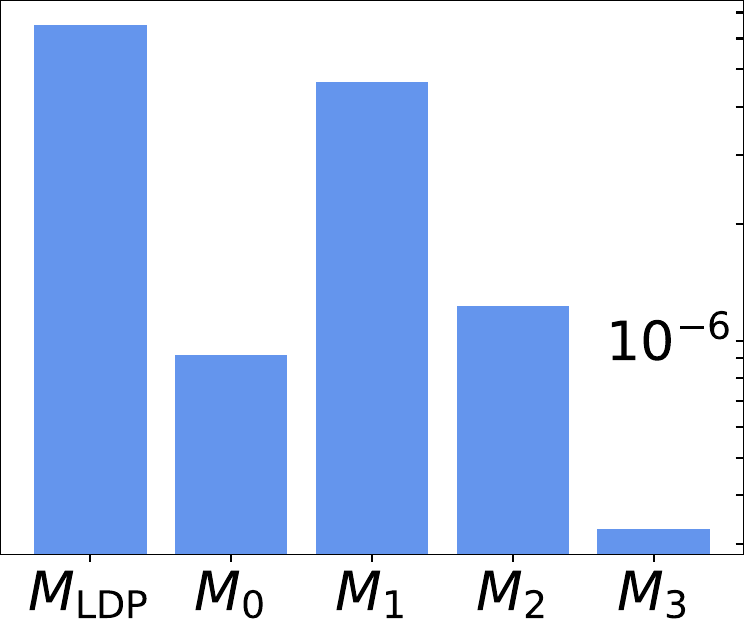}%
            \vskip -.08in
            \subcaption{$h=w$.}
            \label{fig:kde_bar_ref}
         \end{subfigure}%
\;\;
            \begin{subfigure}[t]{0.23\linewidth}%
            \centering
            \includegraphics[width=\textwidth]{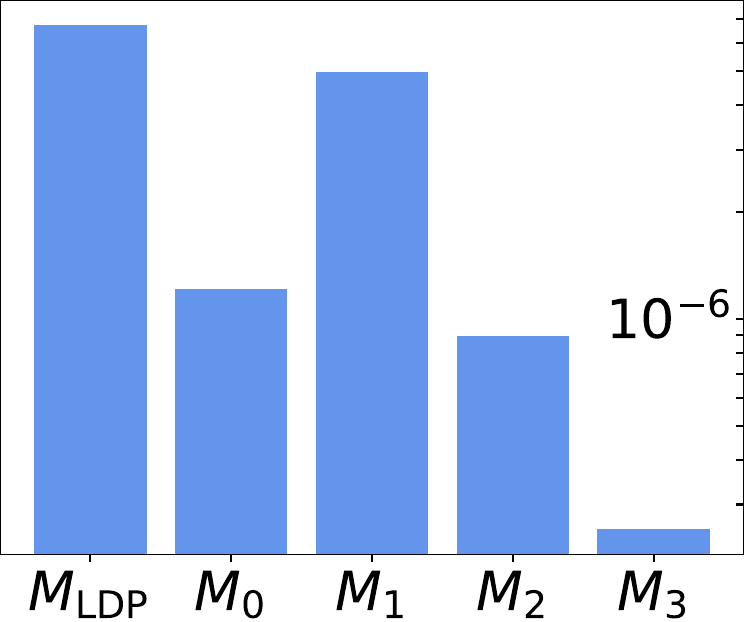}%
            \vskip -.08in
    \subcaption{$h=1.2w$.}
            \label{fig:kde_bar_largeh}
         \end{subfigure}%
         
         \medskip
         
            \begin{subfigure}[t]{0.23\linewidth}%
            \centering
            \includegraphics[width=\textwidth]{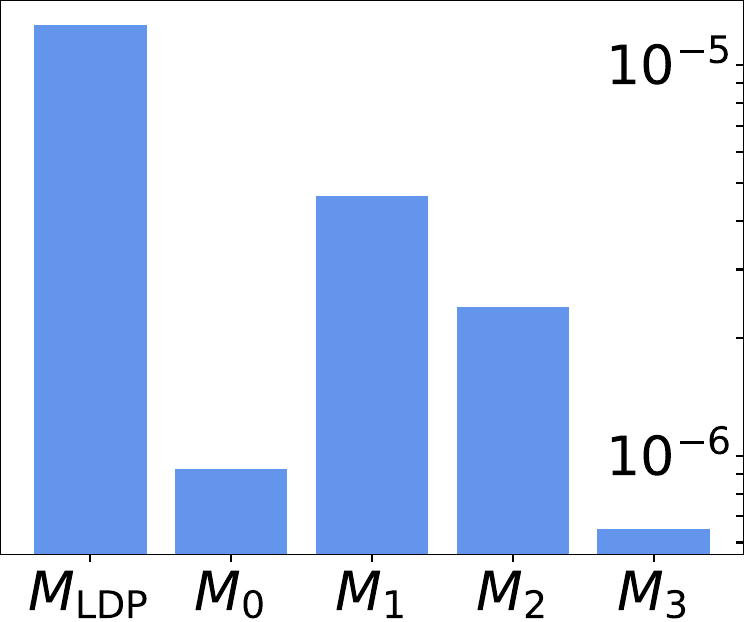}%
            \vskip -.08in
            \subcaption{$n=100000$.}%
            \label{fig:kde_bar_smalln}
         \end{subfigure}
         \;
            \begin{subfigure}[t]{0.23\linewidth}%
            \centering
            \includegraphics[width=\textwidth]{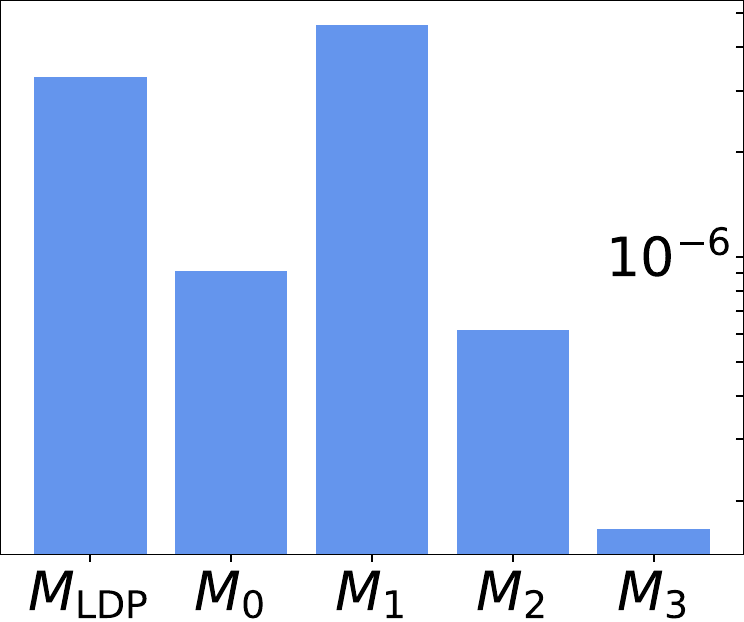}%
            \vskip -.08in
            \subcaption{$n=400000$.}%
            \label{fig:kde_bar_largen}
         \end{subfigure}
         \vskip -.08in
         \caption{Gaussian KDE query. ASE computed on $60\times 60$ grid. Default parameters: $n=200000$, $h=w$ and $\varepsilon=1/1000\mathrm{m}$, corresponding to (\subref{fig:kde_bar_ref}). $C=1000\mathrm{m}$.}
         \label{fig:kde_barplots}
    \vskip -.08in
\end{figure}
In this experiment, $A=\frac{1}{n}\sum_{i\in[n]}\kappa_t(x_i)$, where $\kappa_t:x_i\mapsto e^{-\frac{\|x_i-t\|^2}{2h^2}}$. We draw samples of collision locations (converted from GPS coordinates to $\mathbb{R}^2$ tuples via the Mercator projection) uniformly at random from the New York motor vehicle collisions dataset, in an area of approximately $110\mathrm{km} \times 90\mathrm{km}$. 
We compare the KDE estimates produced by the mechanisms against the true non-private values which are shown as contour plots (based on one run). 
The contour plot corresponding to the default parameters is presented in Fig.~\ref{fig:kde_nymvc_ref}, the remaining contour plots are presented in Appendix~\ref{appendix:add_kde_exp}.
We report the mean of the ASE evaluated on a $60\times60$ grid over 10 repetitions (there is very little variation between repetitions) in the bar plots in Fig.~\ref{fig:kde_barplots}. The first row corresponds to error w.r.t. to changes in $\varepsilon$, the second row corresponds to different values of $h$, and the third row corresponds to that w.r.t. changes in $n$. All mechanisms have reduction in error for the larger value of $\varepsilon$, where the reduction is most prominent for $M_0$ and $M_3$. $M_2$ and $M_3$ have smaller error for the larger value of $h$, while $M_0$ has smaller error for the smaller value of $h$ (see Appendix~\ref{appendix:add_kde_exp} for a further discussion). Also, $M_2$ and $M_3$ benefit from a larger sample size $n$, while $M_0$ and $M_1$ do not. Overall, $M_3$ has the best performance.

\section{Conclusions}
Under DP, mechanisms that add instance-specific noise are known to yield superior utility over worst-case noise-adding mechanisms.  However, such mechanisms are still much under-studied under GP.  This work takes a first step in this direction, by generalizing the smooth sensitivity framework from DP to GP.  It remains an interesting question whether other instance-specific mechanisms from DP can also be applied to GP; some ideas from this work could be useful along this direction.

\bibliographystyle{alpha}
\bibliography{smoothgp}

\newcommand{\etalchar}[1]{$^{#1}$}
\begin{thebibliography}{oGotFRS24}

\bibitem[ABCP13]{andres2013geo}
Miguel~E Andr{\'e}s, Nicol{\'a}s~E Bordenabe, Konstantinos Chatzikokolakis, and Catuscia Palamidessi.
\newblock Geo-indistinguishability: Differential privacy for location-based systems.
\newblock In {\em Proceedings of the 2013 ACM SIGSAC conference on Computer \& communications security}, pages 901--914, 2013.

\bibitem[Abo18]{abowd2018us}
John~M Abowd.
\newblock The us census bureau adopts differential privacy.
\newblock In {\em Proceedings of the 24th ACM SIGKDD International Conference on Knowledge Discovery \& Data Mining}, pages 2867--2867, 2018.

\bibitem[AC23]{aggarwal2023some}
Manisha Aggarwal and {\c{S}}tefan Cobza{\c{s}}.
\newblock On some lipschitz-type functions.
\newblock {\em Journal of Mathematical Analysis and Applications}, 517(2):126631, 2023.

\bibitem[AD20]{asi2020instance}
Hilal Asi and John~C Duchi.
\newblock Instance-optimality in differential privacy via approximate inverse sensitivity mechanisms.
\newblock {\em Advances in neural information processing systems}, 33:14106--14117, 2020.

\bibitem[BBZ05]{bronstein2005quasi}
Alexander~M Bronstein, Michael~M Bronstein, and Michael Zibulevsky.
\newblock Quasi maximum likelihood mimo blind deconvolution: Super-and sub-gaussianity versus consistency.
\newblock {\em IEEE transactions on signal processing}, 53(7):2576--2579, 2005.

\bibitem[BG15]{beer2015locally}
Gerald Beer and M~Isabel Garrido.
\newblock Locally lipschitz functions, cofinal completeness, and uc spaces.
\newblock {\em Journal of Mathematical Analysis and Applications}, 428(2):804--816, 2015.

\bibitem[BS19]{bun2019average}
Mark Bun and Thomas Steinke.
\newblock Average-case averages: Private algorithms for smooth sensitivity and mean estimation.
\newblock {\em Advances in Neural Information Processing Systems}, 32, 2019.

\bibitem[CABP13]{chatzikokolakis2013broadening}
Konstantinos Chatzikokolakis, Miguel~E Andr{\'e}s, Nicol{\'a}s~Emilio Bordenabe, and Catuscia Palamidessi.
\newblock Broadening the scope of differential privacy using metrics.
\newblock In {\em Privacy Enhancing Technologies: 13th International Symposium, PETS 2013, Bloomington, IN, USA, July 10-12, 2013. Proceedings 13}, pages 82--102. Springer, 2013.

\bibitem[CKS19]{cormode2019answering}
Graham Cormode, Tejas Kulkarni, and Divesh Srivastava.
\newblock Answering range queries under local differential privacy.
\newblock {\em Proc. VLDB Endow.}, 12(10):1126–1138, jun 2019.

\bibitem[DFY{\etalchar{+}}22]{dong2022r2t}
Wei Dong, Juanru Fang, Ke~Yi, Yuchao Tao, and Ashwin Machanavajjhala.
\newblock R2t: Instance-optimal truncation for differentially private query evaluation with foreign keys.
\newblock In {\em Proceedings of the 2022 International Conference on Management of Data}, pages 759--772, 2022.

\bibitem[DJW18]{duchi2018minimax}
John~C Duchi, Michael~I Jordan, and Martin~J Wainwright.
\newblock Minimax optimal procedures for locally private estimation.
\newblock {\em Journal of the American Statistical Association}, 113(521):182--201, 2018.

\bibitem[DKY17]{ding2017collecting}
Bolin Ding, Janardhan Kulkarni, and Sergey Yekhanin.
\newblock Collecting telemetry data privately.
\newblock {\em Advances in Neural Information Processing Systems}, 30, 2017.

\bibitem[DL09]{dwork2009differential}
Cynthia Dwork and Jing Lei.
\newblock Differential privacy and robust statistics.
\newblock In {\em Proceedings of the forty-first annual ACM symposium on Theory of computing}, pages 371--380, 2009.

\bibitem[DMNS06]{dwork2006calibrating}
Cynthia Dwork, Frank McSherry, Kobbi Nissim, and Adam Smith.
\newblock Calibrating noise to sensitivity in private data analysis.
\newblock In {\em Theory of cryptography conference}, pages 265--284. Springer, 2006.

\bibitem[EPK14]{erlingsson2014rappor}
{\'U}lfar Erlingsson, Vasyl Pihur, and Aleksandra Korolova.
\newblock Rappor: Randomized aggregatable privacy-preserving ordinal response.
\newblock In {\em Proceedings of the 2014 ACM SIGSAC conference on computer and communications security}, pages 1054--1067, 2014.

\bibitem[FDY22]{fang2022shifted}
Juanru Fang, Wei Dong, and Ke~Yi.
\newblock Shifted inverse: A general mechanism for monotonic functions under user differential privacy.
\newblock In {\em Proceedings of the 2022 ACM SIGSAC Conference on Computer and Communications Security}, pages 1009--1022, 2022.

\bibitem[GGB18]{gonem2018smooth}
Alon Gonem and Ram Gilad-Bachrach.
\newblock Smooth sensitivity based approach for differentially private pca.
\newblock In {\em Algorithmic Learning Theory}, pages 438--450. PMLR, 2018.

\bibitem[HLY21]{huang2021instance}
Ziyue Huang, Yuting Liang, and Ke~Yi.
\newblock Instance-optimal mean estimation under differential privacy.
\newblock {\em Advances in Neural Information Processing Systems}, 34, 2021.

\bibitem[HRW13]{hall2013differential}
Rob Hall, Alessandro Rinaldo, and Larry Wasserman.
\newblock Differential privacy for functions and functional data.
\newblock {\em The Journal of Machine Learning Research}, 14(1):703--727, 2013.

\bibitem[HY21]{huang2021approximate}
Ziyue Huang and Ke~Yi.
\newblock Approximate range counting under differential privacy.
\newblock In {\em 37th International Symposium on Computational Geometry (SoCG 2021)}. Schloss-Dagstuhl-Leibniz Zentrum f{\"u}r Informatik, 2021.

\bibitem[KSS17]{kuhn2017great}
Moritz Kuhn, Moritz Schularick, and Ulrike~I Steins.
\newblock The great american debt boom, 1949-2013.
\newblock {\em Federal Reserve Bank of St. Louis}, 8, 2017.

\bibitem[Luu79]{luukkainen1979rings}
Jouni Luukkainen.
\newblock Rings of functions in lipschitz topology.
\newblock {\em Annales Fennici Mathematici}, 4(1):119--135, 1979.

\bibitem[LY23]{liang2023concentrated}
Yuting Liang and Ke~Yi.
\newblock Concentrated geo-privacy.
\newblock In {\em Proceedings of the 2023 ACM SIGSAC Conference on Computer and Communications Security}, pages 1934--1948, 2023.

\bibitem[Mas18]{mason2018income}
Josh~W Mason.
\newblock Income distribution, household debt, and aggregate demand: A critical assessment.
\newblock {\em Levy Economics Institute, Working Papers Series}, (901), 2018.

\bibitem[MN12]{muthukrishnan2012optimal}
Shanmugavelayutham Muthukrishnan and Aleksandar Nikolov.
\newblock Optimal private halfspace counting via discrepancy.
\newblock In {\em Proceedings of the forty-fourth annual ACM symposium on Theory of computing}, pages 1285--1292, 2012.

\bibitem[MT07]{mcsherry2007mechanism}
Frank McSherry and Kunal Talwar.
\newblock Mechanism design via differential privacy.
\newblock In {\em 48th Annual IEEE Symposium on Foundations of Computer Science (FOCS'07)}, pages 94--103. IEEE, 2007.

\bibitem[NRS07]{nissim2007smooth}
Kobbi Nissim, Sofya Raskhodnikova, and Adam Smith.
\newblock Smooth sensitivity and sampling in private data analysis.
\newblock In {\em Proceedings of the thirty-ninth annual ACM symposium on Theory of computing}, pages 75--84, 2007.

\bibitem[oGotFRS24]{frs2024efa}
Board of~Governors of~the Federal Reserve~System.
\newblock The fed - map: State-level debt-to-income ratio, 1999 - 2023:q3, 2024.
\newblock Available online at \url{https://www.federalreserve.gov/releases/z1/dataviz/household_debt/state/map/}.

\bibitem[Ope24]{nyc2024mvc}
NYC OpenData.
\newblock Motor vehicle collisions [dataset], 2024.
\newblock Available online at \url{https://data.cityofnewyork.us/Public-Safety/Motor-Vehicle-Collisions-Crashes/h9gi-nx95/about_data}.

\bibitem[RFS{\etalchar{+}}24]{census2022cali}
Steven Ruggles, Sarah Flood, Matthew Sobek, Daniel Backman, Annie Chen, Grace Cooper, Stephanie Richards, Renae Rodgers, and Megan Schouweiler.
\newblock Ipums usa: Version 15.0 [dataset], 2024.
\newblock https://doi.org/10.18128/D010.V15.0.

\bibitem[Rid57]{Rider1957GeneralizedCD}
Paul~R. Rider.
\newblock Generalized cauchy distributions.
\newblock {\em Annals of the Institute of Statistical Mathematics}, 9:215--223, 1957.

\bibitem[SZYX20]{sun2020differentially}
Lichao Sun, Yingbo Zhou, Philip~S Yu, and Caiming Xiong.
\newblock Differentially private deep learning with smooth sensitivity.
\newblock {\em arXiv preprint arXiv:2003.00505}, 2020.

\bibitem[Tea17]{appledp2017}
Apple Differential~Privacy Team.
\newblock Learning with privacy at scale.
\newblock Technical report, Apple, 2017.
\newblock Available online at \url{https://machinelearning.apple.com/research/learning-with-privacy-at-scale}.

\bibitem[War65]{warner1965randomized}
Stanley~L Warner.
\newblock Randomized response: A survey technique for eliminating evasive answer bias.
\newblock {\em Journal of the American Statistical Association}, 60(309):63--69, 1965.

\bibitem[WDZ{\etalchar{+}}19]{wang2019answering}
Tianhao Wang, Bolin Ding, Jingren Zhou, Cheng Hong, Zhicong Huang, Ninghui Li, and Somesh Jha.
\newblock Answering multi-dimensional analytical queries under local differential privacy.
\newblock In {\em Proceedings of the 2019 International Conference on Management of Data}, pages 159--176, 2019.

\bibitem[WNM23]{wagner2023fast}
Tal Wagner, Yonatan Naamad, and Nina Mishra.
\newblock Fast private kernel density estimation via locality sensitive quantization.
\newblock In {\em International Conference on Machine Learning}, pages 35339--35367. PMLR, 2023.

\bibitem[WXY{\etalchar{+}}19]{wang2019collecting}
Ning Wang, Xiaokui Xiao, Yin Yang, Jun Zhao, Siu~Cheung Hui, Hyejin Shin, Junbum Shin, and Ge~Yu.
\newblock Collecting and analyzing multidimensional data with local differential privacy.
\newblock In {\em 2019 IEEE 35th International Conference on Data Engineering (ICDE)}, pages 638--649. IEEE, 2019.

\bibitem[ZC20]{zafarani2020differentially}
Farzad Zafarani and Chris Clifton.
\newblock Differentially private naive bayes classifier using smooth sensitivity.
\newblock {\em arXiv preprint arXiv:2003.13955}, 2020.

\end{thebibliography}
\appendix
\onecolumn
\section{Missing Proofs and Derivations}
\label{appendix:proofs}
\subsection{Proofs for smooth sensitivity mechanisms}
\label{appendix:proofs_ssmech}
\begin{corollary}[Corollary \ref{cor:l1_cauchy}]
\label{appendix:cor_l1_cauchy}
    Let $\varepsilon, \gamma, \Lambda > 0$. Let $f:U\rightarrow \mathbb{R}^m$ equipped with the $\ell_1$ metric $\|\cdot\|_1$. Suppose $B(\cdot)$ is a $\gE$-smooth upper bound on $\LC_{f,\Lambda}$. Then the mechanism $M$ which on input $x$ releases $M(x):=f(x)+\frac{B(x)}{\eta}\cdot Z$, where $Z=[Z_1,\dotsb,Z_m]^T$ and each $Z_j\sim_{iid} \mathrm{GenCauchy}(0,1,p>1,\theta\ge 1)$, is $(\varepsilon,0,\Lambda)$-GP, where $\varepsilon=\max(m\gamma,m(p\theta-1)\gamma)+\theta(p-1)^{\frac{p-1}{p}}\eta$.
\end{corollary}
\begin{proof}
    Fix $x\sim_{\Lambda} x'\in U$. Write $a:=f(x)=[a_1,\dotsb,a_m]^T$, $a':=f(x')=[a'_1,\dotsb,a'_m]^T, b:=\frac{B(x)}{\eta}$ and $b':=\frac{B(x')}{\eta}$.
For $y\in \mathbb{R}^m$, let $l(y;x,x')=\prod_{j=1}^m l_{1j}(y_j;b,b')\cdot \prod_{j=1}^m l_{2j}(y_j;a_j,a'_j,b')$ where $l_{1j}(y_j;b,b'):=\frac{\frac{1}{b}h\left(\frac{y_j-a_j}{b}\right)}{\frac{1}{b'}h\left(\frac{y_j-a_j}{b'}\right)}$ and $l_{2j}(y_j;a_j,a'_j,b'):=\frac{\frac{1}{b'}h\left(\frac{y_j-a_j}{b'}\right)}{\frac{1}{b'}h\left(\frac{y_j-a'_j}{b'}\right)}$. As shown in the proof of Lemma~\ref{lm:gp_1d}, for each $j$: $|\ln(l_{2j}(y_j;a_j,a'_j,b'))|\le \theta(p-1)^{\frac{p-1}{p}} \frac{|a_j-a'_j|}{b'}$. Also, each $l_{1j}(y_j;b,b')\le \max\left(\frac{b'}{b},\left(\frac{b}{b'}\right)^{p\theta-1}\right)$. Thus,
\begin{align*}
    \left|\ln(l(y;x,x'))\right| &= \left|\sum_j\ln(l_{1j}(y_j;b,b'))+\sum_j\ln(l_{2j}(y_j;a_j,a'_j,b'))\right|\\
    &\le \sum_j \max(1,p\theta-1)\gamma\cdot\dist(x,x') + \sum_j \theta(p-1)^{\frac{p-1}{p}} \frac{|a_j-a'_j|}{b'}\\
    &= \max(m\gamma,m(p\theta-1)\gamma)\cdot\dist(x,x') + \theta(p-1)^{\frac{p-1}{p}} \frac{\|f(x)-f(x')\|_1}{b'}\\
    &\le \max(m\gamma,m(p\theta-1)\gamma\cdot\dist(x,x') + \theta(p-1)^{\frac{p-1}{p}} \frac{B(x')\cdot\dist(x,x')}{B(x')/\eta}\\
    &= \left(\max(m\gamma,m(p\theta-1)\gamma)+\theta(p-1)^{\frac{p-1}{p}}\eta\right)\cdot\dist(x,x').
\end{align*}
\end{proof}

\begin{lemma} [Lemma~\ref{lm:gp_1d_t}]
    Fix $\varepsilon, \gamma > 0$, let $\Lambda\in \mathbb{R}_{>0} \cup \{\infty\}$. Suppose $B(\cdot)$ is a $\gE(\cdot;\gamma)$-smooth upper bound on $\LC_{f,\Lambda}$ of a function $f:U\rightarrow \mathbb{R}$. Then the mechanism $M$ which on input $x$ releases $M(x):=f(x)+\frac{B(x)}{\eta}\cdot Z$, where $Z\sim \mathcal{T}_\nu(0,1)$ for $\nu > 1$, is $(\varepsilon,0,\Lambda)$-GP, where $\varepsilon=\nu\gamma+\frac{\nu+1}{2\sqrt{\nu}}\eta$.
\end{lemma}

\begin{proof}
As in the proof of Lemma~\ref{lm:gp_1d}, we show that  $l(y;x,x'):=\frac{\frac{1}{b}h\left(\frac{y-a}{b}\right)}{\frac{1}{b'}h\left(\frac{y-a'}{b'}\right)}=\frac{\frac{1}{b}h\left(\frac{y-a}{b}\right)}{\frac{1}{b'}h\left(\frac{y-a}{b'}\right)}\cdot \frac{\frac{1}{b'}h\left(\frac{y-a}{b'}\right)}{\frac{1}{b'}h\left(\frac{y-a'}{b'}\right)} =: l_1(y;b,b')\cdot l_2(y;a,a',b')
$ is bounded pointwise by $e^{\varepsilon \cdot \dist(x, x')}$, where $a, a', b$ and $b'$ are as previously defined. We have
\begin{align*}
l_1(y;b,b')&:=\frac{\frac{1}{b}h\left(\frac{y-a}{b}\right)}{\frac{1}{b'}h\left(\frac{y-a}{b'}\right)} = \frac{\frac{1}{\sqrt{\nu}b}\frac{1}{\left(1+\frac{1}{\nu}(\frac{y-a}{b})^{2}\right)^{(\nu+1)/2}}}{\frac{1}{\sqrt{\nu}b'}\frac{1}{\left(1+\frac{1}{\nu}(\frac{y-a}{b'})^2\right)^{(\nu+1)/2}}} 
= \frac{b'}{b}\frac{\left(1+\frac{1}{\nu}(\frac{y-a}{b'})^2\right)^{(\nu+1)/2}}{\left(1+\frac{1}{\nu}(\frac{y-a}{b})^2\right)^{(\nu+1)/2}}.
\end{align*}
If $b'\ge b$ or $y=a$, then $l_1(y;b,b')\le b'/b$. Otherwise, 
\[l_1(y;b,b')\le \frac{b'}{b}\cdot \frac{(\frac{y-a}{b'})^{2\cdot(\nu+1)/2}}{(\frac{y-a}{b})^{2\cdot(\nu+1)/2}} = \frac{b'}{b}\cdot \left(\frac{b}{b'}\right)^{\nu+1} = \left(\frac{b}{b'}\right)^{\nu}.\]
To bound $l_2(y;a,a',b')$ we again use Claim~\ref{clm:phi_gamma_bound}, which gives for $z_1, z_2\in \mathbb{R}_{\ge 0}$, 
$|\ln(1+z_2^2)-\ln(1+z_1^2)|\le (2-1)^{1/2} |z_2-z_1| = |z_2-z_1|$.
Then for $z_1:=\frac{y-a}{\sqrt{\nu}b'}, z_2:=\frac{y-a'}{\sqrt{\nu}b'}$,
\begin{align*}
l_2(y;a,a',b') &:= \frac{\frac{1}{b'}h\left(\frac{y-a}{b'}\right)}{\frac{1}{b'}h\left(\frac{y-a'}{b'}\right)}=\frac{\left(1+(\frac{y-a'}{\sqrt{\nu}b'})^2\right)^{(\nu+1)/2}}{\left(1+(\frac{y-a}{\sqrt{\nu}b'})^2\right)^{(\nu+1)/2}}\\
    \left|\ln(l_2(y;a,a',b'))\right| &= \frac{\nu+1}{2}\left|\phi(z_2)-\phi(z_1)\right| \le \frac{\nu+1}{2} |z_2-z_1| \\
    &= \frac{\nu+1}{2} \left|\frac{y-a'}{\sqrt{\nu}b'}-\frac{y-a}{\sqrt{\nu}b'}\right|
    = \frac{\nu+1}{2\sqrt{\nu}b'}|a-a'|.
\end{align*}
Thus,
\begin{align*}
    |\ln(l(y;x,x'))| &\le |\ln(|l_1(y;b,b')|)|+\left|\ln(|l_2(y;a,a',b')|)\right|\\
    &\le \max(|\ln(b'/b)|, \nu |\ln(b/b')|)+\frac{\nu+1}{2\sqrt{\nu}b'}|a-a'|\\
    &\le \nu\cdot\ln(e^{\gamma\dist(x,x')}) + \frac{\nu+1}{2\sqrt{\nu}}\cdot \frac{|f(x)-f(x')|}{B(x')/\eta}\\
    &\le \nu \cdot \gamma\dist(x,x') + \frac{\nu+1}{2\sqrt{\nu}} \frac{\eta\cdot B(x')\dist(x, x')}{B(x')}\\
    &= \left(\nu\gamma+\frac{\nu+1}{2\sqrt{\nu}}\eta\right)\cdot\dist(x, x').
\end{align*}
\end{proof}

\subsection{Lipschitzness of $f_{\tau}$ in two-way threshold function}
\label{appendix:sec_2way_lipschitz}
\begin{figure}[htbp]
     \centering
      \begin{subfigure}[t]{0.25\linewidth}
            \includegraphics[width=\textwidth]{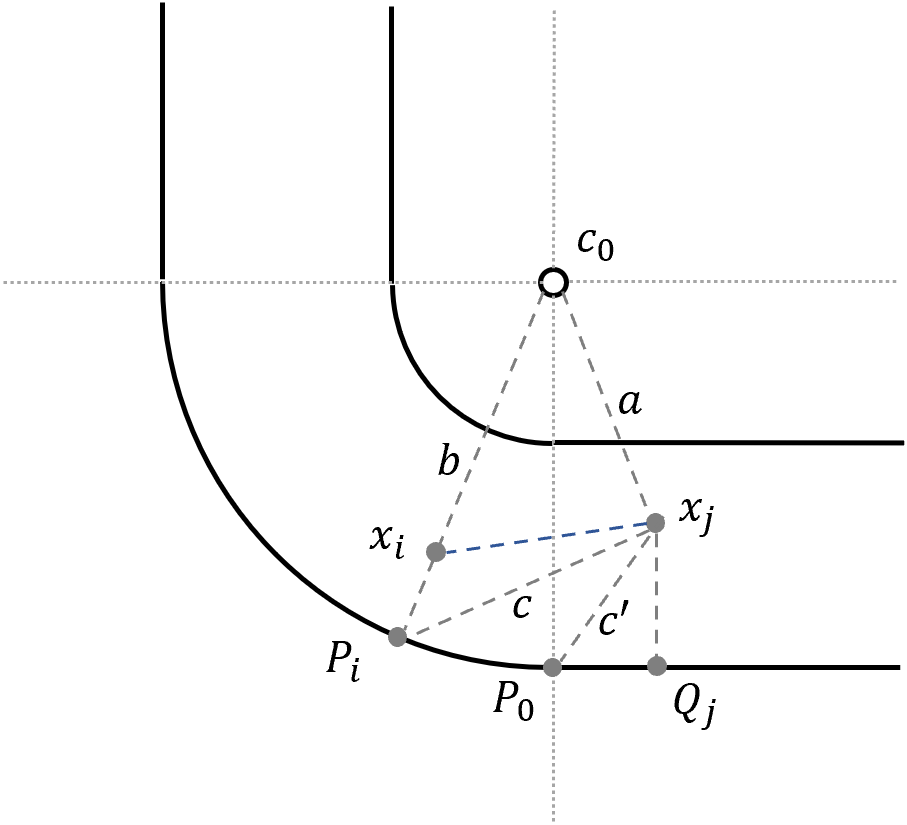}
         \end{subfigure}
         \caption{Illustration of the positions of $P_i$ and $Q_j$ for Claim~\ref{clm:2way_thres_lipschitz}.}
         \label{fig:2way_thres_lipschitz}
    \vskip -.01in
\end{figure}
\label{appendix:2way_thres_lipschitz}
\begin{claim}[Claim~\ref{clm:2way_thres_lipschitz}]
\label{appendix:cml_2way_lipschitz}
The function $f_{\tau}$ defined above is $\frac{1}{\tau}$-Lipschitz on $\mathbb{R}^2$ (w.r.t. $\|\cdot\|$).
\end{claim}
\begin{proof}
    Let $x_i, x_j\in \mathbb{R}^2$. If both $x_i,x_j\in S_{1,k}$, or $S_{0,k}$ for a valid sub-index $k\in\{\alpha,1,2,3\}$, then $f_{\tau}(x_i)-f_{\tau}(x_j)=0$ and the claim trivially holds. If one of $x_i, x_j$ is in $S_{1,r}$ while the other is in $S_{0,r}$, then $\|x_i-x_j\|\ge \tau = |f(x_i)-f(x_j)|\cdot \tau$ and the claim also holds. We thus need to consider the following cases. Let $k, k'\in\{\alpha,1,2,3\}$ be valid sub-indices.
    \begin{enumerate}
        \item[(1)] $x_i, x_j$ in the same region $S_{\tau,k}$.
        \item[(2)] $x_i, x_j$ in different regions $S_{\tau, \alpha}$ and $S_{\tau, 2}$ (or $S_{\tau, \alpha}$ and $S_{\tau, 1}$).
        \item[(3)] $x_i, x_j$ in different regions $S_{\tau, 1}$ and $S_{\tau, 2}$.
        \item[(4)] $x_i, x_j$ in different regions $S_{\tau, k}$ and $S_{1, k'}$ (or $S_{\tau, k}$ and $S_{0, k'}$).
    \end{enumerate}

{Case (1).} Suppose $x_i, x_j\in S_{\tau,\alpha}$. Then 
\begin{align*}
|f_{\tau}(x_i)-f_{\tau}(x_j)|&=\left|\frac{R_2-\|x_i-c_0\|}{\tau}-\frac{R_2-\|x_j-c_0\|}{\tau}\right|
=\left|\frac{\|x_j-c_0\|-\|x_i-c_0\|}{\tau}\right|\le \frac{1}{\tau}\|x_j-x_i\|.
\end{align*}
Similarly, 
\begin{align*}|f_{\tau}(x_i)-f_{\tau}(x_j)|&=\left|\frac{x_{i,k}-(T_k-\tau/2)}{\tau}-\frac{x_{j,k}-(T_k-\tau/2)}{\tau}\right|
=\left|\frac{x_{i,k}-x_{j,k}}{\tau}\right|\le \frac{1}{\tau}\|x_j-x_i\|
\end{align*}
for $x_i, x_j\in S_{\tau,k}$ where $k=1,2$.

{Case (2).} Suppose $x_i\in S_{\tau,\alpha}$ and $x_j\in S_{\tau,2}$ (the reverse direction follows by symmetry). We will show $f_{\tau}(x_i)-f_{\tau}(x_j)\le \frac{\|x_i-x_j\|}{\tau}$ and $f_{\tau}(x_j)-f_{\tau}(x_i)\le \frac{\|x_i-x_j\|}{\tau}$, whence $|f_{\tau}(x_i)-f_{\tau}(x_j)|\le \frac{1}{\tau}\cdot \|x_i-x_j\|$.
\begin{align*}
    &{}\tau\cdot\left(f_{\tau}(x_i)-f_{\tau}(x_j)\right) =\left( R_2-\|x_i-c_0\|\right)-\left(x_{j,2}-(T_2-\tau/2)\right)\\
    & = T_2+(R_2-\tau/2) - x_{j,2} - \|x_i-c_0\| \\
    &= (T_2+\tau/2+R_1) - x_{j,2} - \|x_i-c_0\|\\
    &= \sqrt{\left((T_2+\tau/2+R_1)-x_{j,2}\right)^2} - \|x_i-c_0\|\\
    &\le \sqrt{\left((T_2+\tau/2+R_1)-x_{j,2}\right)^2+\left((T_1+\tau/2+R_1)-x_{j,1}\right)^2} - \|x_i-c_0\|\\
    &= \|x_j-c_0\|-\|x_i-c_0\| \le \|x_j-x_i\|.
\end{align*}
For the other direction, let $P_i:=\frac{R_2}{\|x_i-c_0\|}x_i+\left(1-\frac{R_2}{\|x_i-c_0\|}\right)c_0$ denote the point on the arc $\alpha_2$ which lies on the line that goes through $c_0$ and $x_i$. Let $Q_j:=(x_{j,1}, T_2-\tau/2)$ denote the point on the boundary of the transition band obtained by shifting $x_j$ vertically down by a distance of $x_{j,2}-(T_2-\tau/2)$. Then $\tau\cdot\left(f_{\tau}(x_j)-f_{\tau}(x_i)\right) = \left(x_{j,2}-(T_2-\tau/2)\right)-\left( R_2-\|x_i-c_0\|\right)$ is the difference between the lengths of the line segments $\overline{x_jQ_j}$ and $\overline{x_iP_i}$ (see Fig.~\ref{fig:2way_thres_lipschitz}). Then, it suffices to show that the length of $\overline{x_jQ_j}$ is bounded by that of the segment $\overline{x_jP_i}$, where it would then follow that $\|x_j-P_i\|-\|x_i-P_i\|\le \|x_j-x_i\|$. To this end, let $P_0:=(T_1+\tau/2+R_1,T_2-\tau/2)$ denote the point on the arc $\alpha_2$ which lies on the line separating the regions $S_{\tau,\alpha}$ and $S_{\tau_,2}$. Let $a:=\|x_j-c_0\|$ be the length of the line segment $\overline{x_jc_0}$, $b:=\|P_i-c_0\|=R_2=\|P_0-c_0\|$, $c:=\|x_j-P_i\|$ be the length of $\overline{x_jP_i}$ and $c':=\|x_j-P_0\|$ be the length of $\overline{x_jP_0}$. Let $\theta_0$ denote the angle between the segments $\overline{x_jc_0}$ and $\overline{c_0P_0}$, and $\theta_i$ denote that between $\overline{x_jc_0}$ and $\overline{c_0P_i}$. By the law of cosines, $c=\sqrt{a^2+b^2-2ab\cos\theta_i}$, which increases as $\theta_i$ increases from $0$ to $\pi/2$. Thus, as $\theta_0\le \theta_i$,
\begin{align*}
    \|x_j-P_i\|=:c=\sqrt{a^2+b^2-2ab\cos\theta_i} &\ge \sqrt{a^2+b^2-2ab\cos\theta_0} 
    =c':= \|x_j-P_0\|.
\end{align*}
On the other hand,
\begin{align*}
    \|x_j-P_0\|&=\sqrt{\left(x_{j,1}-(T_1+\tau/2+R_1)\right)^2+\left(x_{j,2}-(T_2-\tau/2)\right)^2}
    \ge x_{j,2}-(T_2-\tau/2) = \|x_j-Q_j\|.
\end{align*}
Thus, $\tau\cdot\left(f_{\tau}(x_j)-f_{\tau}(x_i)\right) = \left(x_{j,2}-(T_2-\tau/2)\right)-\left( R_2-\|x_i-c_0\|\right)=\|x_j-Q_j\|-\|x_i-P_i\|\le \|x_j-P_i\|-\|x_i-P_i\|\le \|x_j-x_i\|$, as required. 
The case where $x_i\in S_{\tau,\alpha}$ and $x_j\in S_{\tau,1}$ follows the same argument, where $P_0$ becomes the point on the same arc, but instead lies on the line separating the regions $S_{\tau,\alpha}$ and $S_{\tau,1}$.

Case (3). Suppose $x_i\in S_{\tau,1}$ and $x_j\in S_{\tau,2}$ (the reverse direction follows by symmetry). Note that $f_{\tau}(x_i), f_{\tau}(x_j) \in [0, 1]$, so $|f_{\tau}(x_i)-f_{\tau}(x_j)|\in [0, 1]$.
Thus, 
\[\|x_i-x_j\|\ge \sqrt{2}R_1 = \sqrt{2}\cdot \frac{\tau}{\sqrt{2}} =\tau \ge \tau\cdot|f_{\tau}(x_i)-f_{\tau}(x_j)|.
\]

Case (4). Suppose $x_i\in S_{\tau, k}$ and $x_j\in S_{1, k'}$ (the reverse direction follows by symmetry). Let $z$ be the point on the line segment starting from $x_j$ and ending in $x_i$ which first crosses the transition band. Then
\begin{align*}
    &{}\left|f_{\tau}(x_i)-f_{\tau}(x_j)\right| \le \left|f_{\tau}(x_i)-f_{\tau}(z)\right|+\left|f_{\tau}(z)-f_{\tau}(x_j)\right| \\
    &\le \frac{1}{\tau}\left\|x_i-z\right\|+\frac{1}{\tau}\left\|z-x_j\right\| = \frac{1}{\tau}\left(\|x_i-z\|+\|z-x_j\|\right)
    =\frac{1}{\tau}\left\|x_i-x_j\right\|,
\end{align*}
where the second inequality follows from 1) the previous three cases since $x_i$ and $z$ are both in the transition band; and 2) that $f_{\tau}$ is continuous on the boundary of the transition band, i.e. $f_{\tau}(z)=1=f_{\tau}(x_j)$. The last inequality is due to the fact that $z$ is a point on the line segment with end points $x_i$ and $x_j$.
\end{proof}

\subsection{Lipschitz constant for 1D Gaussian kernel}
\label{appendix:lipschitz_claim_1dgauss}
\begin{claim} [Claim~\ref{clm:gauss_1d_lipschitzconstant}]
    For $t\in \mathbb{R}$, $h>0$, let $\kappa_t:\mathbb{R}\rightarrow \mathbb{R}_{\ge 0}$ be the function $x\mapsto e^{-\frac{(x-t)^2}{2h^2}}$.  Then $|\kappa_t(x) - \kappa_{t}(x')| \le \frac{e^{-1/2}}{h}|x-x'|$ for all $x, x'\in \mathbb{R}$.
\end{claim}
\begin{proof}
    It suffices to show $\sup_{x\in \mathbb{R}}\left|\frac{d}{dx} \kappa_t(x)\right| \le \frac{e^{-1/2}}{h}$, where the inequality then follows from the mean value theorem. We have $\frac{d}{dx} \kappa_t(x)=\frac{d}{dx}e^{-\frac{(x-t)^2}{2b^2}} = -\frac{x-t}{h^2}e^{-\frac{(x-t)^2}{2h^2}}$, which is continuous in $x$. Since $\frac{d}{dx} \kappa_t(x)$ depends on $x$ only through its distance to $t$, let us write $|x-t|=ch$ for some $c\ge 0$. We want to maximize $\left|-\frac{x-t}{h^2}e^{-\frac{(x-t)^2}{2h^2}}\right|=\left|\frac{c}{h}e^{-c^2/2}\right|$. Since
    \[
        \frac{d}{d{c}} \left(\frac{c}{h}e^{-c^2/2}\right) = \frac{1}{h}e^{-c^2/2}-\frac{c^2}{h}e^{-c^2/2} = 0 \iff c^2 = 1,
        \]
    $\frac{c}{h}e^{-c^2/2}$ has critical points at $c=\pm 1$, both of which give $|\frac{c}{h}e^{-c^2/2}| = \frac{e^{-1/2}}{h} > 0$. 
\end{proof}
\subsection{Smooth sensitivity for KDE}
\label{appendix:smoothsens_KDE}
\begin{lemma} [Lemma~\ref{lm:gaussker_md_to_1d}] 
Let $U\subseteq \mathbb{R}^d$ be equipped with the $\ell_2$ metric. Fix $t\in \mathbb{R}^d$. For $f:U \rightarrow V \subseteq \mathbb{R}$, suppose there is $f_0: \mathbb{R}_{\ge 0} \rightarrow \mathbb{R}$ such that, $\forall x\in U: f(x) = f_0(\|x-t\|)$. Suppose for $x \neq t$, we have $(w^*,z^*)$ such that $B^*(x) = \frac{|f(w^*)-f(z^*)|}{\|w^*-z^*\|g(x,z^*)}$. Then there is a pair $(w_L,z_L)$ on the line connecting $x$ and $t$ such that $\frac{|f(w_L)-f(z_L)|}{\|w_L-z_L\|g(x,z_L)}\ge B^*(x)$.
\end{lemma}
\begin{proof} Let $z_L:= c_z x + (1-c_z)t$, $w_L:= c_w x + (1-c_w)t$ where $c_z := \frac{\|z^*-t\|}{\|x-t\|}$ and $c_w := \frac{\|w^*-t\|}{\|x-t\|}$. We will show $\|z_L-t\|=\|z^*-t\|$, $\|w_L-t\|=\|w^*-t\|$, $\|w_L-z_L\|\le \|w^*-z^*\|$ and $\|z_L-x\|\le \|z^*-x\|$, whence the stated inequality follows.
    \begin{align*}
        \|z_L-t\| &= \|c_z x + (1-c_z)t - t\| = \|c_z(x-t)\| = |c_z|\cdot \|x-t\| 
        = \frac{\|z^*-t\|}{\|x-t\|}\cdot \|x-t\|=\|z^*-t\|,\\
        \|w_L-t\| &=  \|c_w x + (1-c_w)t - t\| = \|c_w(x-t)\| = |c_w|\cdot \|x-t\| 
        = \frac{\|w^*-t\|}{\|x-t\|}\cdot \|x-t\|=\|w^*-t\|,\\
        \|w_L-z_L\| &= \|w_L-t-(z_L-t)\|
        = \|c_w x + (1-c_w)t - (c_z x + (1-c_z)t)\| = \|(c_w-c_z)(x-t)\|\\
        &= |c_w-c_z|\cdot \|x-t\| = \left|\frac{\|w^*-t\|}{\|x-t\|}- \frac{\|z^*-t\|}{\|x-t\|}\right|\cdot\|x-t\|
        =\left|\|w^*-t\|-\|z^*-t\|\right|\le \|w^*-z^*\|,\\
        \|z_L-x\| &= \|c_z x+(1-c_z)t -x \| = \|(c_z-1)(x-t)\| = |c_z-1|\cdot \|x-t\|
        = \left\|\frac{\|z^*-t\|}{\|x-t\|}-1\right\|\cdot \|x-t\| \\
        &= \left\|\frac{\|z^*-t\|}{\|x-t\|}\cdot \|x-t\|-1\cdot \|x-t\|\right\|
        = \left|\|z^*-t\|-\|x-t\|\right| \le \|z^*-x\|.
    \end{align*}
\end{proof}

\begin{lemma} [Lemma~\ref{lm:gauss_smoothsens_xneqt}]
\label{appendix:lm_gauss_smoothsens_xneqt}
Fix $t\neq x \in \mathbb{R}^d$, $\gamma > 0$ and $h > 0$. For the function $\kappa_t: x\mapsto e^{-\frac{\|x-t\|^2}{2h^2}}$ and smooth growth function $\gE: (x,x')\mapsto e^{\gamma\|x-x'\|}$, we have for $x\neq t$, the $\gE$-smooth sensitivity
    \[
        B^*(x) =  \max \left(\max_{(c_w,c_z)\in S_1} \frac{c}{h}\cdot\frac{c_w e^{-c_w^2c^2/2}}{e^{\gamma|c_z-1|ch}}, \max_{c_w\in S_2}\frac{|e^{-c_w^2c^2/2}-e^{-c^2}|}{|c_w-1|ch}, \max_{c_z\in S_3}\frac{c}{h}\cdot \frac{c_z e^{-c_z^2c^2/2}}{e^{\gamma|c_z-1|ch}}, \frac{c}{h}e^{-c^2/2}\right),
        \]
where
\begin{align*}
    S_1 = &\left\{(c_w,c_z>1)\in \mathbb{R}^2_{\ge 0}: c_w = \phi_1(-\gamma), \phi_0(c_w,c_z) = 0\right\} \cup \left\{(c_w,c_z<1)\in \mathbb{R}^2_{\ge 0}: c_w = \phi_1(\gamma), \phi_0(c_w,c_z) = 0\right\}\\
    S_2 = &\left\{c_w\in \mathbb{R}_{\ge 0}: \phi_0(c_w,1)=0 \right\} \cup \left\{c_w=0\right\}\\
    S_3 = &\left\{c_z\in \mathbb{R}_{\ge 0}: c_z = \phi_1(-\gamma), c_z > 1 \right\} \cup \left\{c_z\in \mathbb{R}_{\ge 0}: c_z = \phi_1(\gamma), c_z < 1 \right\},
\end{align*}
and
\[    \phi_0: (c_w, c_z) \mapsto e^{-c_w^2c^2/2}-(1+c^2(c_w-c_z)c_w)e^{-c_w^2c^2/2}, \;\;\;\;\;\;\;
\phi_1: \gamma \mapsto \frac{h}{2c}\left(\gamma + \sqrt{\gamma^2+4/h^2}\right).\]
\end{lemma}
\begin{proof}
Let $c:=\|x-t\|/h>0$. Write $z_L=c_zx+(1-c_z)t$, $w_L=c_wx+(1-c_w)t$, for $c_z, c_w\in \mathbb{R}$. Then $\|z_L-t\|=\|c_z(x-t)\|=|c_z|\cdot\|x-t\|=|c_z|ch$, $\|w_L-t\|=|c_w|ch$, $\|w_L-z_L\|=|c_w-c_z|ch$ and $\|x-z_L\|=|c_z-1|ch$. Thus, instead of working with $w, z\in \mathbb{R}^d$, we can work with $c_w, c_z\in \mathbb{R}$ since 
\[
    \frac{|\kappa_t(w_L)-\kappa_t(z_L)|}{\|w_L-z_L\| g(x,z_L)} = \frac{|e^{-c_w^2c^2/2}-e^{-c_z^2c^2/2}|}{(|c_w-c_z|ch)e^{\gamma|c_z-1|ch}}.
\]
We want to find $c_w, c_z\in \mathbb{R}$ such that the right hand side expression is maximized. Recall that we can find the points corresponding to the maximum of a function by finding the critical points and points where the function is not differentiable. 
\paragraph{1. $w\neq z$, $z\neq x$.}
Note $c_z\neq 1$ in this case. Write $c_w = c_z+a$ for some $a\in \mathbb{R}$. 
\begin{align}
    \nonumber
\frac{\partial \Psi_0(c_z+a)}{\partial a} &= \frac{\partial}{\partial a}\left(\frac{e^{-(c_z+a)^2c^2/2}-e^{-c_z^2c^2/2}}{ach}\right)\\
\nonumber
&= \frac{-(c_z+a)c^2e^{-(c_z+a)^2c^2/2}}{ach} - \frac{e^{-(c_z+a)^2c^2/2}-e^{-c_z^2c^2/2}}{a^2ch}\\
\label{eqn:critical_a}
&= \frac{1}{a^2ch}\left(e^{-c_z^2c^2/2}-(1+ac^2(c_z+a))e^{-(c_z+a)^2c^2/2}\right).
\end{align}
Setting the above to zero, we have $e^{-c_z^2c^2/2}=(1+ac^2(c_z+a))e^{-(c_z+a)^2c^2/2}$, and the objective function becomes
\begin{align*}
    \frac{|e^{-c_w^2c^2/2}-e^{-c_z^2c^2/2}|}{(|c_w-c_z|ch)e^{\gamma|c_z-1|ch}} &= \left|\frac{ac^2(c_z+a)e^{-(c_z+a)^2c^2/2}}{ach}\right|e^{-\gamma |c_z-1|ch} = \frac{c}{h}c_w e^{-c_w^2 c^2/2} e^{-\gamma|c_z-1|ch}.
\end{align*}
For $c_z > 1$,
\begin{align}
    \nonumber
    &{}\frac{d}{dc_z}\left(\frac{c}{h}(c_z+a)e^{-(c_z+a)^2c^2/2} e^{-\gamma|c_z-1|ch}\right) \\
    \nonumber
    &= \frac{c}{h}e^{-(c_z+a)^2c^2/2}e^{-\gamma|c_z-1|ch} - (c_z+a)^2\frac{c^3}{h}e^{-(c_z+a)^2c^2/2}e^{-\gamma|c_z-1|ch} - \gamma c^2 (c_z+a)e^{-(c_z+a)^2c^2/2}e^{-\gamma|c_z-1|ch}\\
    &= -\left(\frac{c^3}{h}(c_z+a)^2 + \gamma c^2(c_z+a)-\frac{c}{h}\right)e^{-(c_z+a)^2c^2/2}e^{-\gamma|c_z-1|ch},
\end{align}
which gives $c_w = c_z+a = \frac{-\gamma c^2 \pm \sqrt{\gamma^2c^4+4c^4/h^2}}{2c^3/h} = \frac{h}{2c}\left(- \gamma \pm \sqrt{\gamma^2+4/h^2}\right)$ for $c_z > 1$. Similarly, for $c_z < 1$, we get  $c_w = c_z+a =\frac{h}{2c}\left(\gamma \pm \sqrt{\gamma^2+4/h^2}\right)$. Now, plugging the values of $c_z+a$ into expression~\eqref{eqn:critical_a} and setting that to zero, we get
\[
    e^{-c_z^2c^2/2}-(1+c^2(c_w-c_z)c_w)e^{-c_w^2c^2/2} = 0
\]
which we can solve to get $c_z$ (e.g. with bisection), where we also need to check for consistency (i.e. whether $c_z > 1$, etc.)

\paragraph{2. $w\neq z, z=x$.} In this case $c_z=1$ and we want to compute $\max_{c_w\ge 0}\frac{|e^{-c_w^2c^2/2}-e^{-c_z^2c^2/2}|}{|c_w-c_z|ch}$. We can find the critical points for $c_w$ by plugging $c_z=1$ into expression~\eqref{eqn:critical_a} and setting that to zero, i.e.,
\[
    e^{-c^2/2}-(1+c^2(c_w-1)c_w)e^{-c_w^2c^2/2} = 0.
\] 
We also need to check the boundary point $c_w=0$, which yields candidate value $\frac{|1-e^{-c^2/2}|}{ch}$.
\paragraph{3. $w=z$, $z\neq x$.} This corresponds to setting $a=0$ in the first case. Thus, we want to compute
\[\max_{c_z\ge 0} \left|\frac{c}{h}c_z e^{-c_z^2c^2/2} e^{-\gamma |c_z-1|ch}\right|.\]
For $c_z > 1$:
\begin{align}
    \nonumber
   \frac{d}{dc_z}\left(\frac{c}{h}c_ze^{-c_z^2c^2/2} e^{-\gamma |c_z-1|ch}\right) &=-\frac{c^3}{h}c_z^2 e^{-c_z^2c^2/2} e^{-\gamma|c_z-1|ch} + \frac{c}{h} e^{-c_z^2c^2/2} e^{-\gamma|c_z-1|ch} - \gamma c^2 c_z e^{-c_z^2c^2/2} e^{-\gamma|c_z-1|ch}\\
   &=-\left(\frac{c^3}{h}c_z^2 + \gamma c^2 c_z - \frac{c}{h}\right)e^{-c_z^2c^2/2} e^{-\gamma|c_z-1|ch}.
\end{align}
Setting the above to zero (and eliminating the negative root), we obtain $c_z = \frac{h}{2c^3}\left(-\gamma c^2 + \sqrt{\gamma^2 c^4 + 4\frac{c^4}{h^2}}\right)=\frac{h}{2c}\left(-\gamma + \sqrt{\gamma^2 + 4/h^2}\right)$.
Similarly, for $c_z < 1$, we get $c_z =\frac{h}{2c}\left(\gamma + \sqrt{\gamma^2 + 4/h^2}\right)$. Note that at most one of these values for $c_z$ is consistent. 
The boundary point $c_z=0$ yields objective value $0$ and thus can be ignored.

\paragraph{4. $w=z=x$.} In this case, the candidate value is 
\[
    \left\|\frac{(x-t)}{h^2}e^{-\frac{\|x-t\|^2}{2h^2}}\right\|=\frac{\|x-t\|}{h^2}e^{-\frac{\|x-t\|^2}{2h^2}} = \frac{c}{h}e^{-c^2/2},
\]
which corresponds to the case $c_w=c_z=1$.
\end{proof}

\begin{lemma} \label{lm:gaussker_md_to_1d_xet}
    Let $(w^*,z^*)$ be such that $B^*(t) = \frac{|\kappa_t(w^*)-\kappa_t(z^*)|}{\|w^*-z^*\|g(t,z^*)}$. Assume $z^*\neq t$. Then there is a $w_L$ on the line connecting $z^*$ and $t$ such that $\frac{|\kappa_t(w_L)-\kappa_t(z^*)|}{\|w_L-z^*\|g(t,z^*)}\ge B^*(t)$.
\end{lemma}

\begin{proof} Let $w_L:= c_w z^* + (1-c_w)t$ where $c_w:=\frac{\|w^*-t\|}{\|z^*-t\|}$. As in the proof of Lemma~\ref{lm:gaussker_md_to_1d}, it can be verified that $\|w_L-t\| = \|w^*-t\|$ and $\|w_L-z^*\|\le \|w^*-z^*\|$.
\end{proof}

\begin{lemma} [Lemma~\ref{lm:gauss_smoothsens_xeqt}]
\label{appendix:lm_gauss_smoothsens_xeqt}
Fix $t \in \mathbb{R}^d$, $\gamma > 0$ and $h > 0$. For the function $\kappa_t: x\mapsto e^{-\frac{\|x-t\|^2}{2h^2}}$ and smooth growth function $\gE: (x,x')\mapsto e^{\gamma\|x-x'\|}$, we have for $x = t$, the $\gE$-smooth sensitivity
    \[
        B^*(t) =  \max \left(\max_{(c_w,l_z)\in S_1} \frac{l_z}{h}\cdot\frac{c_w l_z e^{-c_w^2l_z^2/2}}{e^{\gamma l_z h}}, \max_{l_w\in S_2}\frac{|e^{-l_w^2/2}-1|}{l_w h}, \max_{l_z\in S_3}\frac{l_z}{h}\cdot \frac{e^{-l_z^2/2}}{e^{\gamma l_z h}}\right),
        \]
where
\begin{align*}
    S_1 = &\left\{(c_w,l_z>0)\in \mathbb{R}^2: \phi_0(c_w,\phi_1(c_w)) = 0, l_z=\phi_1(c_w) \right\} \\
    S_2 = &\left\{l_w\in \mathbb{R}_{\ge 0}: l_w = \sqrt{-2 W^{-1}(-e^{-1/2}/2)-1}\right\}\\
    S_3 = &\left\{l_z\in \mathbb{R}_{\ge 0}: l_z = \phi_1(1) \right\},
\end{align*}
$W^{-1}(\cdot)$ is the Lambert $W$ function and
\[    \phi_0: (c_w, l_z) \mapsto e^{-l_z^2/2}-(1+l_z^2(c_w-1)c_w)e^{-c_w^2l_z^2/2}, \;\;\;\;\;\;\;
\phi_1: c_w \mapsto \frac{h}{2c_w^2}\left(-\gamma + \sqrt{\gamma^2+4c_w^2/h^2}\right).\]
\end{lemma}

\begin{proof}
Let $l_z := \frac{\|z-t\|}{h} = \frac{\|z-x\|}{h}$. Write $w = c_w z +(1-c_w)t$ for some $c_w\in \mathbb{R}$. Then $\|w-t\|=\|c_w(z-t)\|=|c_w|l_zh$, $\|w-z\|=\|(c_w-1)(z-t)\|=|c_w-1|l_zh$. We want to compute
\[
    \max_{(w,z)}\frac{|\kappa_t(w)-\kappa_t(z)|}{\|w-z\| g(t,z)} = \max_{(l_z, c_w)}\frac{|e^{-c_w^2l_z^2/2}-e^{-l_z^2/2}|}{(|c_w-1|l_z)e^{\gamma l_z h}}.
\]

\paragraph{1. $w \neq z, z\neq x$.}
In this case $l_z > 0$. Rewrite $c_w$ as $1+a$ for some $a\in \mathbb{R}$
\begin{align}
    \nonumber
    \frac{\partial}{\partial a}\left(\frac{e^{-(1+a)^2l_z^2/2}-e^{-l_z^2/2}}{a l_zh} \right) &= \frac{-(1+a)l_z^2 e^{-(1+a)^2l_z^2/2}}{a l_z h}-\frac{e^{-(1+a)^2l_z^2/2}-e^{-l_z^2/2}}{a^2 l_z h}\\
    \label{eqn:critical_a_xet}
    &= \frac{1}{a^2 l_z h}\left(e^{-l_z^2/2}-(1+a(1+a)l_z^2)e^{-(1+a)^2l_z^2/2}\right).
\end{align}
Setting the above to zero,  we have $e^{-l_z^2/2}=(1+a(1+a)l_z^2)e^{-(1+a)^2l_z^2/2}$, then
\[
    \left|\frac{e^{-c_w^2l_z^2/2}-e^{-l_z^2/2}}{(c_w-1)l_zh}\right|e^{-\gamma l_z h}=\left|\frac{a(1+a)l_z^2 e^{-(1+a)^2l_z^2/2}}{a l_z h}\right|e^{-\gamma l_z h} = \left|\frac{(1+a)l_ze^{-(1+a)^2l_z^2/2}}{h}e^{-\gamma l_z h}\right|.
\]
\begin{align*}
    \frac{\partial}{\partial l_z} \left(\frac{(1+a)l_ze^{-(1+a)^2l_z^2/2}}{h}e^{-\gamma l_z h}\right) &= \left((1+a)-(1+a)^3l_z^2-\gamma h (1+a) l_z\right)\frac{e^{-(1+a)^2l_z^2/2}e^{-\gamma l_z b}}{h}\\
    &= -\left((1+a)^2l_z^2+\gamma h l_z - 1\right)\frac{(1+a)e^{-(1+a)^2l_z^2/2}e^{-\gamma l_z h}}{h},
\end{align*}
which gives $l_z = \frac{1}{2(1+a)^2}\left(-\gamma h + \sqrt{\gamma^2 h^2 + 4(1+a)^2}\right)$. Plugging into expression~\eqref{eqn:critical_a_xet} and setting that to zero allows us to solve for $a$.

\paragraph{2. $w \neq z, z = x$.}
In this case $l_z = 0$, we have to solve $\max_{l_w> 0} \left|\frac{e^{-l_w^2/2}-1}{l_w h}\right|$, where $l_w=\|w-t\|/h=\|w-x\|/h=\|w-z\|/h$. 
\begin{align*}
    0 = \frac{\partial}{\partial l_w} \left(\frac{e^{-l_w^2/2-1}}{l_w h}\right) &= \frac{-l_w e^{-l_w^2/2}}{l_w h} - \frac{e^{-l_w^2/2}-1}{l_w^2 h} = \frac{1}{l_w^2 h}\left(1-(1+l_w^2)e^{-l_w^2/2}\right),\\
    -(1+l_w^2)e^{-l_w^2/2} & = -1\\
    -\frac{1+l_w^2}{2}e^{-(l_w^2+1)/2} &= -\frac{e^{-1/2}}{2},
\end{align*}
which can be solved via the the Lambert $W$ function $W^{-1}(\cdot)$ as $-(l_w^2+1)/2 = W^{-1}(-e^{-1/2}/2)$. I.e. $l_w = (2(-W^{-1}(-e^{-1/2}/2))-1)^{1/2}>0$.

\paragraph{3. $w = z, z \neq x$.}
Writing $c_w = 1+a$ for some $a\in \mathbb{R}$, then 
\[\lim_{c_w\rightarrow 1}\frac{e^{-c_w^2l_z^2/2}-e^{-l_z^2/2}}{(c_w-1)l_z h} =  \lim_{a\rightarrow 0}\frac{e^{-(1+a)^2l_z^2/2}-e^{-l_z^2/2}}{a l_z} = \left|\frac{d}{du} \left( \frac{e^{-u^2l_z^2/2}}{l_z h}\right)\right|_{u=1}=\left|\frac{u}{h}l_ze^{-u^2l_z^2/2}\right|_{u=1}=\frac{1}{h}l_ze^{-l_z^2/2}\]

Thus, we want to compute $\max_{l_z>0} \frac{1}{h}l_z e^{-l_z^2/2}e^{-\gamma l_z h}$.
\begin{align*}
    \frac{d}{d l_z} \left(\frac{1}{h}l_z e^{-l_z^2/2}e^{-\gamma l_z h}\right) &= \left(e^{-l_z^2/2}e^{-\gamma l_z h} - l_z^2 e^{-l_z^2/2}e^{-\gamma l_z h} -\gamma h l_z e^{-l_z^2/2}e^{-\gamma l_z h}\right)\frac{1}{h}\\
    &= -(l_z^2+\gamma h l_z - 1)\frac{1}{h}e^{-l_z^2/2}e^{-\gamma l_z h}.
\end{align*}
Setting the above to zero, we get $l_z = \frac{1}{2}(-\gamma h + \sqrt{\gamma^2 h^2 + 4})>0$. 

\paragraph{4. $w = z, z = x$.}
This corresponds to setting $l_z = 0$ in the case above, which yields objective value $0$.

\end{proof}

\subsection{Smooth sensitivity for other functions on $\mathbb{R}^d$}
\label{appendix:lm_compute_smooth_others}

\begin{lemma} [Lemma~\ref{lm:compute_smooth_others}]
Fix $\beta_0\in \mathbb{R}$, $\beta = [\beta_1,\beta_2,\dotsb,\beta_d]\in \mathbb{R}^d$ such that $\|\beta\|\neq 0$. 
Let $f:\mathbb{R}^d \rightarrow \mathbb{R}$ be defined by $f(x)=f_0(\varphi(x))$, where $f_0: \mathbb{R} \rightarrow \mathbb{R}$ and $\varphi(x):=\beta_0 + \langle \beta, x\rangle$.
    Let $(w^*,z^*)$ be such that $B^*(x)=\frac{|f(w^*)-f(z^*)|}{\|w^*-z^*\|g(x,z^*)}$. 
    Then, there is $(w_a, z_c)$ where $z_c:=x+\frac{c}{\|\beta\|}\beta$, $w_a:=z_c+\frac{a}{\|\beta\|}\beta$ for some $a, c\in \mathbb{R}$ such that $\frac{|f(w_a)-f(z_c)|}{\|w_a-z_c\|g(x,z_c)}\ge B^*(x)$.
\end{lemma}

\begin{proof}
    It suffices to show there are $a, c\in \mathbb{R}$ such that $\varphi(w_a) = \varphi(w^*)$, $\varphi(z_c) = \varphi(z^*)$ (i.e. $f(w_a)=f(w^*)$, $f(z_c)=f(z^*)$), $\|w_a-z_c\|\le \|w^*-z^*\|$ and $\|z_c-x\|\le \|z^*-x\|$. Let $a := \frac{\varphi(w^*)-\varphi(z^*)}{\|\beta\|}$, $c:=\frac{\varphi(z^*)-\varphi(x)}{\|\beta\|}$. Then,
    \begin{align*}
        \varphi(z_c) &= \beta_0 + \langle \beta, z_c \rangle = \beta_0 + \langle \beta, x \rangle +\langle \beta,\frac{c}{\|\beta\|}\beta \rangle = \varphi(x) +\frac{c}{\|\beta\|}\langle \beta,\beta \rangle\\
        &= \varphi(x) +{c}{\|\beta\|} 
        = \varphi(x) + \frac{\varphi(z^*)-\varphi(x)}{\|\beta\|}\cdot \|\beta\| = \varphi(z^*), \\
        \varphi(w_a) &= \beta_0 + \langle \beta, w_a \rangle = \beta_0 + \langle \beta, z_c \rangle +\langle \beta,\frac{a}{\|\beta\|}\beta \rangle = \varphi(z_c) + \frac{a}{\|\beta\|} \langle \beta, \beta \rangle \\
        &= \varphi(z^*) + a\|\beta\| = \varphi(z^*) + \frac{\varphi(w^*)-\varphi(z^*)}{\|\beta\|}\cdot \|\beta\| = \varphi(w^*),
        \\
        \|w_a-z_c\| &=\left\|\frac{a}{\|\beta\|}\beta\right\|=\left\| \frac{\varphi(w^*)-\varphi(z^*)}{\|\beta\|^2}\beta\right\|= \frac{|\varphi(w^*)-\varphi(z^*)|}{\|\beta\|}
        = \frac{|\langle \beta, w^*\rangle -\langle \beta, z^*\rangle|}{\|\beta\|} \\
        &= \frac{|\langle \beta, w^*-z^*\rangle |}{\|\beta\|} 
        \le \frac{\|\beta\|\cdot\|w^*-z^*\|}{\|\beta\|} = \|w^*-z^*\|,\\
        \|z_c-x\| &= \left\|\frac{c}{\|\beta\|}\beta\right\| = \left\| \frac{\varphi(z^*)-\varphi(x)}{\|\beta\|^2}\beta\right\|= \frac{|\varphi(z^*)-\varphi(x)|}{\|\beta\|}
        = \frac{|\langle \beta, z^*\rangle -\langle \beta, x\rangle|}{\|\beta\|} \\
        &= \frac{|\langle \beta, z^*-x\rangle |}{\|\beta\|} 
        \le \frac{\|\beta\|\cdot\|z^*-x\|}{\|\beta\|} = \|z^*-x\|.
    \end{align*}
\end{proof}
\newpage
\section{Additional Experiments}
\subsection{One-way Threshold}
\label{appendix:add_1way_exp}
\begin{figure}[htbp]
     \centering
         \begin{subfigure}[t]{0.323\linewidth}%
            \centering
            \includegraphics[width=\textwidth]{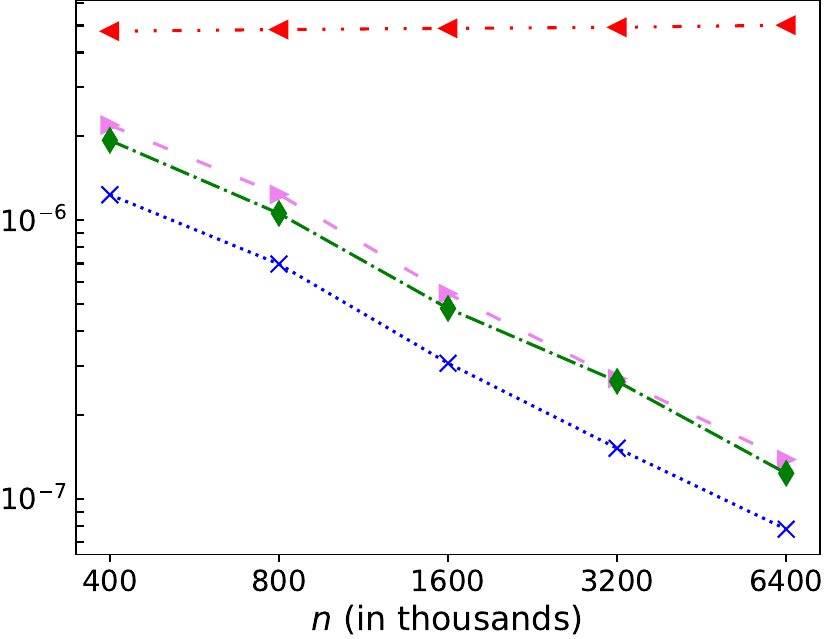}
             \vskip -.08in
            \subcaption{\scriptsize{$T=\$100000, \varepsilon =1/\$8000, \\
            {\;\;\;\;\;\;\;} C=\$12000$.}}
            \;
            \label{fig:1way_thres_exp_T1100klp}
         \end{subfigure}
        \hfill
         \begin{subfigure}[t]{0.323\linewidth}%
            \centering
            \includegraphics[width=\textwidth]{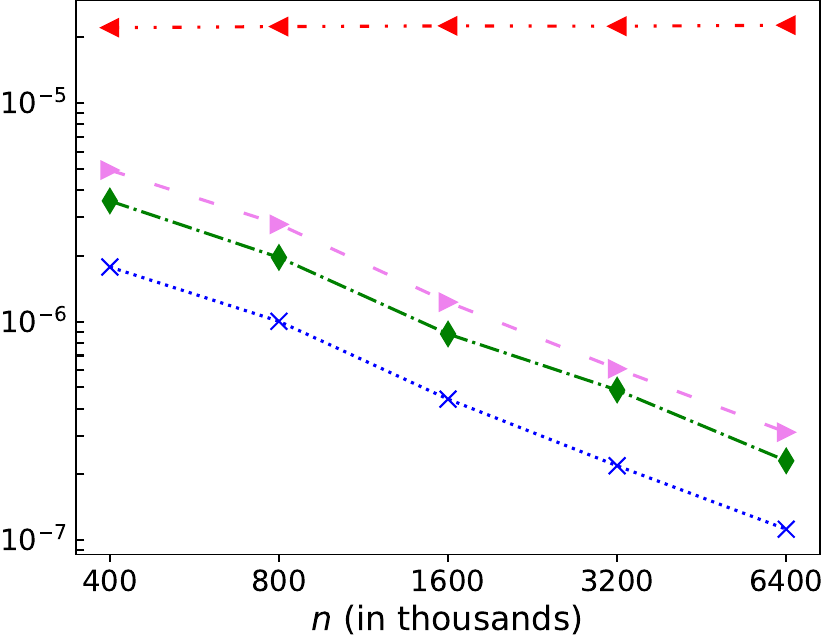}
            \vskip -.08in
            \subcaption{\scriptsize{$T=\$100000, \varepsilon =1/\$12000,\\
            {\;\;\;\;\;\;\;} C=\$12000$.}}
            \;
            \label{fig:1way_thres_exp_T1100kmp}
         \end{subfigure}
         \hfill
         \begin{subfigure}[t]{0.323\linewidth}%
            \centering
            \includegraphics[width=\textwidth]{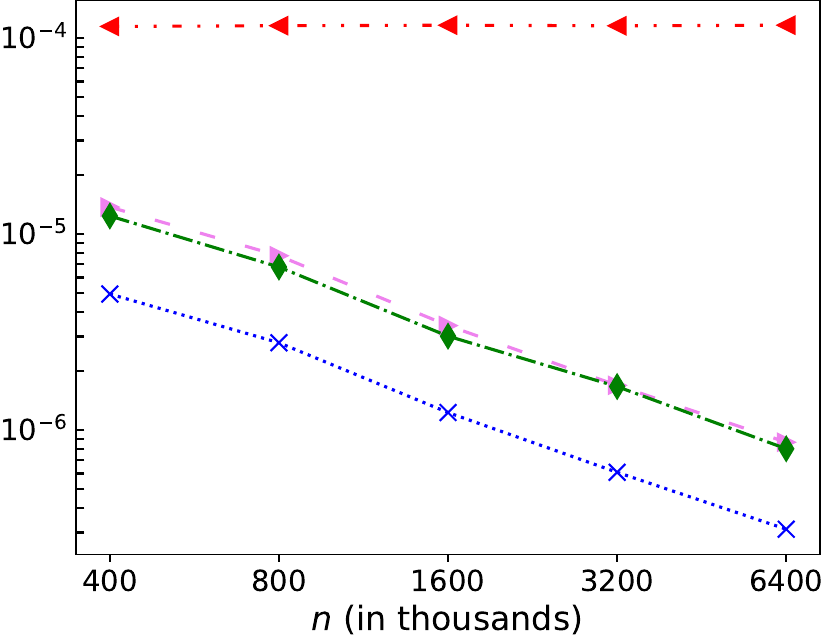}
            \vskip -.08in
            \subcaption{\scriptsize{$T=\$100000, \varepsilon =1/\$20000, \\
            {\;\;\;\;\;\;\;} C=\$12000$.}}
            \label{fig:1way_thres_exp_T1100khp}
         \end{subfigure}
\vfill
         \begin{subfigure}[t]{0.323\linewidth}%
            \centering
            \includegraphics[width=\textwidth]{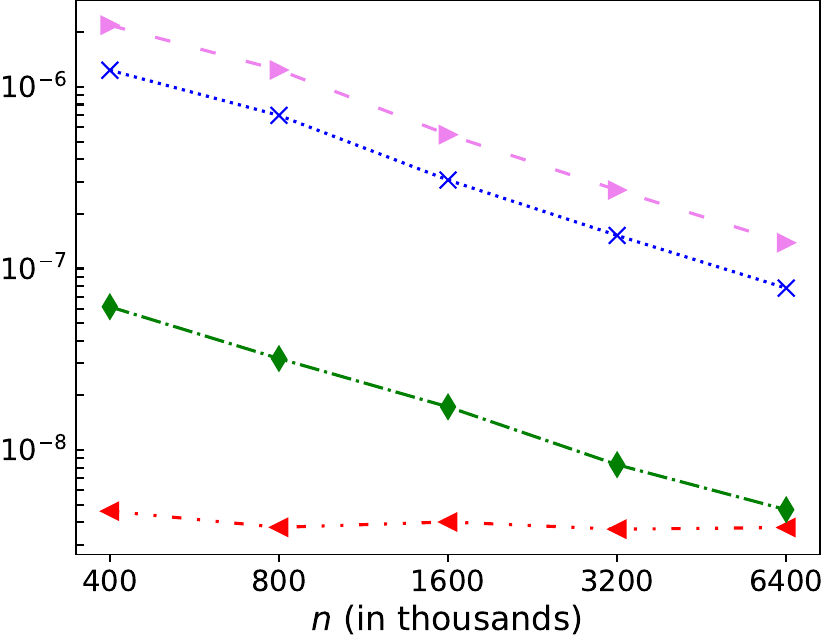}
             \vskip -.08in
            \subcaption{\scriptsize{$T=\$1000000, \varepsilon =1/\$32000,\\ {\;\;\;\;\;\;\;} C=\$48000$.}}
            \;
            \label{fig:1way_thres_exp_T1millp}
         \end{subfigure}
        \hfill
         \begin{subfigure}[t]{0.323\linewidth}%
            \centering
            \includegraphics[width=\textwidth]{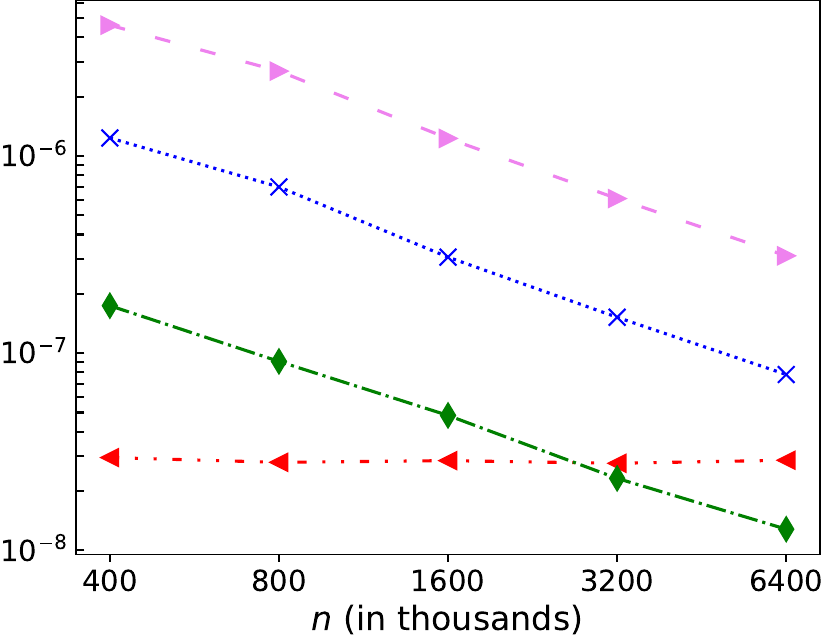}
            \vskip -.08in
            \subcaption{\scriptsize{$T=\$1000000, \varepsilon =1/\$48000, \\
            {\;\;\;\;\;\;\;} C=\$48000$.}}
            \;
            \label{fig:1way_thres_exp_T1milmp}
         \end{subfigure}
         \hfill
         \begin{subfigure}[t]{0.323\linewidth}%
            \centering
            \includegraphics[width=\textwidth]{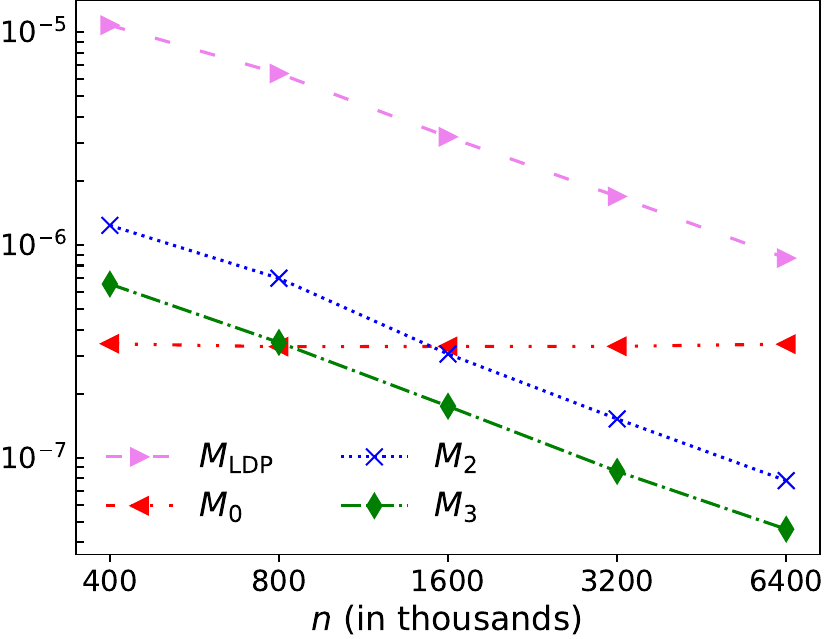}
            \vskip -.08in
            \subcaption{\scriptsize{$T=\$1000000, \varepsilon =1/\$80000, \\
            {\;\;\;\;\;\;\;} C=\$48000$.}}
            \label{fig:1way_thres_exp_T1milhp}
         \end{subfigure}
         \vskip -.08in
         \caption{MSE for one-way threshold query. Threshold amounts in USD.}
         \label{fig:1way_thres_hhincome2}
    \vskip -.01in
\end{figure}

For $T=\$100000$ in Fig.~\ref{fig:1way_thres_exp_T1100klp}-\ref{fig:1way_thres_exp_T1100khp}, there are many $x_i$'s around this level, and we see $M_0$ does not perform well even for larger values of $\varepsilon$. 
When many points are near the area of action, on average smooth sensitivity is similar to the global Lipschitz constant (leading to an overall error in $M_3$ that is to a constant factor worse than that in $M_2$).

For $T=\$1000000$ in Fig.~\ref{fig:1way_thres_exp_T1millp}-\ref{fig:1way_thres_exp_T1milhp}, most of $x_i$'s are far from the area of action, and we see similar relative performances in the mechanisms as those for $T=\$10000$.

\newpage
\subsection{Gaussian KDE}
\label{appendix:add_kde_exp}

\begin{figure}[H]
    \vskip -.1in
     \centering
         \begin{subfigure}[t]{0.9\textwidth}
            \centering
            \includegraphics[width=\textwidth]{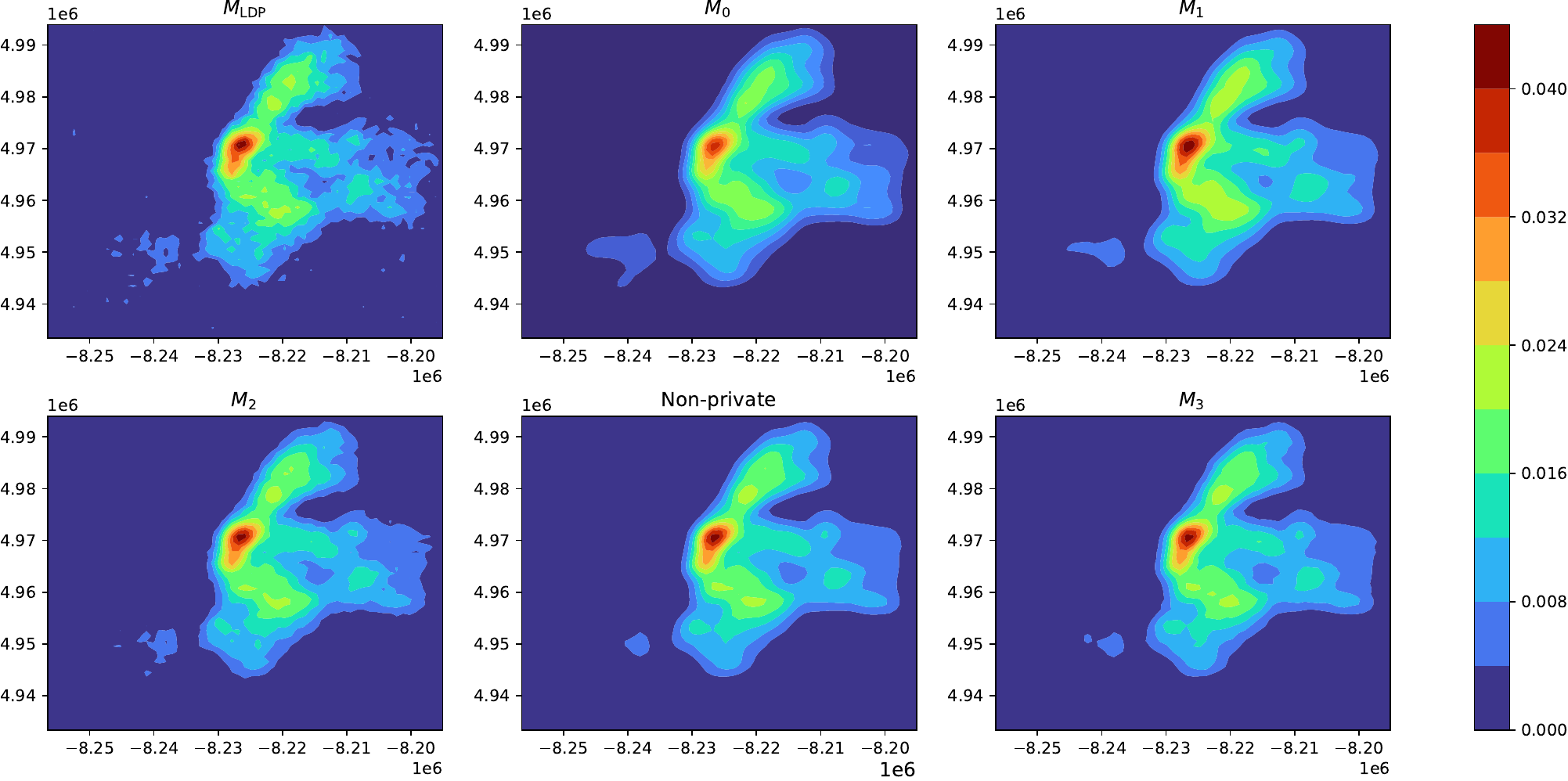}
             \vskip -.08in
            \subcaption{$\varepsilon=1/500\mathrm{m}$.}
            \label{fig:kde_large_eps}
            \;
         \end{subfigure}
        \vfill
         \begin{subfigure}[t]{0.9\textwidth}
            \centering
            \includegraphics[width=\textwidth]{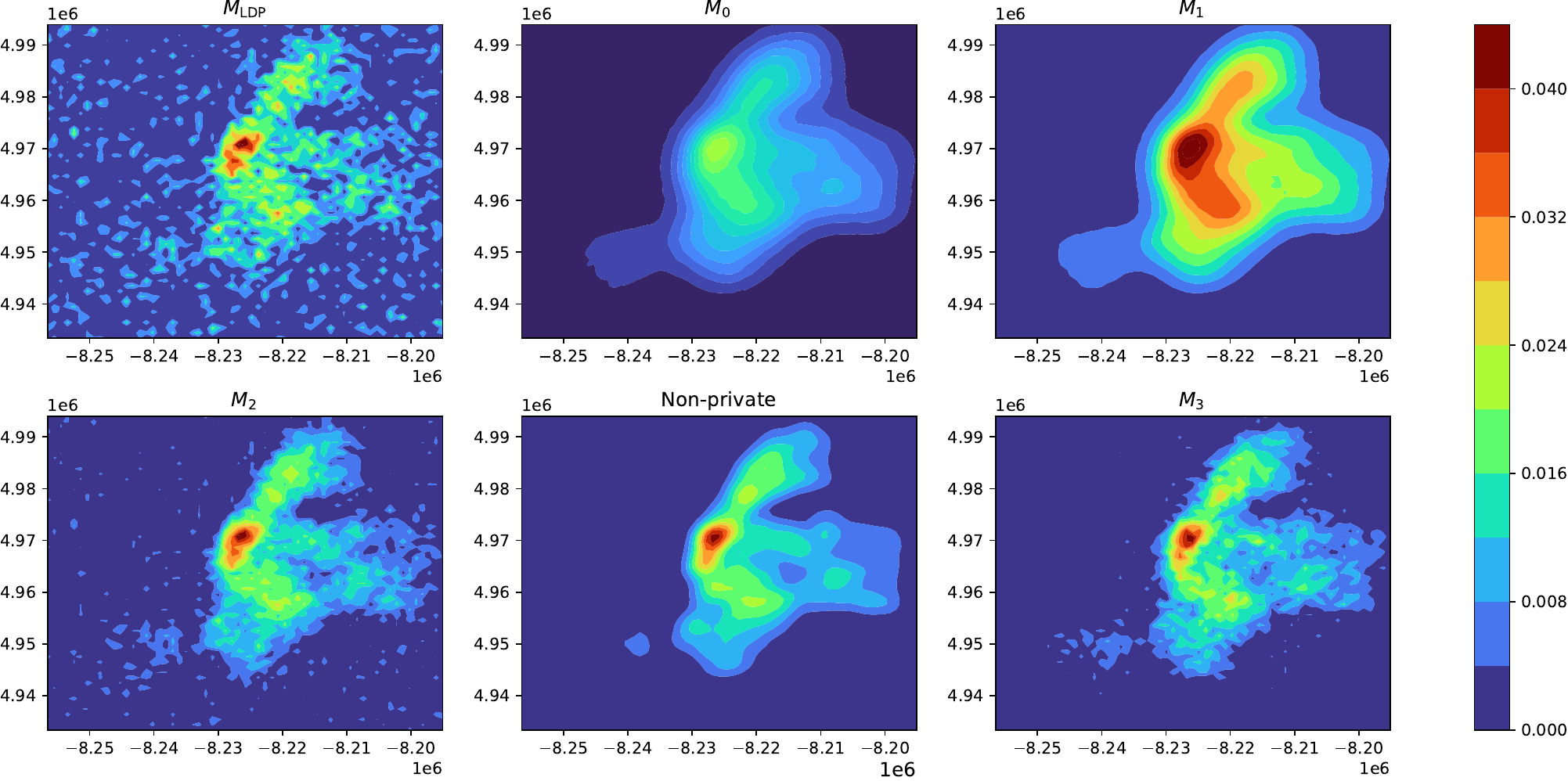}
            \vskip -.08in
            \subcaption{$\varepsilon=1/2000\mathrm{m}$.}
            \label{fig:kde_small_eps}
         \end{subfigure}
         \vskip -.08in
         \caption{KDE on New York motor vehicle collision dataset, computed on $60\times 60$ grid. $h=w, n=200000$}
         \label{fig:kde_nymvc_epsilon}
    \vskip -.05in
\end{figure}

In Fig.~\ref{fig:kde_nymvc_epsilon} (with corresponding error plots Fig.~\ref{fig:kde_bar_lp}-\ref{fig:kde_bar_hp}), we examine the error w.r.t. to the privacy level $\varepsilon$, where we fix $h=w, n=200000$. We see that all mechanisms have improved error for the larger value of $\varepsilon$ in Fig.~\ref{fig:kde_large_eps}. For the smaller value of $\varepsilon$ in Fig.~\ref{fig:kde_small_eps}, we see that the weaknesses of the mechanisms become more pronounced: $M_0$ is more negatively biased, $M_1$ is more positively biased, while $M_2$ and $M_3$ become more noisy. Overall, $M_3$ still has the best performance.

\begin{figure}[H]
    \vskip -.1in
     \centering
         \begin{subfigure}[t]{0.9\linewidth}
            \centering
            \includegraphics[width=\textwidth]{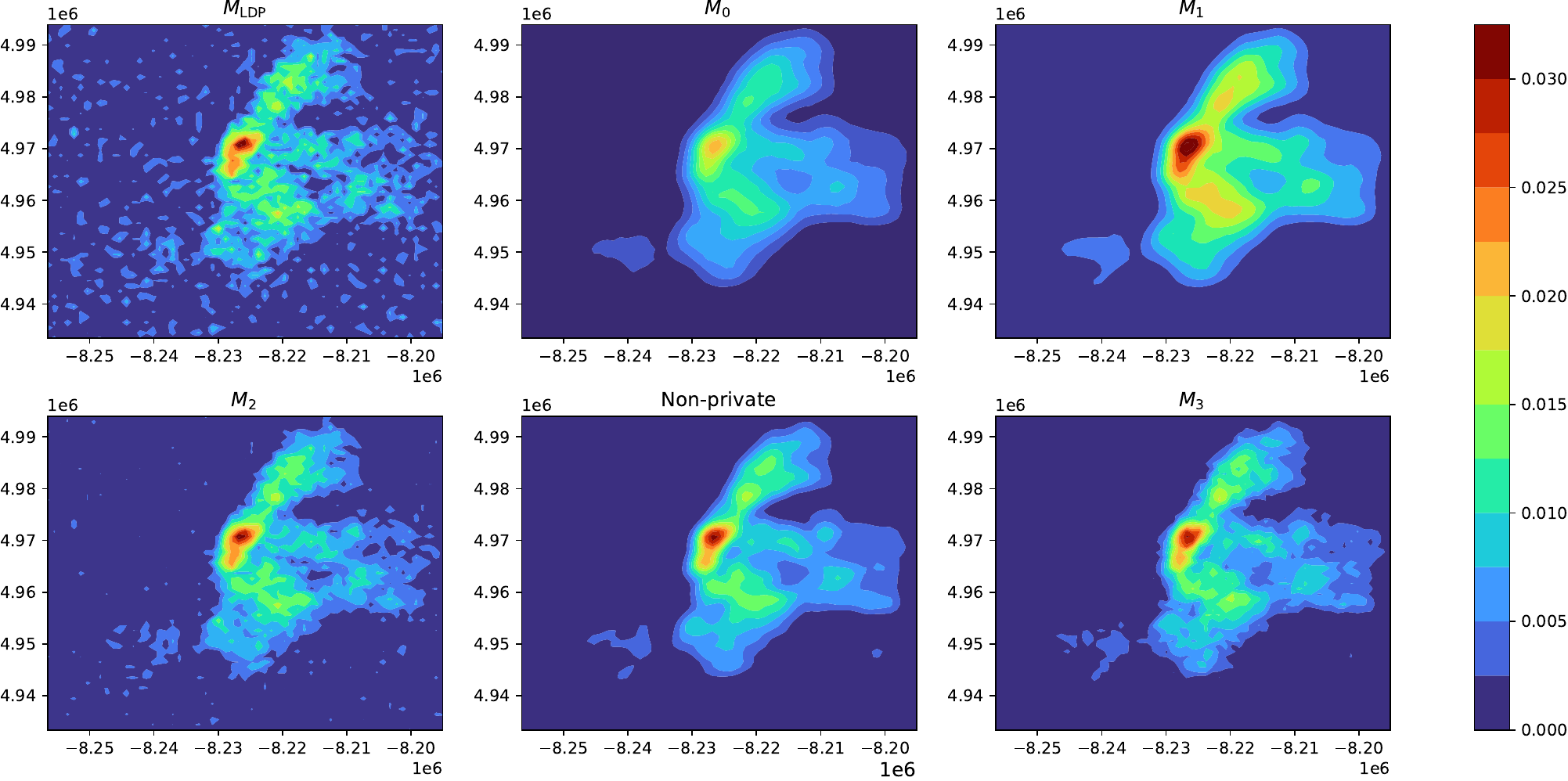}%
             \vskip -.08in
            \subcaption{$h = 0.8w$.}
            \label{fig:kde_smallh}
            \;
         \end{subfigure}
         \vfill
         \begin{subfigure}[t]{0.9\linewidth}
            \centering
            \includegraphics[width=\textwidth]{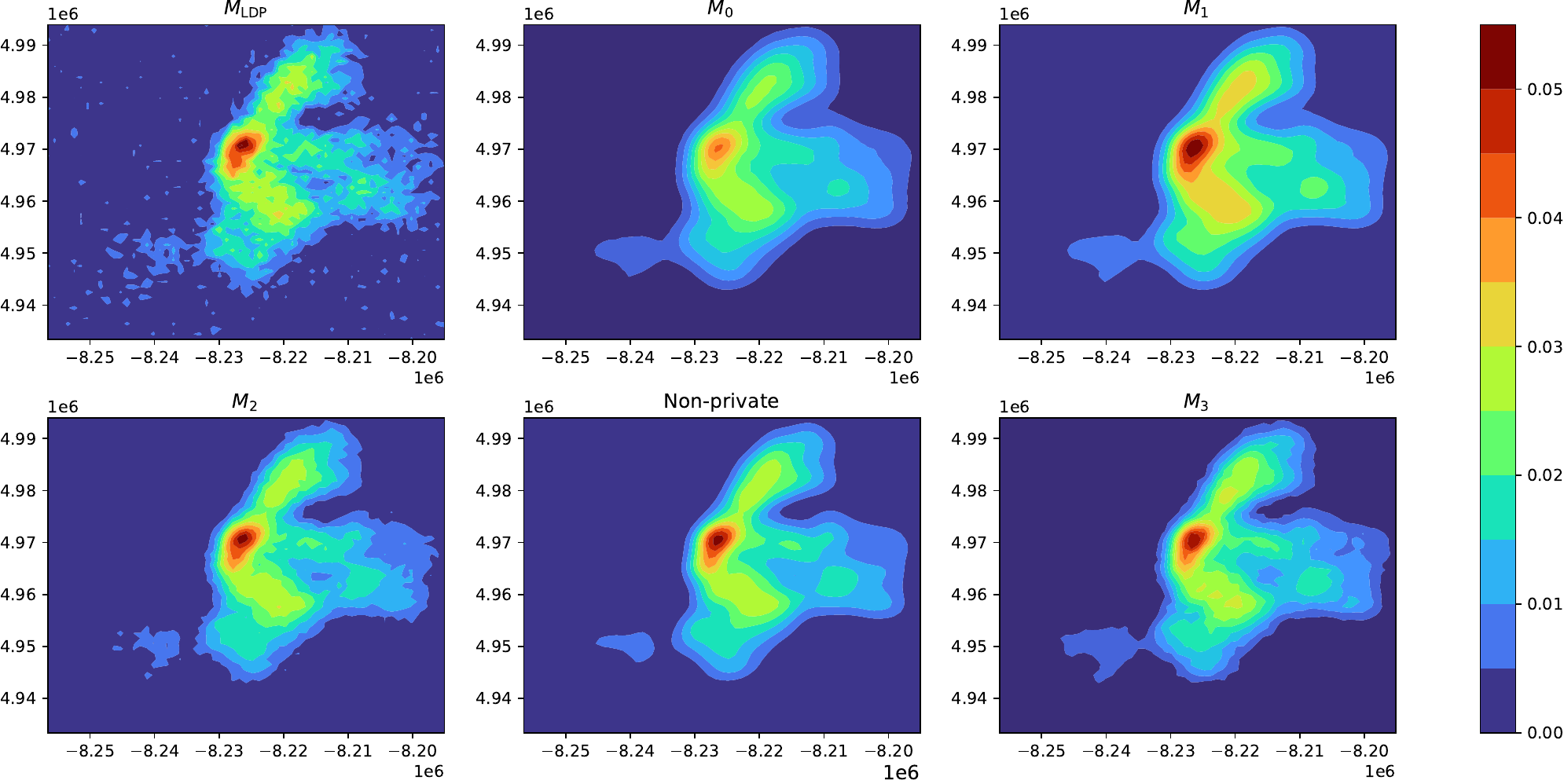}%
            \vskip -.08in
            \subcaption{$h=1.2w$.}
            \label{fig:kde_largeh}
         \end{subfigure}
         \vskip -.1in
         \caption{KDE on New York motor vehicle collision dataset, computed on $60\times 60$ grid.  $\varepsilon=1/1000\mathrm{m}, n=200000$}
         \label{fig:kde_nymvc_h}
    \vskip -.01in
\end{figure}

In Fig.~\ref{fig:kde_nymvc_h} (with corresponding error plots Fig.~\ref{fig:kde_bar_smallh}-\ref{fig:kde_bar_largeh}), we examine the error w.r.t. to the bandwidth size $h$, where we fix $ \varepsilon=1/1000\mathrm{m}, n=200000$. We observe that $M_2$ and $M_3$ have smaller error for the larger value of $h$, while $M_0$ and $M_1$ have smaller error for the smaller value of $h$.

\begin{figure}[htbp]
    \vskip -.1in
     \centering
         \begin{subfigure}[t]{0.9\textwidth}
            \centering
            \includegraphics[width=\textwidth]{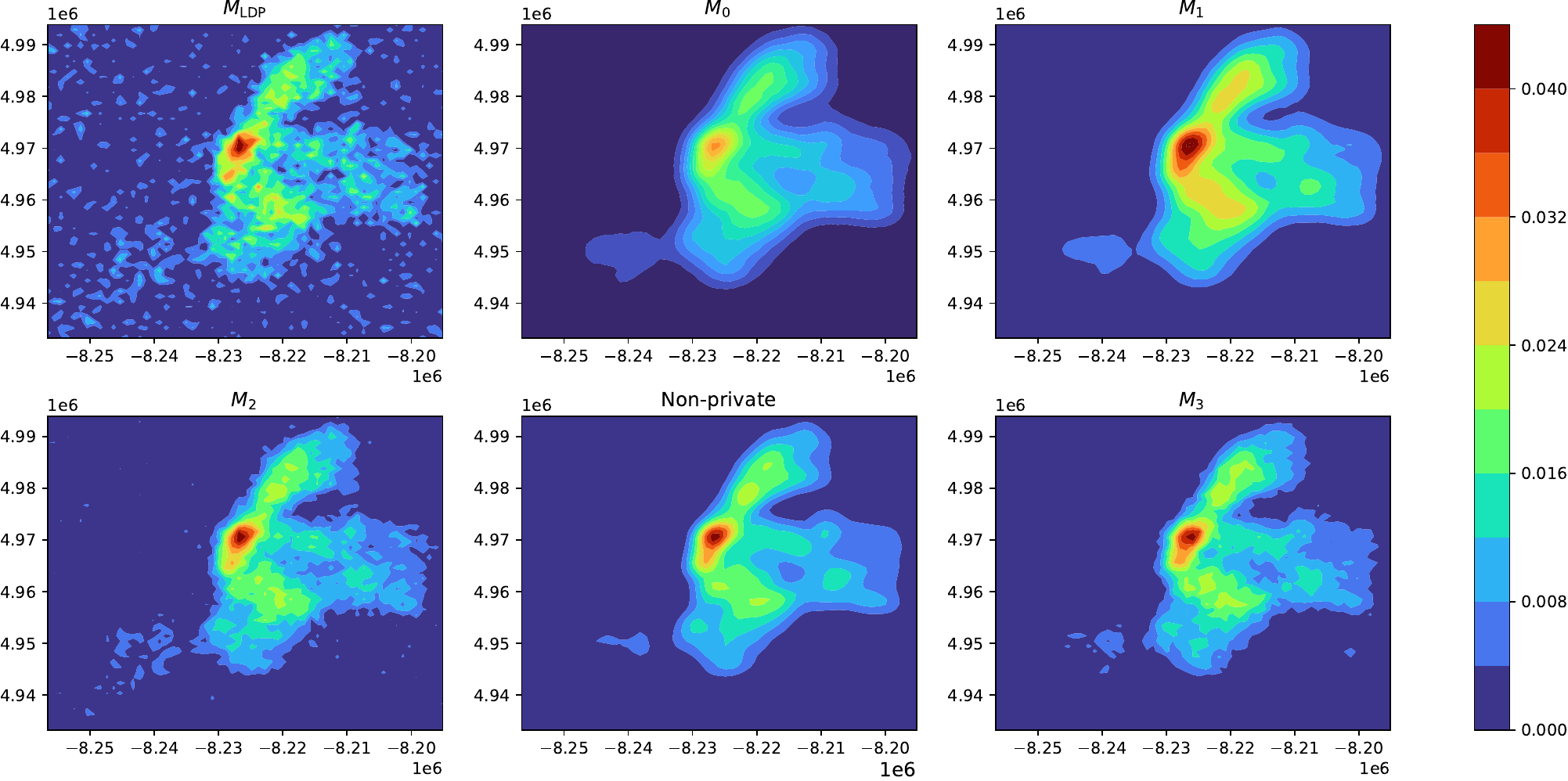}
             \vskip -.08in
            \subcaption{$n=100000$.}
            \;
         \end{subfigure}
        \vfill
         \begin{subfigure}[t]{0.9\textwidth}
            \centering
            \includegraphics[width=\textwidth]{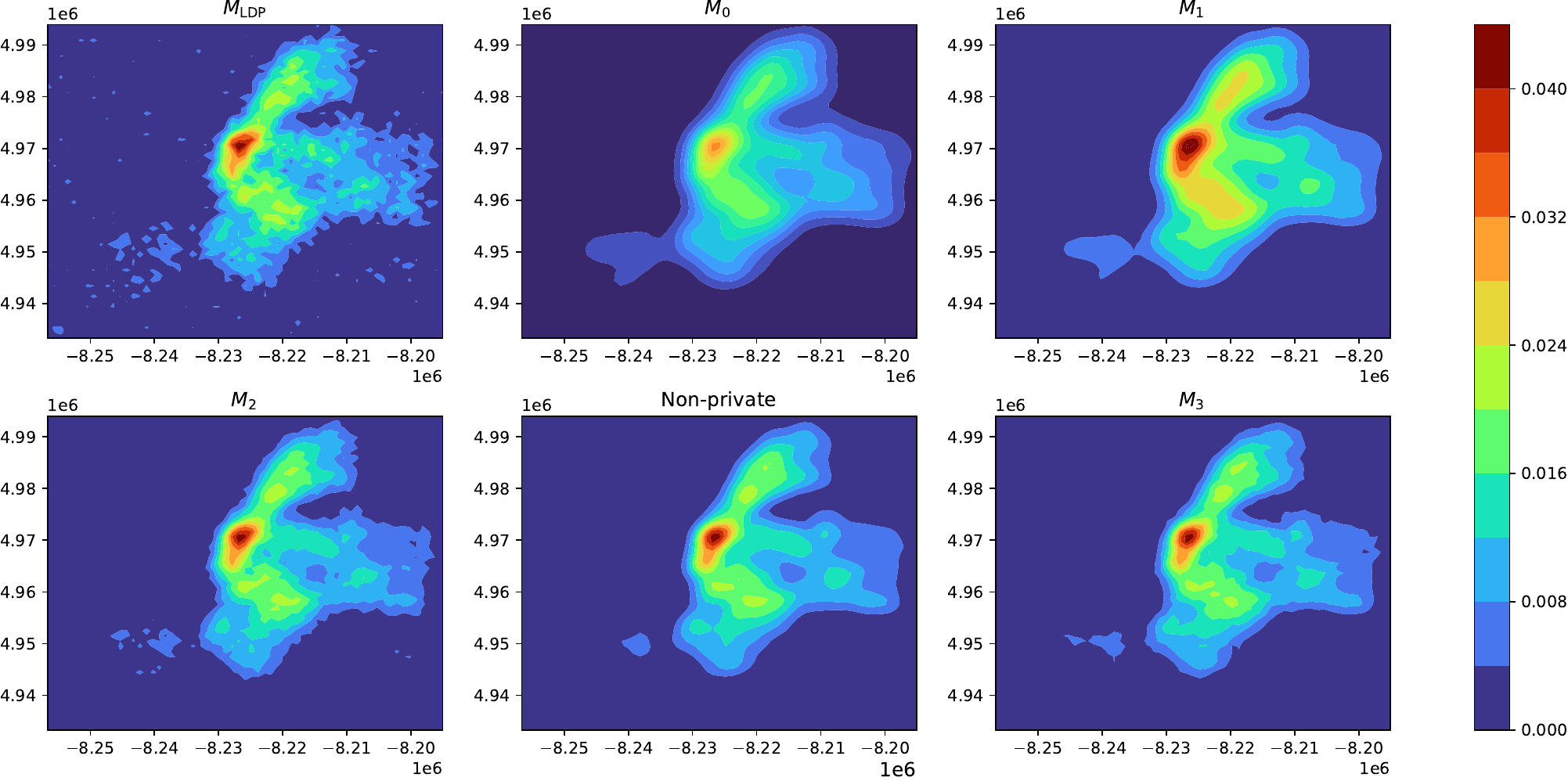}
            \vskip -.08in
            \subcaption{$n=400000$.}
         \end{subfigure}
         \vskip -.08in
         \caption{KDE on New York motor vehicle collision dataset, computed on $60\times 60$ grid. $h=w, \varepsilon=1/1000\mathrm{m}$.}
         \label{fig:kde_nymvc_n}
    \vskip -.05in
\end{figure}

In Fig.~\ref{fig:kde_nymvc_n} (with corresponding error plots Fig.~\ref{fig:kde_bar_smalln}-\ref{fig:kde_bar_largen}), we examine the error w.r.t. the sample size $n$, where we fix $\varepsilon=1/1000\mathrm{m}, h=w$. We see that as $n$ increases, both $M_2$ and $M_3$ have reduced error. $M_0$ and $M_1$ do not benefit from a larger $n$ as their errors are dominated by the bias.

\textbf{A high-level explanation for the results.} 
(1). The contour plot for $M_2$ appears quite noisy in Fig.~\ref{fig:kde_smallh} which corresponds to a smaller value of $h$. Since $M_2$ applies noise of the same magnitude to every point, when the noise magnitude is large compared to the actual KDE value, the privatized output will be noisy. This happens exactly when the KDE values are small ($x_i$'s are far from the query point $t$). On the other hand, $M_3$ would have smaller smooth sensitivity for $x_i$'s far away, and thus gives outputs that are less noisy.

(2). The contour plots for $M_0$ appear to show consistently smaller KDE values. Consider applying $M_0$ to a point $x\in \mathbb{R}^d$, $M_0(x)=e^{-\frac{\|x+Z-t\|^2}{2h^2}}=e^{\frac{-\|x-t\|^2-2\langle x-t,Z\rangle -\|Z\|^2}{2h^2}}$, where $\mathbb{E}[\|Z\|]=\theta(d/\varepsilon)$ since $Z$ is drawn from the Planar Laplace mechanism with scale $1/\varepsilon$. Thus, $M_0(x)$ is negatively biased. Also, note that $\mathbb{E}[M_0(x)-f(x)]=e^{\frac{-\|x-t\|^2}{2h^2}}\left(e^{\frac{-2\langle x-t,Z\rangle -\|Z\|^2}{2h^2}}-1\right)=f(x)\left(e^{\frac{-2\langle x-t,Z\rangle -\|Z\|^2}{2h^2}}-1\right)$, so the error will be proportional to the actual value of $f(x)$. Thus, we see in Fig.~\ref{fig:kde_nymvc_h} (error plots Fig.~\ref{fig:kde_bar_smallh}-\ref{fig:kde_bar_largeh}) a small increase in the error as $h$ increases since $f(x)$ increases with $h$.

(3). The contour plots for $M_1$ appear to show consistently larger KDE values. Consider $M_1(x)=e^{-\frac{(\|x-t\|+Z)^2}{2h^2}}=e^{\frac{-\|x-t\|^2+2Z\cdot\|x-t\|+Z^2}{2h^2}}$ where $Z\sim\mathrm{Lap}(0,1/\varepsilon)$. $\mathbb{E}\left[e^{-\frac{2Z\cdot\|x-t\|}{2h^2}}\right]=\frac{1}{1-\frac{\|x-t\|^2}{\varepsilon^2 h^4}}$ for $\frac{\|x-t\|}{\varepsilon h^2} < 1$, while $\mathbb{E}\left[e^{-\frac{Z^2}{2h^2}}\right] \ge e^{-\frac{1}{h^2 \varepsilon^2}}$. Thus $\mathbb{E}\left[M_1(x)\right] \ge e^{-\frac{\|x-t\|^2}{2h^2}}e^{\frac{1}{\varepsilon^2 h^2}(\frac{\|x-t\|^2}{h^2}-1)}$, i.e. $M_1(x)$ is positively biased for $\|x-t\|>h$ and $\varepsilon h > 1$. $\mathbb{E}[M_1(x)-f(x)]$ is also proportional to the value of $f(x)$, and we also see in Fig.~\ref{fig:kde_nymvc_h} (error plots Fig.~\ref{fig:kde_bar_smallh}-\ref{fig:kde_bar_largeh}) a small increase in the error as $h$ increases.

\section{Other Technical Details}

\subsection{Generating $\mathrm{GenCauchy}(a,b,4,1)$ random variables.}
\label{appendix:gen_cauchy4}
Recall that for any random variable $Z$ with cdf $F$, we can use the inverse transformation method for drawing samples. Specifically, the random variable $F^{-1}(U)$ has the same cdf $F$, where $U\sim \mathrm{Unif}(0,1)$.
For $Z\sim \mathrm{GenCauchy}(0,1,4,1)$, its pdf is givey by $h(z) = \frac{\sqrt{2}}{\pi}\cdot \frac{1}{1+z^4}$ for $z\in \mathbb{R}$.
Let 
\[H(q):=\int_0^q \frac{1}{1+z^4} dz = \frac{1}{4\sqrt{2}}\left(\ln\left(\frac{q^2+\sqrt{2}q+1}{q^2-\sqrt{2}q+1}\right)+2\tan^{-1}\left(1+\sqrt{2}q\right)-2\tan^{-1}\left(1-\sqrt{2}q\right)\right).\]
Then its cdf of $Z$ at $q$ is
\[
F(q) = \Pr[Z\le q] = \int_{-\infty}^q \frac{\sqrt{2}}{\pi}\cdot \frac{1}{1+z^4} dz = \begin{cases}
\frac{1}{2} - \frac{\sqrt{2}}{\pi}\cdot H(|q|), \;\;\; &{q<0}\\
\frac{1}{2} + \frac{\sqrt{2}}{\pi}\cdot H(q), \;\;\; &\text{otherwise}.
\end{cases}
\]
Since the cdf function on the reals is monotonically increasing, it admits a unique inverse. Thus, for $u\in (0,1)$ we can compute $F^{-1}(u)$ by solving for $q$ in $F(q)=u$. 
Then, $a+b\cdot F^{-1}(U) \sim \mathrm{GenCauchy}(a,b,4,1)$.

\subsection{On the smooth upper bound}
\label{appendix:smooth_upper_finite}
Condition (2) of Definition~\ref{def:smoothupper_plc} can be also replaced with the following, where $\lambda \ge \Lambda$.
\begin{enumerate}
    \item[(2)] $\forall x\sim_{\lambda} x'\in U: B(x)\le g(x,x')\cdot B(x')$. 
\end{enumerate}
Since $\lambda \ge \Lambda$, this is sufficient for the smooth sensitivity mechanisms. However, this relaxation impacts the the proof of Lemma~\ref{lm:smoothsens_smallest_bound}; with this relaxed condition, it's unclear whether smooth sensitivity defined in equation~\eqref{eqn:smooth_sens_GP} with a general growth function $g$ would still give the smallest smooth upper bound for $\LC_{f,\Lambda}$. Fortunately, for $\gE$, this is still the case.

\begin{lemma} Let  $0 < \Lambda < \infty$. For $\Lambda$-locally Lipschitz function $f:U\rightarrow \mathbb{R}$, let $B$ be a $\gE$-smooth upper bound (with the relaxed condition above) for $\LC_{f,\Lambda}$ and let $B^*$ be the smooth sensitivity function given in equation~\eqref{eqn:smooth_sens_GP} computed with $\gE$. Then $B(x)\ge B^*(x)$.
\end{lemma}
\begin{proof} Fix any $x\in U$. It suffices to show that  
    $B(x)\ge \frac{\LC_{f,\Lambda}(z)}{\gE(x,z)}$ for all $z\in U$. Let any $z\in U$ be given and let $r:=\dist(z,x)\ge 0$. For $r=0$, we have $B(x)\ge \LC_{f,\Lambda}(x) = \frac{\LC_{f,\Lambda}(z)}{\gE(z,z)}$. For $r>0$, let $z_0:=x,z_1,\dotsb,z_k:=z$ be points in $U$ such that $\dist(z,x)=\sum_{j=1}^k \dist(z_{j},z_{j-1})$ and $\dist(z_{j},z_{j-1})\le \lambda$, for some $k\ge 1$. (i.e., the $z_j$'s are points on the line connecting $x$ and $z$, at most $\lambda$ apart.). Since adjacent $z_j$'s are at most $\lambda$ apart, $B(z_{j-1}) \ge \frac{B(z_{j})}{\gE(z_j,z_{j-1})}$ for $j\in [k]$. Also, $\prod_{j=1}^k\gE(z_j,z_{j-1})=\prod_{j=1}^k e^{\gamma\dist(z_j,z_{j-1})}=e^{\gamma \dist(z,x)}$.
    Then,
    \begin{align*}
        B(x) = B(z_0) &\ge \frac{B(z_{1})}{\gE(z_1,z_{0})} \ge \frac{B(z_{2})}{\gE(z_2,z_{1})\cdot \gE(z_1,z_{0})}\ge \dotsb 
        \ge \frac{B(z_{k})}{\prod_{j=1}^k \gE(z_j,z_{j-1})}=\frac{B(z)}{e^{\gamma\dist(z,x)}}\ge \frac{\LC_{f,\Lambda}(z)}{\gE(z,x)}.
    \end{align*}
\end{proof}
\subsection{Smooth sensitivity for 1D bounded range queries}
\label{appendix:smoothsens_1Drange}
Give $l<r\in \mathbb{R}$, let $f:x\mapsto \mathbb{1}\{l\le x \le r\}$. Then for $\tau > 0$, $f$ can be approximated by the following Lipschitz function:
\begin{align*}
f_{\tau}(x)\begin{cases}
0, &{}x < l-\tau/2\\
0, &{}x > r+\tau/2\\
\frac{x-l+\tau/2}{(r-l)/2+\tau/2}, &{}l-\frac{\tau}{2}\le x \le m\\
\frac{r+\tau/2-x}{(r-l)/2+\tau/2}, &{}m\le x \le r+\tau/2.
\end{cases}
\end{align*}
Using arguments similar to those in the proof of Lemma~\ref{lm:1way_thres_sens}, one can show that, for $\gE:(x,x')\mapsto e^{\gamma\dist(x,x')}$, the $\gE$-smooth sensitivity for $\LC_{f,\infty}$ at $x$ can by computed as
\begin{align*}
B^*(x)=\begin{cases}
\max\left(\frac{1}{\left|(l+r)/2-x\right|}, \frac{1}{(r-l)/2+\tau/2}\cdot e^{-\gamma|l-\tau/2-x|}\right), &{}x<l-\tau/2\\
\max\left(\frac{1}{\left|(l+r)/2-x\right|}, \frac{1}{(r-l)/2+\tau/2}\cdot e^{-\gamma|r+\tau/2-x|}\right), &{}x>r+\tau/2\\
\frac{1}{(r-l)/2+\tau/2}, &{x\in[l-\tau/2,r+\tau/2]}.
\end{cases}
\end{align*}
\end{document}